\documentclass{aa}
\usepackage{natbib}
\usepackage{color}
\usepackage{mathtools}
\usepackage[colorlinks=true,citecolor=blue]{hyperref} 
\usepackage{multirow}
\usepackage{graphicx}
\usepackage{txfonts}

\def    \apjl           {\rm {ApJL}}

\def    \apjl           {\rm {ApJL}}

\def    \cm             {\,{\rm {cm}}}
\def    \K              {\,{\rm {K}}}
\def    \g              {\,{\rm {g}}}
\def    \mum    {\,{\mu \rm{m}}}

\def \bea {\begin{eqnarray}}
\def \ena {\end{eqnarray}}

\def    \cm     {\,{\rm cm}}

\def    \g      {\,{\rm g}}

\def    \G      {{\rm G}}

\def    \pc     {\,{\rm pc}}

\def    \s      {\,{\rm s}}

\def    \au     {\,{\rm au}}

\graphicspath{{./Figures/}}  

\begin{document}

\title{Accuracy of ALMA estimates of young disk radii and masses}
\subtitle{Predicted observations from numerical simulations}
\author{Ngo-Duy Tung \inst{1,2}
        \and Leonardo Testi \inst{3,4}
        \and Ugo Lebreuilly \inst{1}
        \and Patrick Hennebelle \inst{1}
        \and Ana\"{e}lle Maury \inst{1}
        \and Ralf S.\ Klessen \inst{5,6}
        \and Luca Cacciapuoti \inst{7,4,8}
        \and Matthias González \inst{2}
        \and Giovanni Rosotti \inst{9,10}
        \and Sergio Molinari \inst{11}}

\institute{Université Paris-Saclay, Université Paris Cité, CEA, CNRS, AIM, 91191, Gif-sur-Yvette, France\\
\email{duy-tung.ngo@cea.fr}
\and Université Paris Cité, Université Paris-Saclay, CEA, CNRS, AIM, F-91191, Gif-sur-Yvette, France
\and
Alma Mater Studiorum Università di Bologna, Dipartimento di Fisica e Astronomia (DIFA), Via Gobetti 93/2, I-40129, Bologna,
Italy
\and INAF-Osservatorio Astrofisico di Arcetri, Largo E. Fermi 5, I-50125, Firenze, Italy
\and Universit\"{a}t Heidelberg, Zentrum f\"{u}r Astronomie, Institut f\"{u}r Theoretische Astrophysik, Albert-Ueberle-Str. 2, 69120 Heidelberg, Germany  
\and Universit\"{a}t Heidelberg, Interdisziplin\"{a}res Zentrum f\"{u}r Wissenschaftliches Rechnen, Im Neuenheimer Feld 205, 69120 Heidelberg, Germany
\and European Southern Observatory, Karl-Schwarzschild-Strasse 2, 85748 Garching bei M\"{u}nchen, Germany
\and Fakultat f\"{u}r Physik, Ludwig-Maximilians-Universit\"{a}t M\"{u}nchen, Scheinerstr. 1, 81679, M\"{u}nchen, Germany
\and Dipartimento di Fisica, Università degli Studi di Milano, Via Giovanni Celoria, 16, 20133, Milano, Italy
\and Leiden Observatory, Leiden University, Niels Bohrweg 2, NL-2333 CA Leiden, The Netherlands
\and INAF-Istituto di Astrofisica e Planetologia Spaziali (INAF-IAPS), Via Fosso del Cavaliere 100, I-00133, Roma, Italy}

\date{}

 
\abstract
{Protoplanetary disks, which are the natural consequence of the gravitational collapse of the dense molecular cloud cores, host the formation of the known planetary systems in our universe. Substantial efforts have been dedicated to investigating the properties of these disks in the more mature Class II stage, either via numerical simulations of disk evolution from a limited range of initial conditions or observations of their dust continuum and line emission from specific molecular tracers.  The results coming from these two standpoints have been used to draw comparisons. However, few studies have investigated the main limitations at work when measuring the embedded Class 0/I disk properties from observations, especially in a statistical fashion.}
{In this study, we provide a first attempt to compare the accuracy of some critical disk parameters in Class 0/I systems, as derived on real ALMA observational data, with the corresponding physical parameters that can be directly defined by theoreticians and modellers in numerical simulations. The approach we follow here is to provide full post-processing of the numerical simulations and apply it to the synthetic observations the same techniques used by observers to derive the physical parameters.}
{We performed 3D Monte Carlo radiative transfer and mock interferometric observations of the disk populations formed in a magnetohydrodynamic (MHD) simulation model of disk formation through the collapse of massive clumps with the tools \textsc{Radmc-3d} and \textsc{Casa}, respectively, to obtain their synthetic observations. With these observations, we re-employed the techniques commonly used in disk modelling from their continuum emissions to infer the properties that would most likely be obtained with real interferometers. We then demonstrated how these properties may vary with respect to the gas kinematics analyses and dust continuum modelling.}
{Our modelling procedure, based on a two-component model for the disk and the envelope, shows that the disk sizes can be properly recovered from observations with sufficient angular resolutions, with an uncertainty of a factor $\approx 1.6-2.2$, whereas their masses cannot be accurately measured.  Overall, the masses are predominantly underestimated for larger, more massive disks by a median factor of $\approx 2.5,  $ and even up to $10$ in extreme cases, with the conversion from flux to dust mass under the optically thin assumption. We also find that the single Gaussian fittings are not a reliable modelling technique for young, embedded disks characterised by a strong presence of the envelopes. Thus, such an approach is to be used with caution.}
{The radiative transfer post-processing and synthetic observations of MHD simulations offer genuine help in linking important observable properties of young planet-forming disks to their intrinsic values in simulations. Further extended investigations that tackle the caveats of this study, such as the lack of variation in the dust composition and distribution, dust-to-gas ratio, and other shortcomings in the numerical models, would be essential for setting constraints on our understanding of disk and planet formations.}

\keywords{Synthetic observations --
            Protoplanetary disks --
            Star: formation
           }
           
\titlerunning{Synthetic Observations of Protoplanetary Disk Simulations}
\maketitle
%
\section{Introduction}\label{sec:intro}
The formation of stars and planetary systems starts within a dense molecular cloud core: the high-density region of the cloud where the majority of star formation activities take place thanks to the dominance of self-gravity over gas pressure. This dominance also gives rise to gravitational collapse of the core. In a simplistic picture, the angular momentum conservation of the core during the collapse causes most of the infalling matter to form a circumstellar disk around the protostar, instead of falling directly onto it (see \citealt{1987ARA&A..25...23S}; \citealt{2011ARA&A..49...67W}; \citealt{2023ASPC..534..317T} for reviews). This disk of gas and dust is often referred to as the protostellar disk in the broad sense and also referred to as protoplanetary disk (hereafter PPD) in reference to its role as a (proto-)planet formation site. Throughout all the evolutionary stages of PPDs (and most evident in the later ones), dust grains play a central role in the evolution of PPDs as they grow into larger aggregates, then centimeter-sized pebbles, which eventually become planetesimals, followed by kilometer-sized bodies that make up the building blocks of planetary systems (\citealt{2014prpl.conf..339T}).

Based on their spectral energy distributions (SEDs), the evolution of such star-disk systems can be classified into different stages, ranging from the cold, dense pre-stellar core phase to Class 0-I protostellar phase, where the transition from cores to disks happens. This is followed by the pre-main sequence Class II-II disks, where most of the planet formation activities are widely believed to take place (\citealt{2000prpl.conf...59A}).

So far, much of our understanding of the evolution of PPDs and planet formation is based on theory and observations of isolated, more evolved Class II objects. However, our Sun, as most stars, was not initially isolated; it formed in a stellar cluster environment (\citealt{2010ARA&A..48...47A}; \citealt{2015PhyS...90f8001P}), where the radiation field (photoevaporation) and (to a much lesser extent) tidal interactions from other forming stars strongly influenced the formation of the Earth and other planets in the Solar System (\citealt{2018MNRAS.478.2700W}). 
Thus, the need for modelling star and planet formation in massive star-forming complexes becomes ineluctable. Added to this is the fact that isolated simulations of disk formation and early evolution of disks from single core collapses in different environments often yield different outcomes regarding the disk properties (\citealt{2020A&A...635A..67H}; \citealt{2021A&A...648A.101L}). Therefore, it is of great importance to model these processes starting from the large-scale environment and compare the outputs of these simulations with observations of young stellar objects (YSOs), with an aim to answer the question of whether the physical ingredients in these simulations are sufficient to represent the reality of disks in different galactic environments. In this light, \cite{2021ApJ...917L..10L, 2023arXiv231019672L} recently performed simulations of protoplanetary disk formation, starting from the collapse of massive star-forming clumps to the formation of a cluster of YSOs in the Class 0/early Class I stage. These simulations open up new horizons for the comparison between models and observations of young disk populations, as they not only investigate the influence of various initial conditions on the self-consistent disk population within their parent environments, they also provide large samples of disks that form the basis for insightful statistical analyses.

On the observational side, the disk properties inferred from dust continuum and molecular line emission often suffer from uncertainties arising from noise, resolution limitations, and instrumental effects, as well as the modelling techniques to derive the physical parameters (see, e.g. \citealt{2017ApJ...849....3K}). Another complication in the modelling of Class 0/I disks comes from the fact that these small embedded objects are still surrounded by a circumstellar envelope, which makes it difficult to separate the emission of  the two components in sub-millimetric observations (\citealt{2015ApJ...805..125T}; \citealt{2018ApJ...866..161S}; \citealt{2019A&A...621A..76M}). As a result, the comparison between simulated disk populations and their synthetic observations is also expected to help to test how closely the disk properties that we obtain from these modelling processes resemble their true properties.

The aim of the work presented in this paper is the first step toward our overarching effort of the {\it Synthetic Populations of Protoplanetary Disks} project (see also \citealt{2023arXiv231019672L}). This endeavour is aimed at connecting the disk population properties from simulations of cluster-forming clouds to observations of young disk populations in our Galaxy obtained with the Atacama Large Millimeter/submillimeter Array (ALMA). We centre our study on the comparison methodology as well as the issue of how to design observations and data analyses to derive the disk parameters in embedded protostars. For our purposes, we post-processed an MHD simulation of protoplanetary disk population through the collapse of  a $1000~M_{\odot}$ clump in \cite{2023arXiv231019672L} by means of continuum radiative transfer using the Monte Carlo radiative transfer code \textsc{Radmc-3d} (\citealt{2012ascl.soft02015D}). We produced realistic interferometric observation simulations using the Common Astronomy Software Applications (\textsc{Casa}) tool (\citealt{2007ASPC..376..127M}; \citealt{2022PASP..134k4501C}), which provides us with a simulator for ALMA. We then treated them as real observational data and employed the modelling methods from recent studies and surveys (\citealt{2019A&A...621A..76M}; \citealt{2021MNRAS.506.2804T}; \citealt{2021MNRAS.506.5117T}) to extract the disks' parameters.

The paper is structured as follows. In Sect. \ref{sec:simu}, we present the simulation model and snapshot used for the demonstration of our methodology. Sections \ref{sec:radmc} and \ref{sec:casa} provide the technical details for the radiative transfer and interferometric post-processing, respectively. The modelling of the synthetic observations is described in Sect. \ref{sec:method}, followed by an in-depth analysis of the modelled data regarding the disk properties in Sect. \ref{sec:analyses}. We then spend Sect. \ref{sec:discussion} to discuss various implications of our results on the modelling of Class 0 disks from real observations. A summary of our finding is given in Sect. \ref{sec:summary}.

\section{Simulation models}\label{sec:simu}

\begin{figure}
    \includegraphics[width=0.45\textwidth]{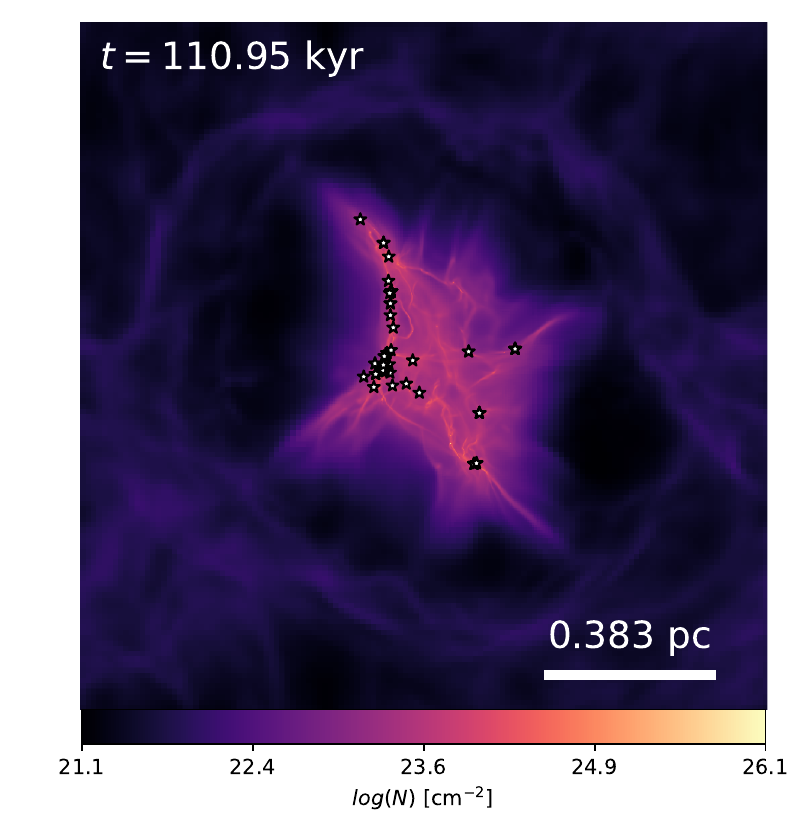}
    \caption{Column density map integrated along the $z-$direction of the NMHD-F01 model from \cite{2023arXiv231019672L} at the simulation snapshot used in this study, seen at a full simulation box's length. The star symbols indicate the sink particle positions.}
    \label{fig:coldens}
\end{figure}

We begin by briefly presenting the simulation model (NMHD-F01 of \citealt{2023arXiv231019672L}) used for the study in this paper. More details can be found in the corresponding paper as well as in \cite{2021ApJ...917L..10L}. The simulation is performed with the adaptive mesh refinement (AMR) (\citealt{1984JCoPh..53..484B}) code \textsc{Ramses} (\citealt{2002A&A...385..337T}; \citealt{2006A&A...457..371F}), which solves the MHD equations using the finite volume method and enables us to resolve up to the disk formation scale with a maximum resolution of $\Delta x_{\rm max} \sim 1.2\au$. The starting point is a $1000~M_{\odot}$ clump of temperature $10\K$, initial radius $\sim 0.38\pc$, and uniform density $3\times10^{-19}\g\cm^{-3}$, initialised with supersonic turbulent velocities at $\mathcal{M} = 7$ with random phases and a Kolmogorov power spectrum of $k^{-11/3}$. The initially vertical magnetic field is set according to the mass-to-flux to critical-mass-to-flux ratio such as $\mu=10$, which corresponds to a magnetic field of $\sim 9.4\times10^{-5}~\G$. Sink particles (\citealt{2014MNRAS.445.4015B}), which form when the density reaches $n_{\rm thre} = 10^{13} \cm^{-3}$, were used as sub-grid models to account for the presence of fully formed stars. These sinks were not allowed to merge.

There are two types of luminosity associated with these sinks. The first one, accretion luminosity, arises when a star of mass, $M_{\star}$, and radius, $R_{\star}$, accretes mass and the incoming kinetic energy of the gas is radiated away. This luminosity has a value of
\begin{equation}
    L_{\rm acc} = f_{\rm acc}\frac{\mathcal{G}M_{\star}\dot{M}_{\star}}{R_{\star}},
\end{equation}
where $f_{\rm{acc}}$ is the fraction of the accretion gravitational energy radiated away and $\mathcal{G}$ is the gravitational constant. We adopt $f_{\rm{acc}}=0.1$ in the model to investigate the influence of the radiative feedback. The second source of luminosity comes from the stellar radiation. The star radius $R_{\star}$ and its internal luminosity $L_{\rm int}$ are inferred using the stellar models from \cite{2013ApJ...772...61K}.

The simulation is evolved until the total mass in all the sinks of the simulations reaches $M_{\rm sink}=150M_{\odot}$. For the calculations presented in the following sections, we use a particular output of the simulations where the star formation efficiency (SFE) is $\approx 10\%$, which corresponds to $M_{\rm sink}=100M_{\odot}$, and whose physical time is $110.95$ kyr. Figure \ref{fig:coldens} shows the column density map of the simulation box ($\sim1.2\pc$ in length) integrated along the $z-$direction at the chosen output.

\section{Radiative transfer post-processing}\label{sec:radmc}

From the simulation output, we performed 3D Monte Carlo (MC) radiative transfer calculations on the native AMR grid using the \textsc{Radmc-3d} code (\citealt{2012ascl.soft02015D}) to ultimately simulate the sky to be observed and obtain the intensity maps of the disk population at a resolution equivalent to the maximum resolution the simulation.

One of the most important properties of the dust and gas components in the simulation grid is their temperature, which is critical for the later derivation of their emissions. On the one hand, \textsc{Ramses}'s extension to radiative transfer in the flux limited diffusion (FLD) approximation (\citealt{1981ApJ...248..321L}; \citealt{2011A&A...529A..35C}; \citealt{2014A&A...563A..11C}) provides us with the temperature profile of the gas and dust in thermal equilibrium. On the other hand, \textsc{Radmc-3d} enables us to re-calculate the dust temperature using a full Monte Carlo approach, without the FLD approximation limitation. Therefore, it is useful to compare the results of the two methods to assess the accuracy of the temperature computation.

For the \textsc{Radmc-3d} MC thermal calculation (as well as the imaging process through ray-tracing), a dust-to-gas ratio of 0.01 is assumed, similarly to what is done in the simulation, to infer the dust density profile for \textsc{Radmc-3d} from the gas density calculated with \textsc{Ramses}. Then, multiple sources of photons are input to \textsc{Radmc-3d} from the sink properties. In particular, the sources' luminosity are taken to be the sum of their internal and accretion luminosities $L_{\rm tot} = L_{\rm acc} + L_{\rm int}$. These luminosities, along with other stellar properties namely the effective temperature and the star's radius, are computed in the simulation run. Another important input for the radiative transfer modelling is the dust opacity. Since \textsc{Ramses} only works with Planck and Rosseland mean opacity which are not compatible with the wavelength-dependent opacity required for \textsc{Radmc-3d}, here we adopted the DIANA standard dust model and computed the opacity file for \textsc{Radmc-3d} using the tool for dust particle opacities calculations, \textsc{OpTool} (\citealt{2021ascl.soft04010D}). This model comprises a specific pyroxene ($70\%$ Mg) and carbon, in a mass ratio of $0.87/0.13$, and with a porosity of $25\%$ (\citealt{2016A&A...586A.103W}). An MRN power-law sized distribution with the minimum grain size $a_{\min} = 1~{\rm nm}$ and the maximum size $a_{\max} = 10\mum$ was assumed for this dust model. To ease the computation cost, we treated this size distribution as one single dust population in the opacity file.

We then used both temperature profiles to carry out the disk size analyses. We give a comparison of the results in Appendix~\ref{app:temperature}. In general, we find that the two temperature profiles do not differ significantly in the outer region of the disks (where $r \gtrsim 10\au$), which is expected if the temperature is mainly determined by the reprocessing of radiation, unless there is substantial dissipation via turbulence; namely, high turbulence could result in strong local heating. In a standard model of an accretion disk, most of the viscous heating takes place in the inner region of the disk and the outer disk is always dominated by passive heating. As a result, the disk parameters obtained with both profiles are quite similar. However, the temperature obtained with \textsc{Radmc-3d} sometimes introduces over-heated region surrounding the photon source which results in a spike in the central flux and causes difficulties in the subsequent modelling of the post-processed results, whereas in \textsc{Ramses}, the inclusion of a smoothing kernel for the photon flux within the $\approx 5\au$ radius of the sink particle and dust sublimation maintains the consistency of the central intensity. Therefore, we chose to continue with the dust temperature from \textsc{Ramses}.

Finally, we imaged all the disks present in the simulation snapshot by performing the ray-tracing through the grid to get their continuum emissions at ALMA's band 7 wavelength of $\lambda=0.89~{\rm mm}$. The imaging on the plane of the sky is done through 2D regularly-spaced cameras observing the 3D AMR grids and, thus, for practical computational reasons, we proceeded with the second-order ray-tracing by zooming on individual sinks particles whose disks are detected. We carried out the imaging procedure in sub-regions of the grid around the sink positions with an $1\au$ pixel resolution, which allowed us to emulate the maximum numerical resolution of $\approx 1.2\au$ in the most resolved regions of the disks. For each of these regions, $10^8$ scattering photon packages were used for the MC scattering calculations before the ray-tracing to get the intensity maps.
We accounted for the variation of the viewing angle in observations by varying the camera positions to image the disks in three different planes of the internal simulation grid: $xy$, $xz$, and $yz$. We note that these projections do not necessarily correspond to the direction of the angular momentum vector of the disks, which are randomly oriented by nature.

\section{Mock interferometric observations}\label{sec:casa}

The \textsc{Radmc-3d} maps provide us with the ideal images of different regions of the sky surrounding each individual protostar. From these maps, by introducing proper instrumentational effects, we can produce realistic observations of the simulated sky as would otherwise be obtained with interferometers. To that end, we use thde ALMA simulator of the \textsc{Casa} software (\citealt{2007ASPC..376..127M}; \citealt{2022PASP..134k4501C}) to `observe' each of the stars and simulate a real observing session.

In particular, each emission map from \textsc{Radmc-3d} was used as the sky model for the \texttt{simobserve} task to get three simulated measurement sets for compact and extended ALMA antenna configurations, C43-3, C43-4, and C43-6, in Band 7 (at $0.89{\rm ~mm}$, or $345{\rm ~GHz}$); this corresponds to three angular resolutions of $\theta_{\rm res} = 0.41,~0.266, ~0.0887''$, respectively.
For C43-3 and C43-4, which are relatively lower resolution, we integrated, in a "survey" fashion, for $300\s; $ this is comparable to the integration time of ALMA Band 7 observations in \cite{2015ApJ...814...22E} and \cite{2022ApJ...933...23O}. For the high-resolution observations with C43-6, we increased the integration time as inversely proportional to $\theta_{\rm res}^2$ to $\sim2700\s$ to achieve the same signal-to-noise ratio (S/N) per resolution element.
A precipitable water vapor (PWV) value of 0.7 mm (as under good atmospheric conditions) was assumed. Thermal noise (default ATM model; \citealt{2019adw..confE..36P}) was added to the simulated data. The measurement sets simulated by \texttt{simobserve} were then imaged and analyzed using the \texttt{simanalyze} task, which uses the \texttt{tclean} algorithm to perform an iterative cleaning process based on the `clark' deconvolver (\citealt{1980A&A....89..377C}) and the `briggs' weighting scheme (\citealt{1995PhDT.......238B}), with a robust factor of 0.5.

\begin{figure*}
    \centering
    \includegraphics[width=\textwidth]{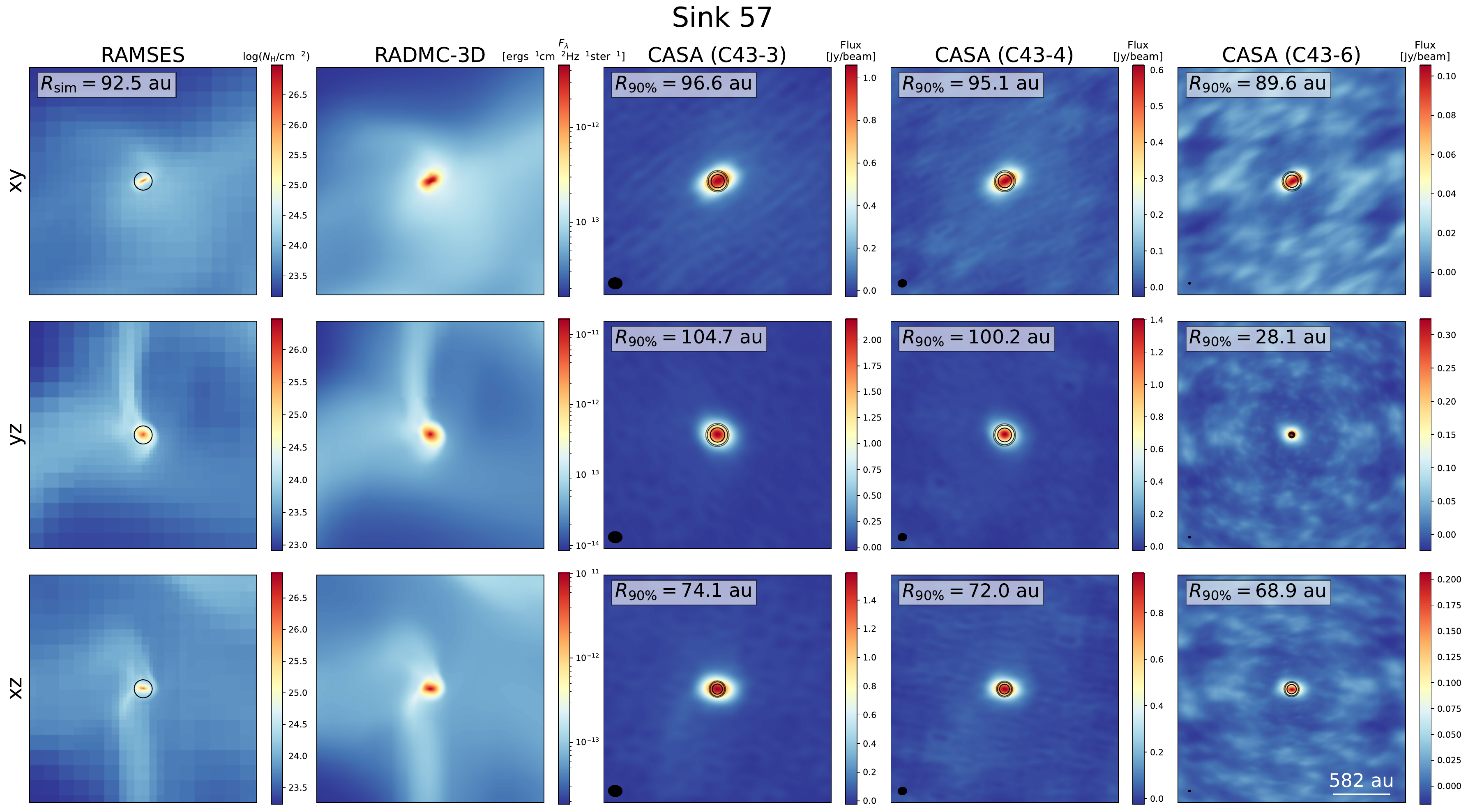}
    \caption{Images of the disk around sink 57 viewed in different projections and through different stages of the pipeline. {\it From left to right:} Column density maps of the $1000\au$ region centred around the sink particle in 3 projections along the internal axes from the \textsc{Ramses} simulation; continuum emission maps obtained with \textsc{Radmc-3d} of the same regions; and \textsc{Casa} images obtained with \texttt{tclean} from the C43-3, C43-4 and C43-6 visibilities. The disk sizes inferred from the gas kinematics, $R_{\rm sim}$, and the modelled disk radii, $R_{90\%}$, (detailed in Sect. \ref{sec:analyses}) are noted for each configuration and projection.}
    \label{fig:ramses2casa}
\end{figure*}

Figure \ref{fig:ramses2casa} illustrates (from left to right) the outcome at each step of the pipeline, namely: the column density integrated along one of the three coordinate axes $x$, $y$, or $z$ through the entire grid, the dust continuum emission maps from \textsc{Radmc-3d}, and the visibility data achieved with C43-3, C43-4, and C43-6 imaged with the \texttt{tclean} algorithm integrated in \textsc{Casa} for one disk in the population. The structures are filtered out in the \textsc{Casa} images due to an incomplete coverage of the $uv-$plane; thus, recovering the disk properties at this stage is not trivial, just like recovering the `real-life' disk properties in real observations.

Based on the images produced by \texttt{tclean} and their radial intensity profiles from the star positions, we selected the isolated disks for our subsequent analyses from the $uv-$data. Since the modelling of binaries and triple systems in the $uv-$space is not trivial without the use of a complex analytical model, such systems (which comprise of 10-11 out of the 26 disks formed in this simulation) were excluded in our current study. However, we note that we selected the single systems by looking at the observations imaged with \texttt{tclean} -- and not at the images of the original disks; this was done to avoid prior knowledge of the real objects. As such, any imaged system in which there is more than one source to model within the primary beam is considered a binary or multiple, provided that the angular resolution of the observation is sufficient to produce a noticeable separation of the multiple sources.

\section{Synthetic modelling}\label{sec:method}

We now treat these measurement sets produced by \textsc{Casa} as if they were real observations and we applied similar techniques to model the "observed" disks to those in recent observational studies of YSOs (\citealt{2019A&A...621A..76M}; \citealt{2021MNRAS.506.2804T}; \citealt{2021MNRAS.506.5117T}). This experiment allowed us to obtain their observable properties, namely:\ disk {sizes, masses}, and inclination angles. 

\subsection{Two-component model for the disk and the envelope}

\begin{figure}
    \centering
    \includegraphics[width=0.45\textwidth]{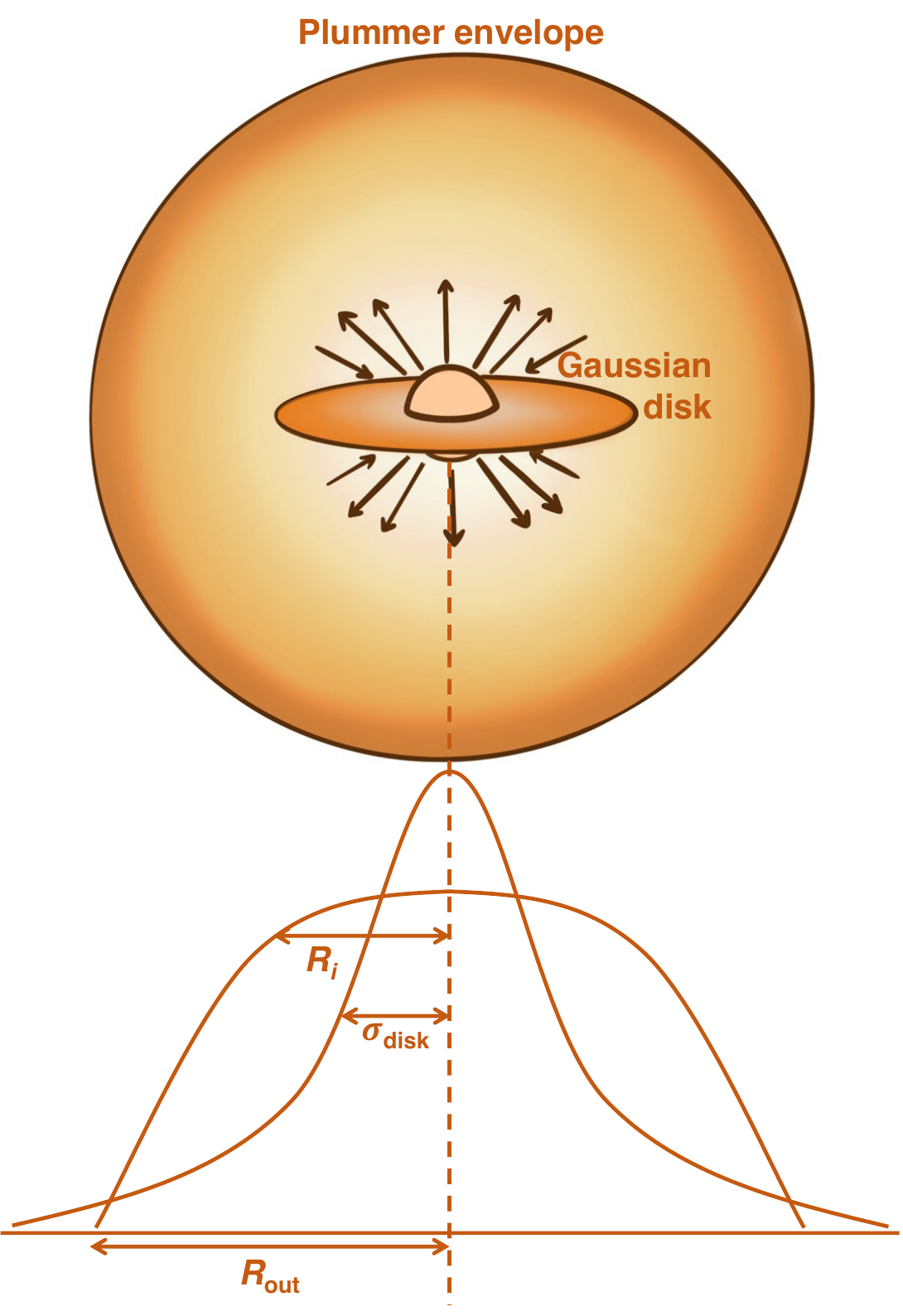}
    \caption{Schematic illustration of the disk and envelope model used to describe the emission of the YSOs formed in the simulation snapshot. A disk whose emission is described by an elliptic 2D Gaussian of semi-major axis $\sigma_{\rm disk}$ is surrounded by a circular envelope following a Plummer-like density profile with rotation-to-infall transitional radius, $R_i$, and truncated outer radius, $R_{\rm out}$.}
    \label{fig:sketch}
\end{figure}

In our synthetic Class 0/I disk sample, due to the strong presence of the envelopes surrounding the young protostars, we opted to model each of these objects as a disk-envelope system, as illustrated in Fig. \ref{fig:sketch}. Thus, in addition to the Gaussian profile for the disk, we adopted the empirical analytical model of a truncated Plummer envelope according to the method presented in \cite{2019A&A...621A..76M} to account for the more extended emission.

We let $I_0$ be the total intensity at $r=0$, of which a fraction, $f,$ is attributed to the disk. We can then express the surface brightness profile of the 2D disk of semi-major axis, $\sigma_{\rm disk}$, as:
\begin{eqnarray}
    I_{\nu,{\rm disk}}(\bar{r}) = f I_0\exp\left(-\frac{\bar{r}^2}{2\sigma_{\rm disk}^2}\right),
\end{eqnarray}
and that of the circularly symmetric Plummer envelope truncated at outer radius, $R_{\rm out}$, with a power-law distribution of the density of $\rho(\bar{r}) \propto \bar{r}^{-m}$ and temperature of $T(\bar{r}) \propto \bar{r}^{-n}$ (\citealt{1991ApJ...382..544A}) as:
\begin{eqnarray}
    I_{\nu,{\rm env}}(\bar{r}) = \left\{
    \begin{array}{l l}
         \frac{(1-f)I_0}{(1+(\bar{r}/R_i)^2)^{-(m+n-1)/2}} & {\rm for~} 0 \leq \bar{r} \leq R_{\rm out}\\
         
         0 & {\rm for~} \bar{r} > R_{\rm out}
    \end{array}\right\}.\label{eq:model}
\end{eqnarray}
Here, $\bar{r}$ is the projected radius in the sky and $R_i$ is the envelope's radius at which the approximately uniform density of the inner region ends, which can be physically interpreted as the transition between envelope motions at $r>R_i$ and rotational motions within $r < R_i$. 

\subsection{Parameter fitting}

Armed with these models, we could then use the \texttt{galario} (\citealt{2018MNRAS.476.4527T}) and \texttt{emcee} (\citealt{2013PASP..125..306F}) libraries for Python to fit the visibilities obtained with \textsc{Casa}. As the system is comprised of an inclined disk and a spherical envelope, for each iteration, we used the radial profiles to construct two separate 2D images of the disk with an inclination angle, $i,$ introduced, along with a flat circular envelope with the above-mentioned parameters; in addition an offset (${\rm dRA}$, ${\rm dDec}$) was added from the centre and position angle ${\rm (PA)}$. Then, we superposed the two images and computed the synthetic visibility data for the model. The value of $\chi^2$ can be obtained by comparing these visibilities with the "observed" ones.
Finally, we ran the affine invariant Markov chain Monte Carlo (MCMC) from \cite{2010CAMCS...5...65G} with 80 walkers to explore the parameter space for each single disk present in the simulation output. A "burn-in" period of 4000 steps is initialised, followed by 20000 main steps to guarantee convergence.

\begin{figure*}
    \centering
    \includegraphics[height=0.4\textwidth]{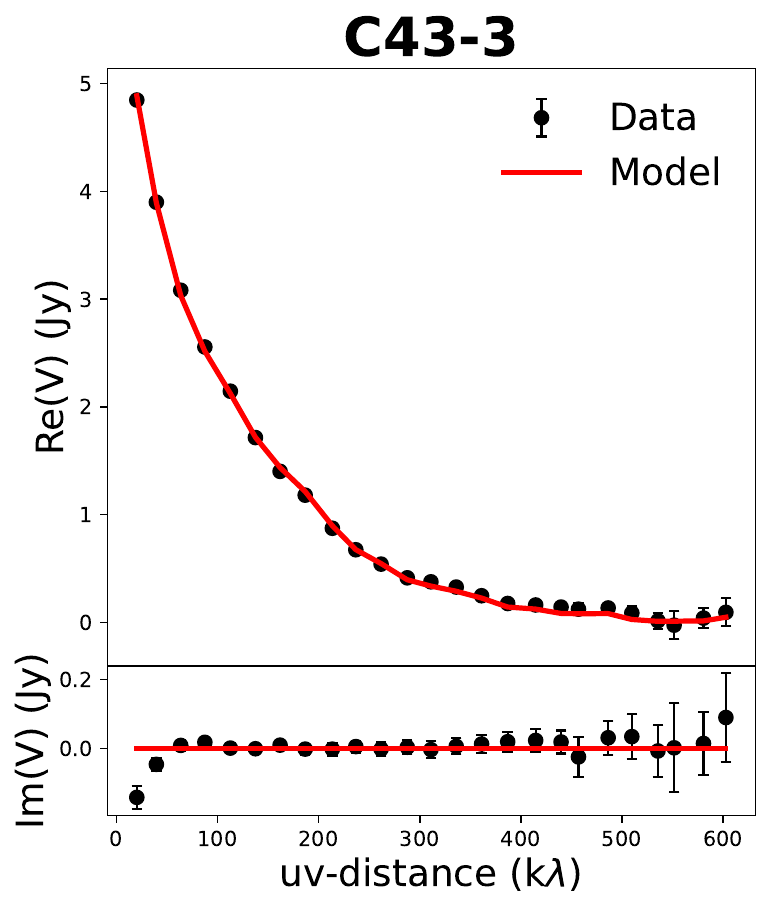}
    \includegraphics[height=0.4\textwidth]{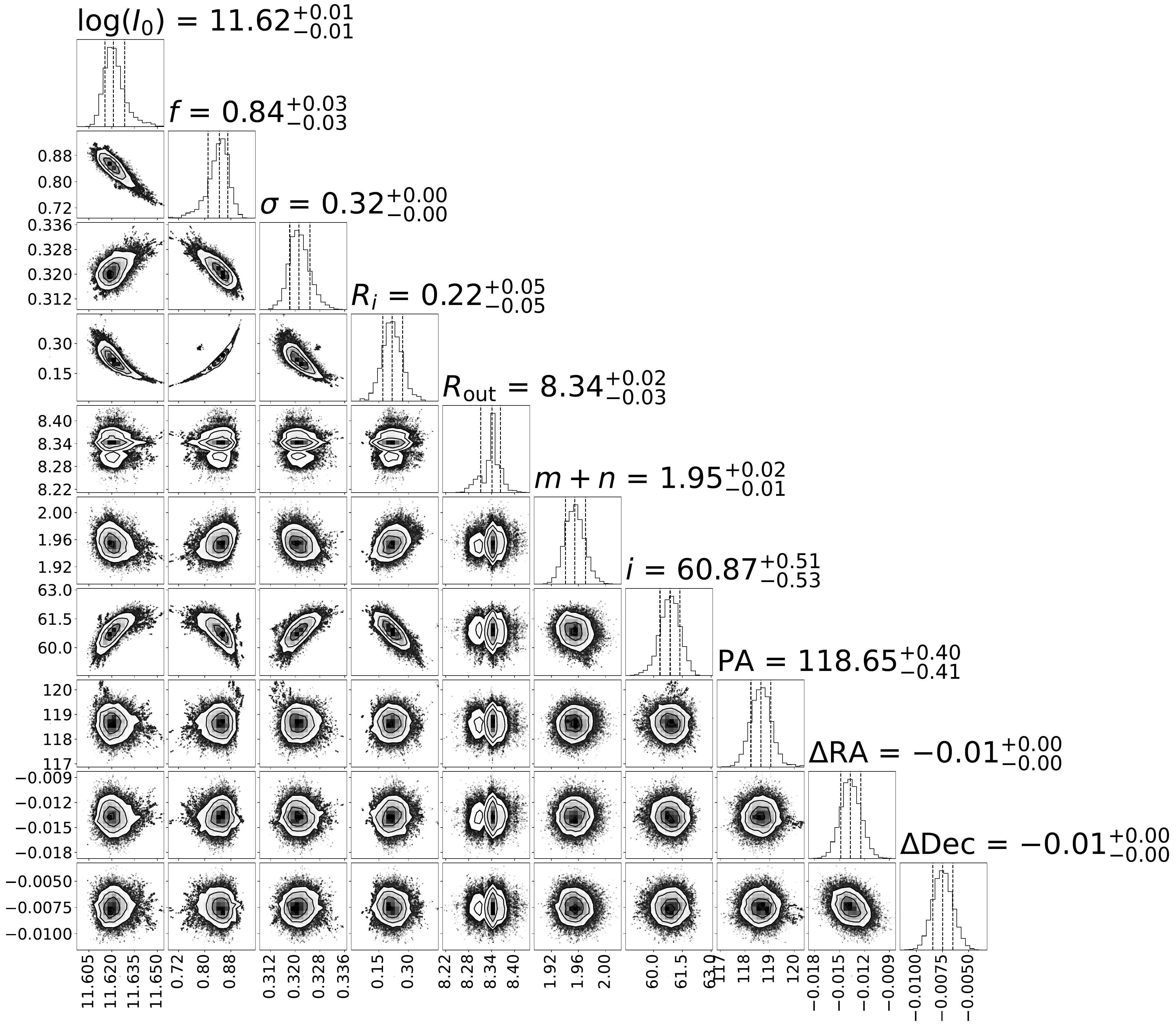}
    \includegraphics[height=0.4\textwidth]{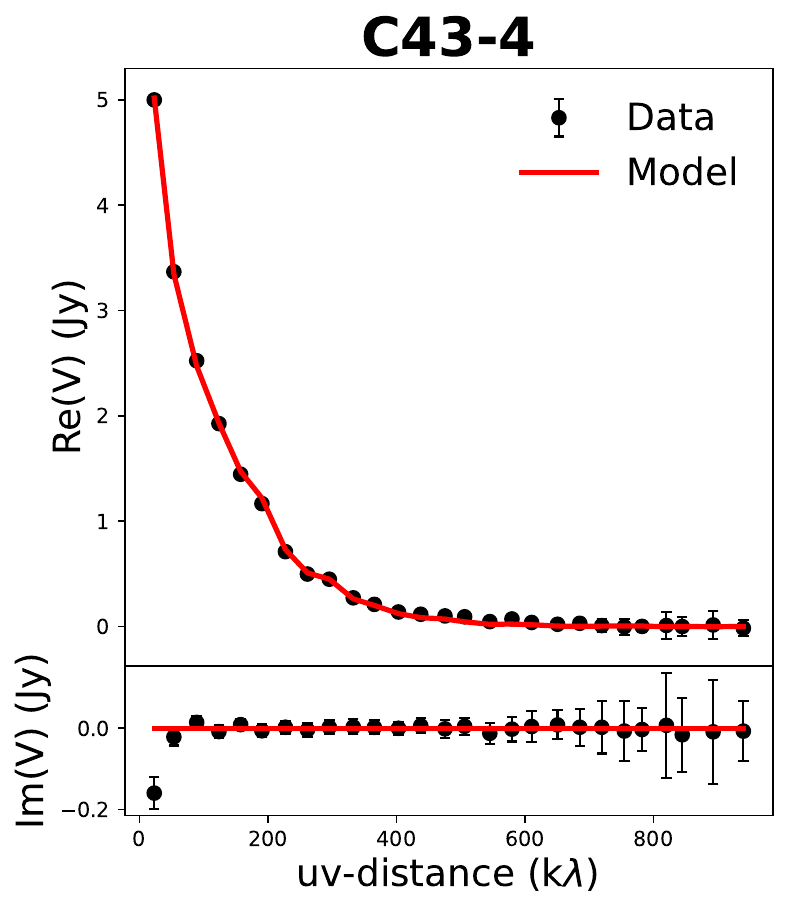}
    \includegraphics[height=0.4\textwidth]{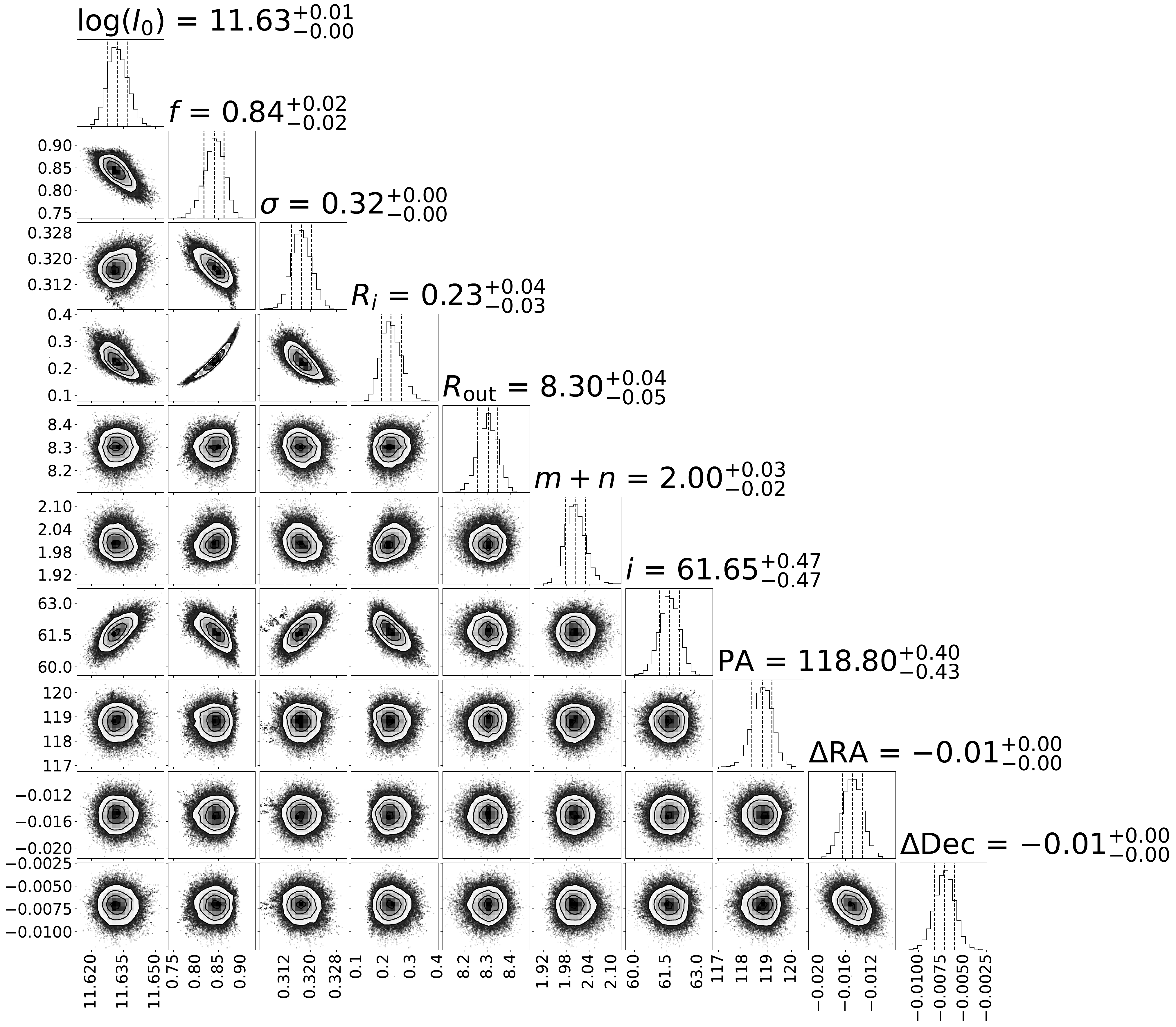}
    \includegraphics[height=0.4\textwidth]{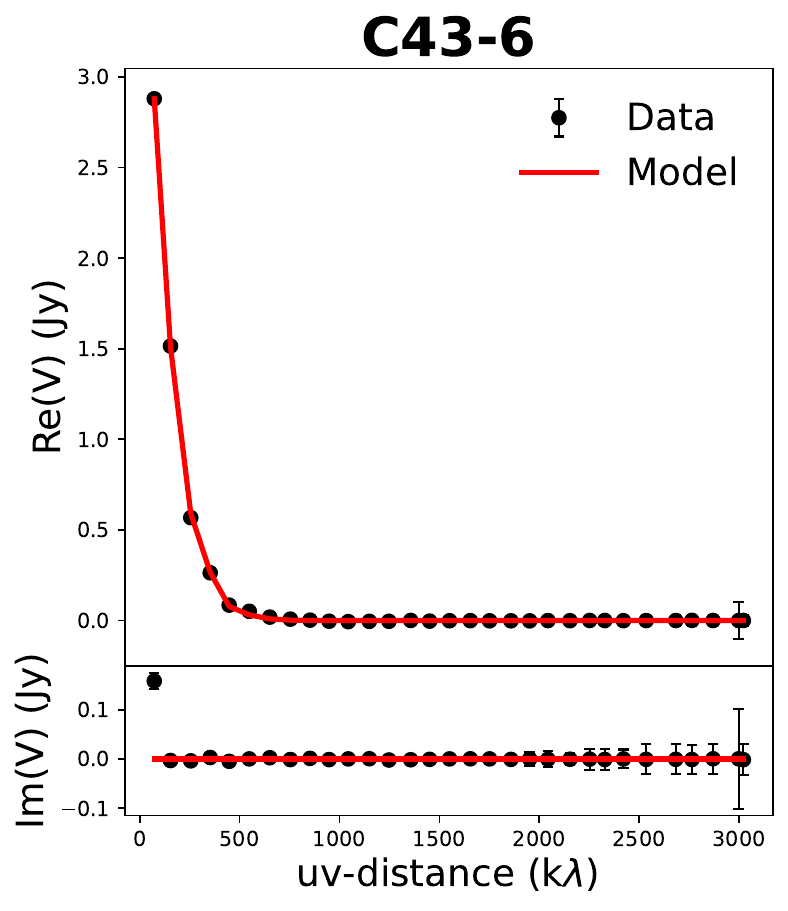}
    \includegraphics[height=0.4\textwidth]{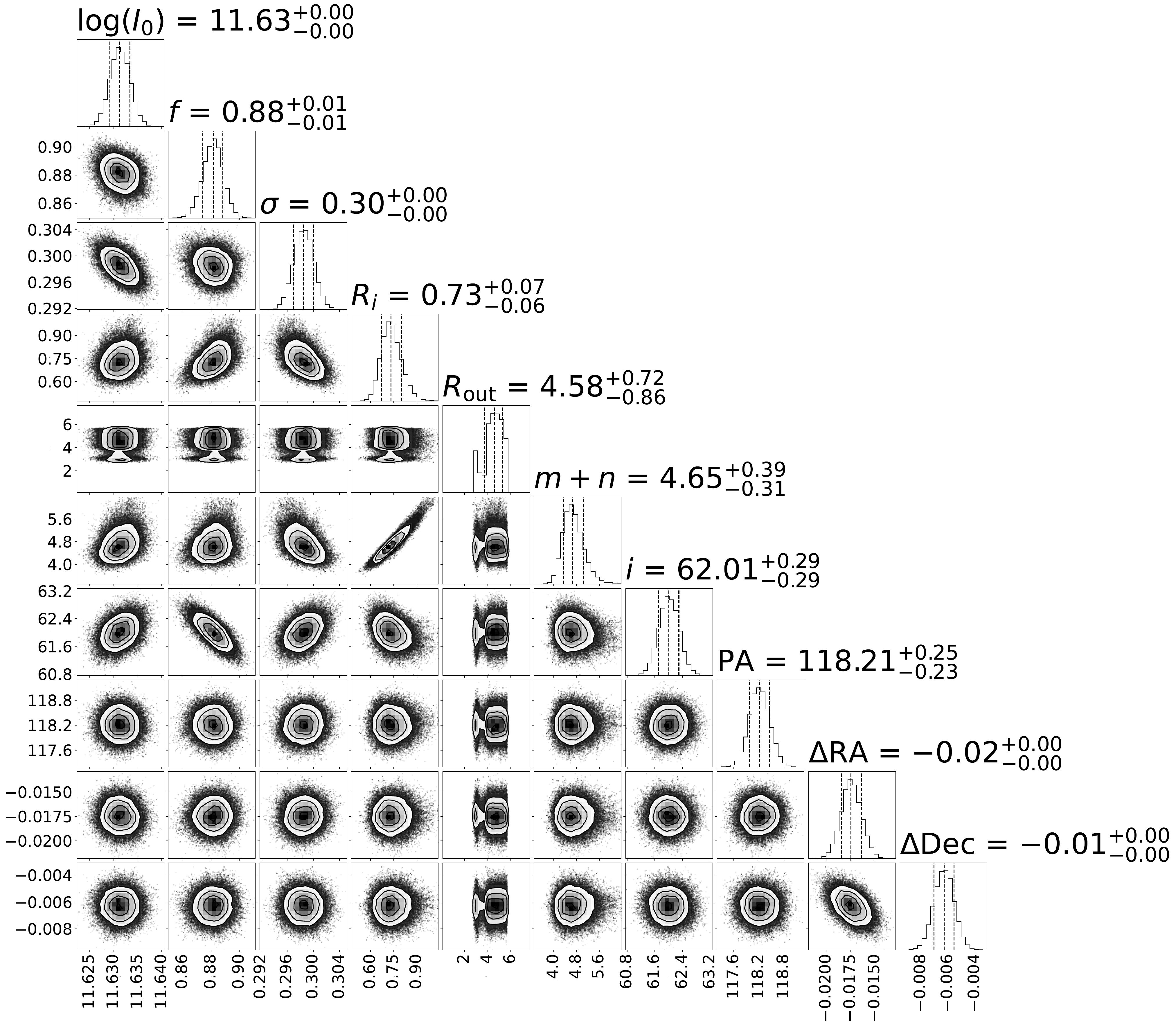}
    \caption{Fitting results from \texttt{galario} for the synthetic observations of the disk around sink 57, seen in the $xy$ projection. {\it Left:} Observed (black dots with error bars) and modelled (red curve) visibilities in real (top) and imaginary parts (lower panel) as a function of the baseline (in ${\rm k}\lambda$) for three configurations C43-3 (upper), C43-4 (middle), and C43-6 (lower). {\it Right:} Corner plots of the \texttt{galario} fittings of the visibility data with the disk and envelope model. The top sub-panels show the 1D histograms of the free parameters from the MCMC chains. The rest of the panels are the 2D histograms between each pair of parameters.}
    \label{fig:corner}
\end{figure*}

\begin{figure*}
    \centering
    \includegraphics[width=0.85\textwidth]{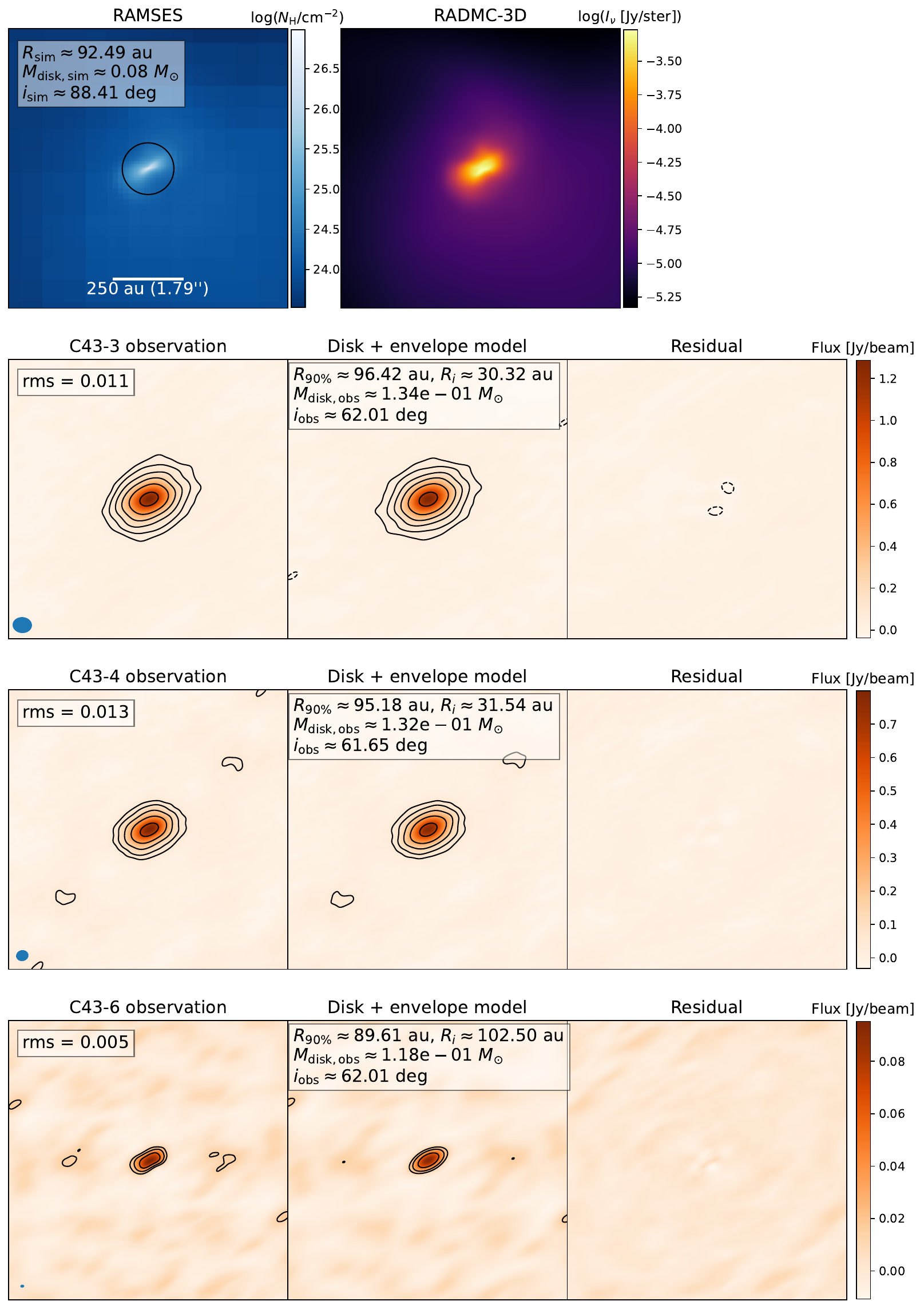}
    \caption{Modelling results imaged for the disk around sink 57. Column density plot from the \textsc{Ramses} output with the circle enclosing the disk size (top panel) derived from simulation (left), and dust continuum image obtained with \textsc{Radmc-3d} (right) of the $1000\au$ region around the star. The{\it } \textsc{Casa} emission map imaged (bottom panel) with \texttt{tclean} from the \texttt{simobserve} visibilities with C43-4 (left), emission map of the best-fit model obtained with \texttt{tclean} (middle), and the corresponding residual (right). Contour lines are plotted at 3, 6, 12, 24, 48, and 96 multiples of the rms, $\sigma$. The disk properties inferred from the simulation and their modelled properties from observations are noted in the corresponding panels.}
    \label{fig:model_images}
\end{figure*}

Figure \ref{fig:corner} shows an example of the results obtained with this fitting method for one of the disks in the simulation snapshot. The fits for the visibility data returned by \texttt{simobserve} for three configurations C43-4, C43-4, and C43-6 with the disk-envelope model are provided in the upper panels, along with the triangular plots showing the correlation between the fitted parameters in the lower ones. The real parts of the $uv$ data are aptly fitted with the models, whereas there is an offset at small $uv$ distance in the imaginary parts, suggesting that the large scale components (i.e. the envelopes) are not perfectly modelled. Nevertheless, it might be more intuitive to check the goodness of the modelling by looking at the easier-to-visualise image plane. Thus, for illustrative and comparative purposes, an image of the model after interferometric filtering can be obtained by \textsc{Casa}'s \texttt{tclean} to compare with the synthetic image of the simulated disks by looking at the residual of the two images. An example of such images is shown in the lower panels of Fig. \ref{fig:model_images}. The residual map is within the $3\sigma$ noise level, showing relatively well modelled systems in the image plane. For the remainder of the disks in our simulation in the same $xy$ projection, we refer to Appendix \ref{app:images} for the model images and the comparison with their corresponding imaged observations from the three antenna configurations. 

The results returned by \texttt{galario} have allowed us to obtain a simple analytical model of the disks. This kind of model has the advantage that it enables us to investigate the disk properties directly from the parameters in a self-consistent way. Moreover, it is also the method that is most widely used in disk modelling from observations (see, e.g. \citealt{2019A&A...621A..76M}; \citealt{2020A&A...633A.114S}); therefore, it is more comparable with the properties extracted from real observations than those directly inferred from simulations.

\section{Analysis}\label{sec:analyses}

The disk properties obtained from the modelling of observations often suffer from numerous uncertainties due to instrumental and interferometric effects. The information regarding their intrinsic properties that we are able to derive from our simulations of their formation allows us to quantify the accuracy of these observational measurements. Thus, in this section, we take the simulation-derived properties as the "ground truth" that forms the basis of the subsequent assessment.

\subsection{Disk sizes}

\subsubsection{Radius measurements}\label{sec:radius_measure}
The most accessible and best studied physical property of the disks from continuum observations is probably the disk size, whose measurements are essential to study the disk evolution driven either by the viscosity or MHD disk wind (\citealt{2021MNRAS.507..818T}; \citealt{2023MNRAS.518L..69T}; \citealt{2019MNRAS.486.4829R}; \citealt{2023ASPC..534..501M}). We therefore use our synthetic observations to assess the relationship between the radii derived by the simple analytical modelling of the observations with the theoretically defined disk radii as measured in numerical simulations.

On the theoretical side, accompanied with the simulations of the PPD formation, \cite{2021ApJ...917L..10L} proposed a method known as "disk finder" to select the disk materials from the gas kinematics based on the criteria from \cite{2012A&A...543A.128J}, which can be highlighted as follows. In cylindrical coordinates ($R$; $\phi$; $z$), co-moving with the central star, the disk materials will be selected if they satisfied the following condition:
\begin{itemize}
    \item Rotating faster than falling radially: $v_{\phi} > 2v_r$,
    \item Rotating faster than falling vertically: $v_{\phi} > 2v_z$,
    \item Composed of dense material of gas number density $n > n_{\rm thre} = 10^9 \cm^{-3}$.
\end{itemize}
It is worth pointing out that while this result is closer to the true properties, this is still more of a statistical method for determining the parameters. As such, it does not aim to represent the physical reality of the object as disks in these simulations are often not well defined and separated from the collapsing flows. This is in line with what was found by \cite{2020ApJ...905..174A} in a more extended investigation of the disk identification methods from both simulations and synthetic observations of an isolated protostellar disk formed in the single core-collapse scenario. Observationally, it is also the realistic picture confirmed by recent studies of Class 0/I systems in which the presence of streamers (see, e.g. \citealt{2023ASPC..534..233P}; \citealt{2023FaDi..245..164B}; \citealt{2023A&A...676A...4C}) or bridges (\citealt{2018ApJ...869..115S}) were detected. With that in mind, we re-applied the same method to derive the disk radii, $R_{\rm sim}$, from the simulations. 

From the observations, the MCMC visibility probability fitting gives us two important parameters in the models: the semi-major axis of the Gaussian disk, $\sigma_{\rm disk}$, and the radius of the inner region of the Plummer envelope, $R_i$. While $R_i$ could be a good estimation of the disk size as inferred from the envelope's geometry, in parallel we used a more robust definition of the disk radius from the Gaussian component, taking  its emission into account. This employs a characterisation of the disk size by the radius within which a certain fraction of the total emission is contained. The most commonly used fractions in the literature are $68\%$ (\citealt{2017ApJ...845...44T}; \citealt{2018ApJ...865..157A}; \citealt{2019A&A...626L...2F}; \citealt{2019ApJ...882...49L}; \citealt{2019A&A...628A..95M}), $90\%$, or $95\%$ (\citealt{2017A&A...606A..88T}). The corresponding radii can be related to $\sigma_{\rm disk}$ as:
\begin{eqnarray}
    R_{68\%} &=& \sigma_{\rm disk}\cdot\sqrt{-2\cdot\ln(1-0.68)} \simeq 1.51\cdot \sigma_{\rm disk},\label{eq:radii}\\\nonumber
    R_{90\%} &\simeq& 1.42\cdot R_{68\%},\\\nonumber
    R_{\rm 95\%} &\simeq& 1.62\cdot R_{\rm 68\%},
\end{eqnarray}

\begin{figure*}
    \hspace{0.001\textwidth}
    \includegraphics[width=0.31\textwidth]{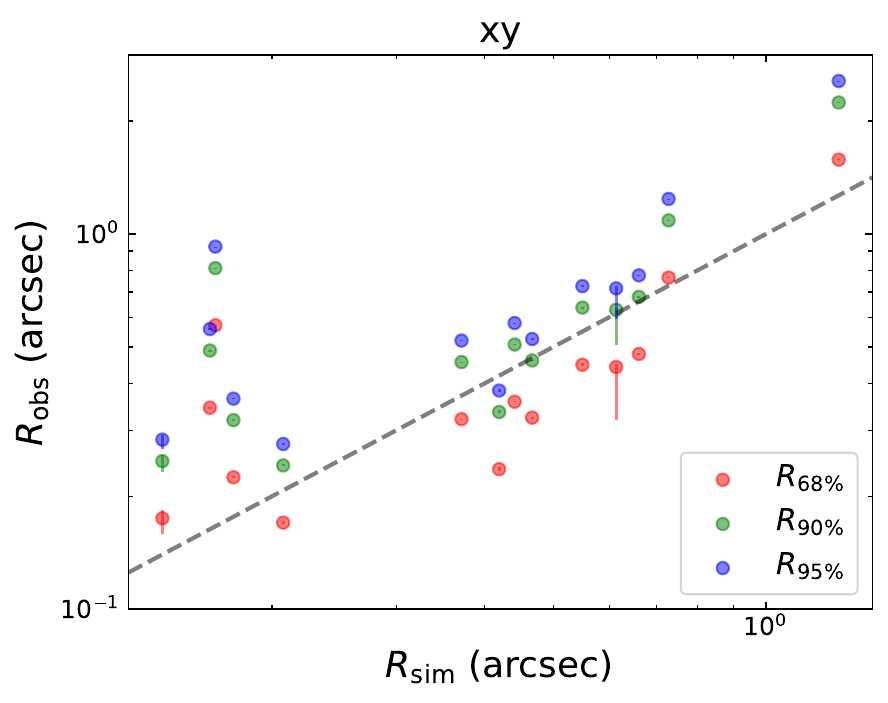}
    \hspace{0.0032\textwidth}
    \includegraphics[width=0.31\textwidth]{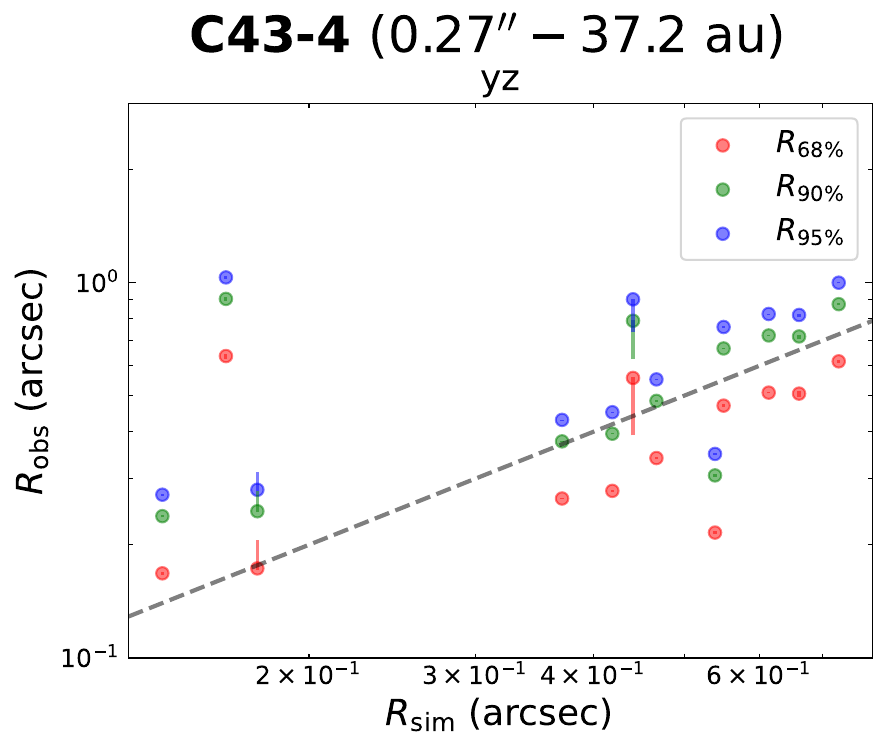}
    \hspace{0.0032\textwidth}
    \includegraphics[width=0.31\textwidth]{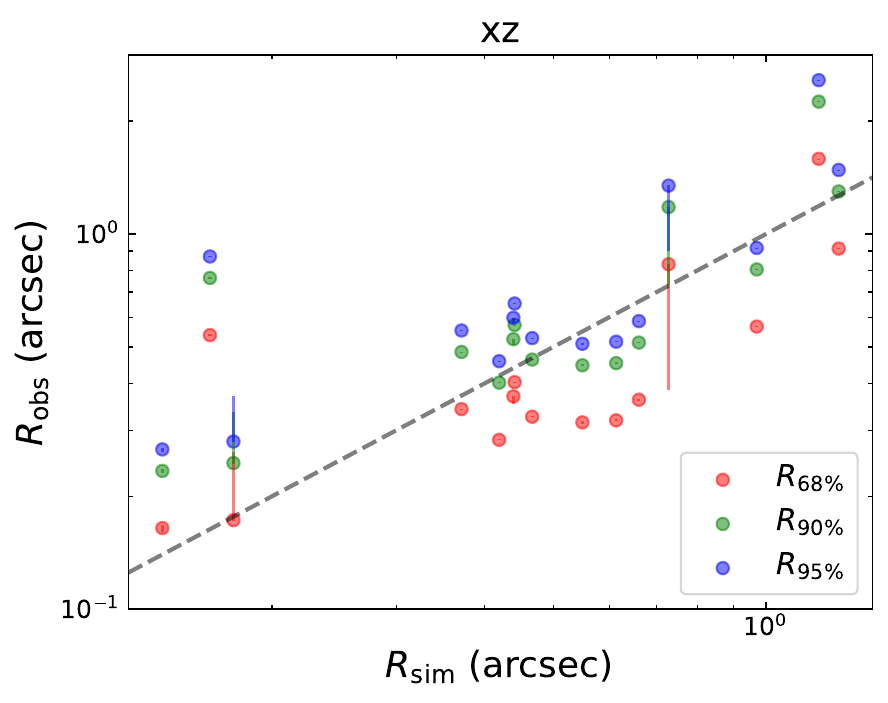}

    \begin{center}
    \includegraphics[width=0.321\textwidth]{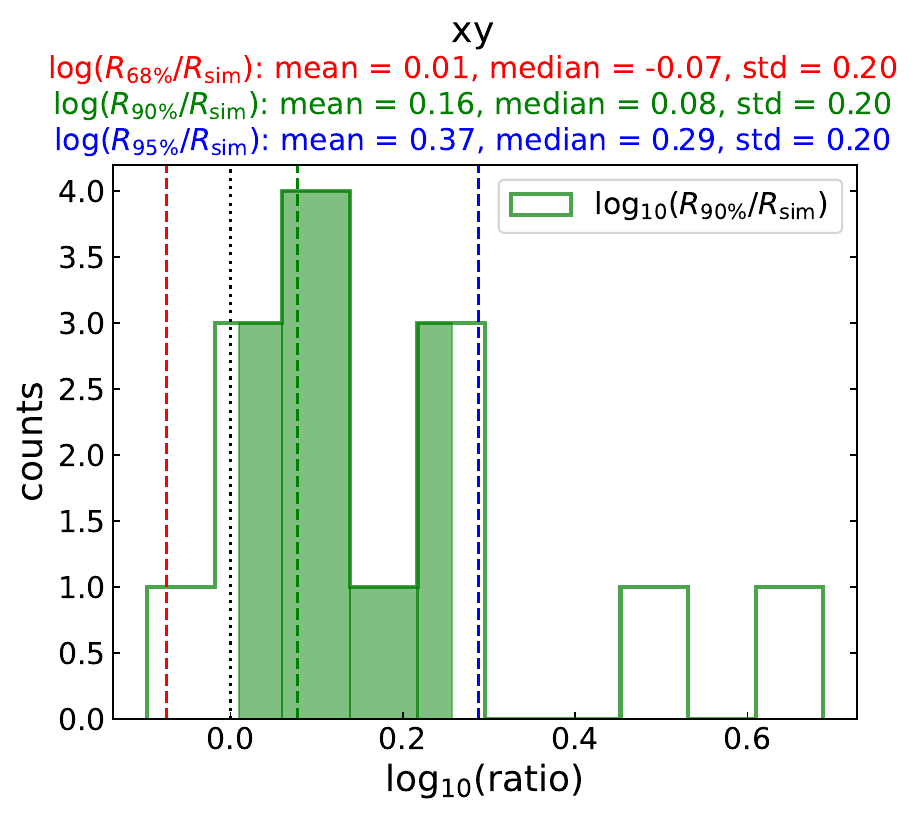}
    \includegraphics[width=0.321\textwidth]{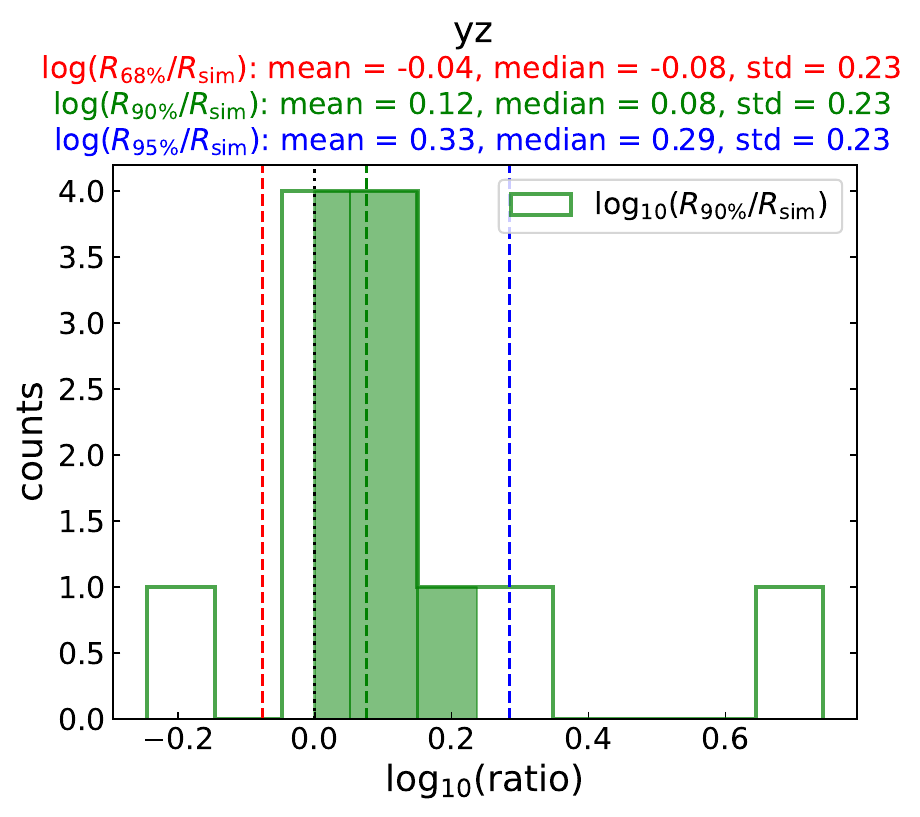}
    \includegraphics[width=0.321\textwidth]{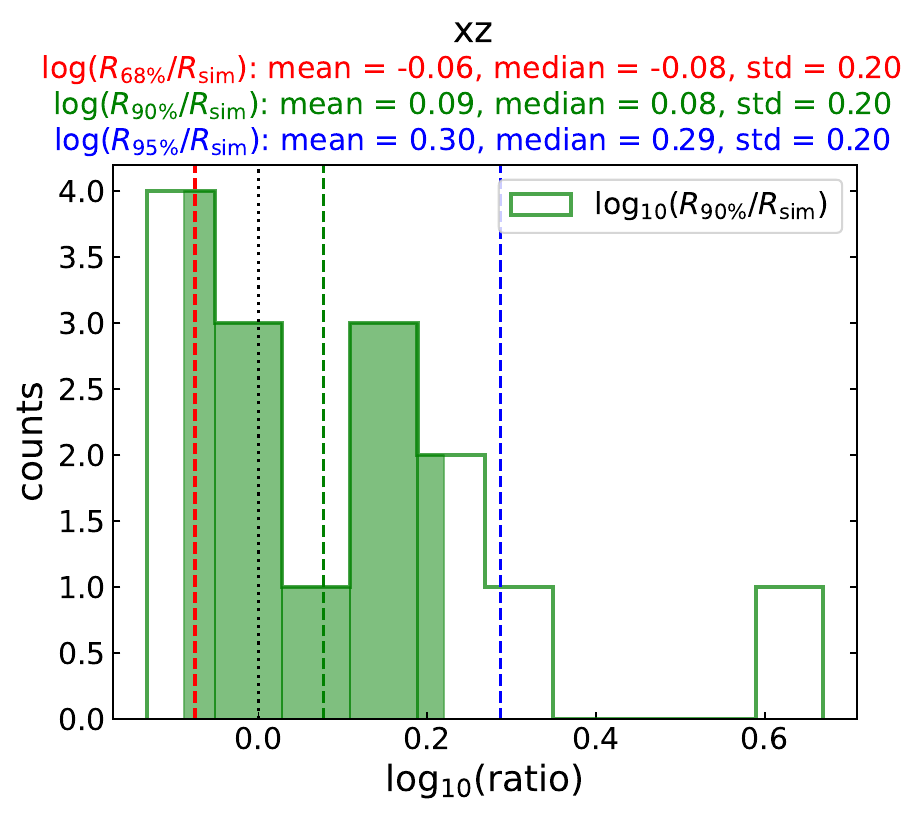}
    \end{center}
    \caption{Comparison of the disk sizes measured from the observations with the C43-4 configuration and the theoretical radii. Disk sizes in arcsecs obtained from the modelling of \textsc{Casa} observations vs. the values derived from the gas kinematics in the simulation for all single star-disk systems present in the output, seen in three projections: $xy$ (left), $yz$ (middle), and $xz$ (right), shown at the top. Black dashed line shows the one-to-one correlation. Histograms of the $R_{90\%}/R_{\rm sim}$ ratio for the three projections, shown at the bottom. Black dotted line marks $R_{90\%}/R_{\rm sim}=1,$ where the two are in full agreement. Vertical dashed lines show the medians of the ratio distributions of three radii $R_{68\%}$ (red), $R_{90\%}$ (green), and $R_{95\%}$ (blue). The values within the 16th to the 84th percentiles are covered by the shaded areas. The statistics (including the mean, median, and standard deviation) for the three radii are indicated above each panel.}
    \label{fig:disk_size}
\end{figure*}

In this two component model, since $R_i$ and $\sigma_{\rm disk}$ are not explicitly related according to our analytical description, there could be four scenarios for the results of the fits regarding the disk sizes:
\begin{enumerate}
    \item The Gaussian disk is within the rotation-to-infall transition marked by $R_i$ and both are above the resolution threshold of the observation, namely, $\theta_{\rm res}/(2\sqrt{2\ln2}) < \sigma_{\rm disk} < R_i$.
    \item The Gaussian disk size is below the resolution of the observation and the rotation-to-infall transition is above, namely, $\sigma_{\rm disk} < \theta_{\rm res}/(2\sqrt{2\ln2}) < R_{i}$. This usually happens for large disks in the more resolved observations (mostly C43-6), whose low-surface-brightness structure is missed by the Gaussian, which instead ends up fitting the narrow central peak.
    \item The Gaussian disk radius is beyond the rotation-to-infall transition and its peak intensity is lower than that of the envelope, namely, $\sigma_{\rm disk} > R_i$ and $f < 0.5$. We consider the Gaussian to be tracing the envelope's more extended emission in such case.
    \item The Gaussian disk radius and the rotation-to-infall transition are both below the resolution threshold of the observation, namely, $\sigma_{\rm disk}, R_i < \theta_{\rm res}/(2\sqrt{2\ln2})$.
\end{enumerate}
In the first scenario, we used the three Gaussian radii given by Eq. (\ref{eq:radii}) to determine the observational disk radius, $R_{\rm obs}$, and subsequently, the disk mass $M_{\rm obs}$ (in combination with the peak intensity of the disk, $fI_0$). In the next two cases, we switched the roles of the two analytical components, taking the inner structure of the Plummer-like component as the disk and the Gaussian the envelope, namely: using $R_i$ and $(1-f)I_0$ as the radius and the peak intensity of the disk instead, so as to minimise the number of outliers. In the final case, we used the greater radius of the two and the corresponding properties. Our pre-testing found that this alternative approach to using the Gaussian's radii mainly helps to improve the results for C43-6 high angular resolution observations. This is because it works to compensate for the weakness of the Gaussian in tracing the low-surface-brightness structure of the larger disks; whereas for the other two lower-resolution configurations, it has a negligible impact on the results.

We started with the C43-4 configuration, which allowed us to sufficiently resolve most of the disks in the sample with its angular resolution of $0.266''$ ($\approx 42\au$), comparable to that of recent surveys, such as \cite{2020ApJ...890..130T} and \cite{2022ApJ...929...76S}. Figure \ref{fig:disk_size} compares the value we get at the two ends. We plot in the upper panels the disk sizes obtained from the synthetic modelling of the $uv-$data `observed' with C43-4 and Eq. (\ref{eq:radii}) versus their sizes from the disk finder method for the \textsc{Ramses} simulations. The uncertainties (marked by the corresponding error bars) are determined by a lower and upper boundary at the 16th and 84th percentiles of the samples, respectively. The $R_{90\%}$ radii follow relatively well the one-to-one correlation shown by the black straight curve. So does $R_{68\%}$, although it does fall slightly below the one-to-one line for the most part.  On the other hand, $R_{95\%}$ tends to overestimate disk size most of the time. 

As a more direct comparison, we divided the three radii by the disk radius values from the simulation and plotted the histograms of the logarithms (base 10) of these ratios in the three projections: $xy$, $yz$, $xz$ according to the simulation grid (shown in the lower panels). Since all these ratios can be related to $\sigma_{\rm disk}$ and to each other by a certain factor, the standard deviation $\sigma_{\rm ratio}$ calculated for $R_{68\%}$, $R_{90\%}$, and $R_{95\%}$ does not differ. Thus, the distributions of the three ratios in logarithmic scale can be related by a horizontal shift equal to the shifts of the median values indicated by the vertical dashed lines; this is also the case for the size of the $1\sigma$ ranges indicated by the shaded areas under the histograms. The values of $\sigma_{\rm ratio} \sim 0.22-0.23$ indicate a factor of $\sim 1.6-1.7$ in the uncertainty of the estimates. For the $xy$ and $yz$ projections, the distributions for $R_{68\%}/R_{\rm sim}$ and $R_{90\%}/R_{\rm sim}$ peaks near unity, indicated by the vertical dotted black lines, as well as the means and the medians of their distributions. Based on their statistics, $R_{68\%}$ and $R_{90\%}$ distributions have medians equally close to $1$ in general, with $R_{68\%}$ slightly underestimates and $R_{90\%}$ slightly overestimates the disk radii. To include as much as possible the disk emission, we use $R_{90\%}$ for the histograms of the ratio of the disk radii, as the values for $R_{68\%}$ and $R_{95\%}$ can be related to it by a certain factor.

\subsubsection{Effect of resolution}

\begin{figure*}
    \includegraphics[width=0.3225\textwidth]{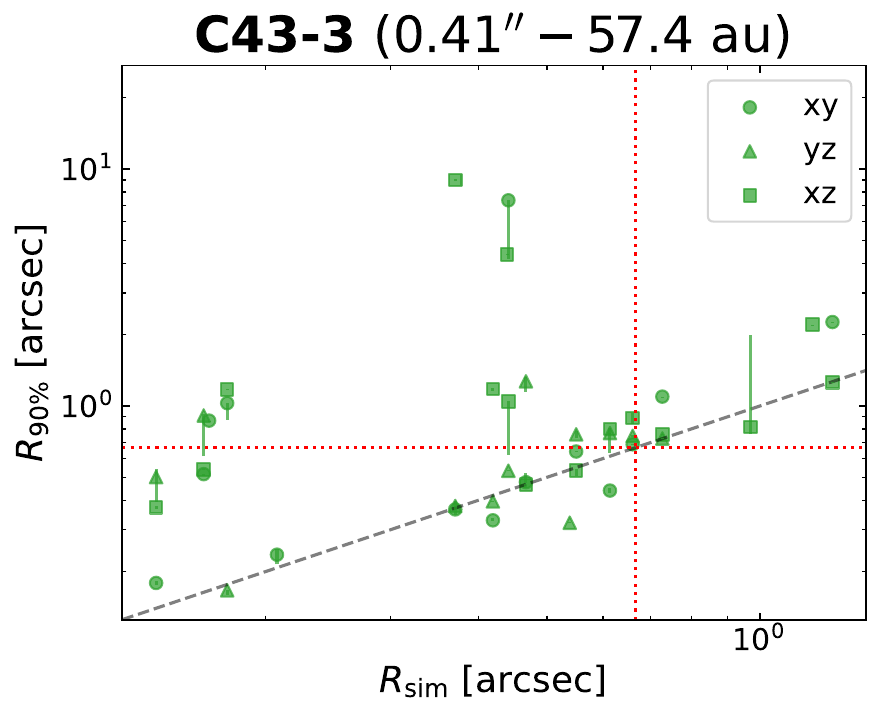}
    \includegraphics[width=0.3225\textwidth]{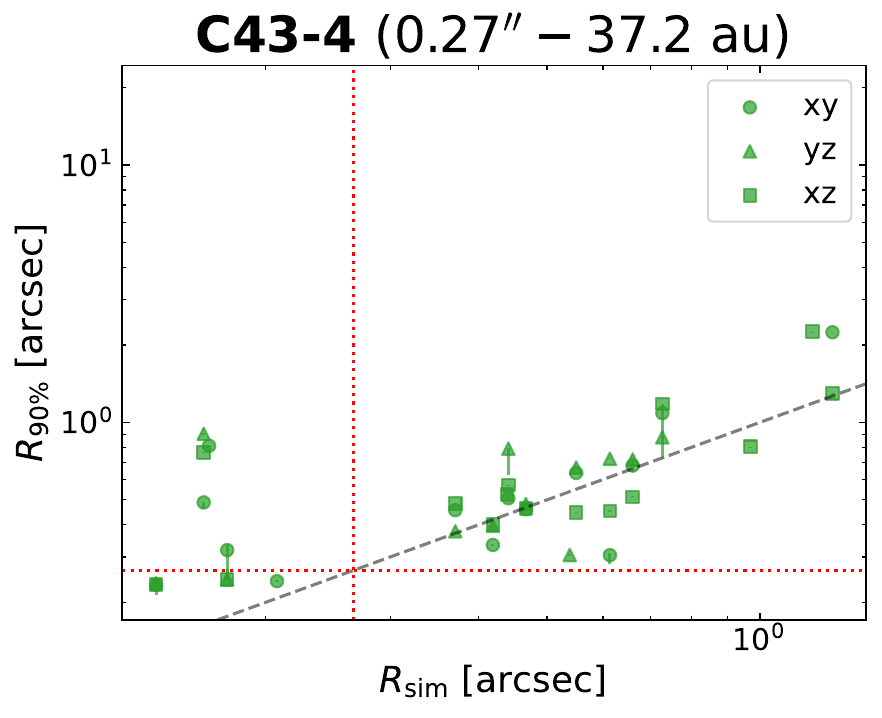}
    \includegraphics[width=0.3225\textwidth]{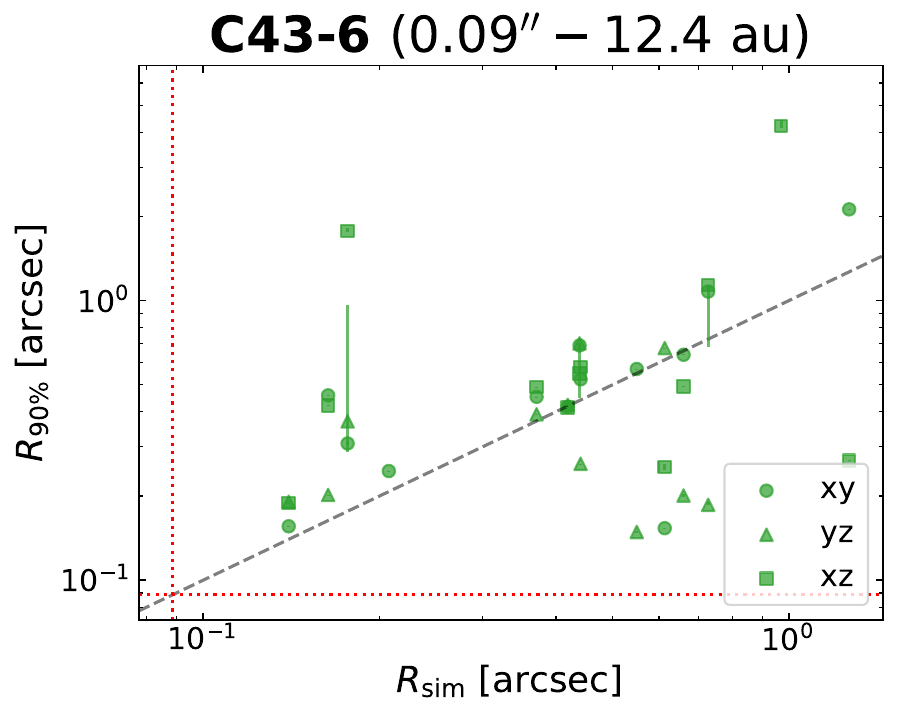}

    \begin{center}
    \includegraphics[width=0.3245\textwidth]{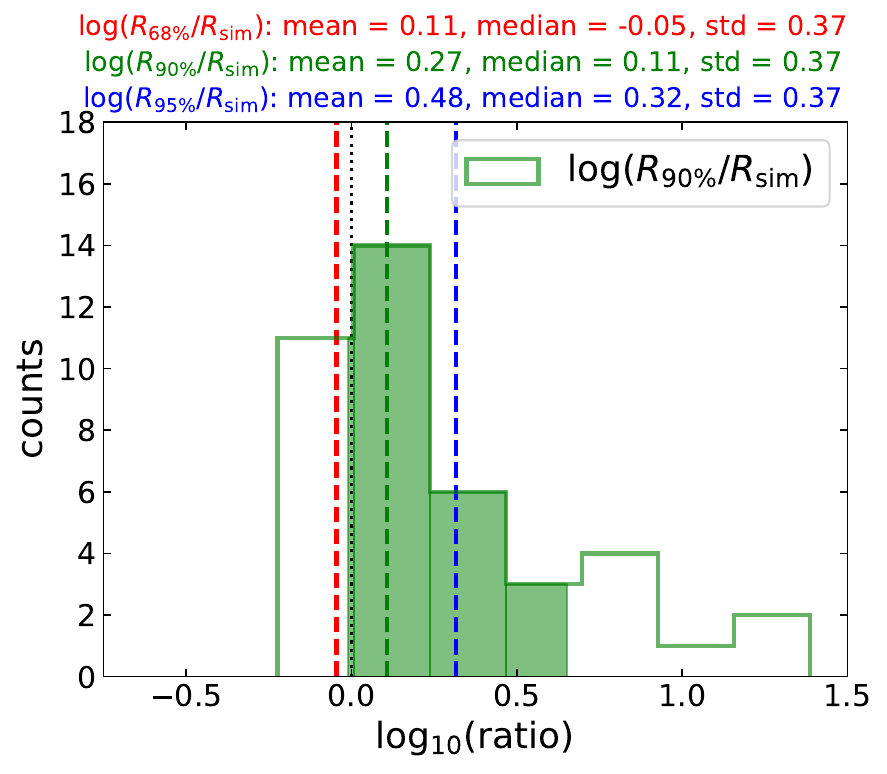}
    \includegraphics[width=0.3245\textwidth]{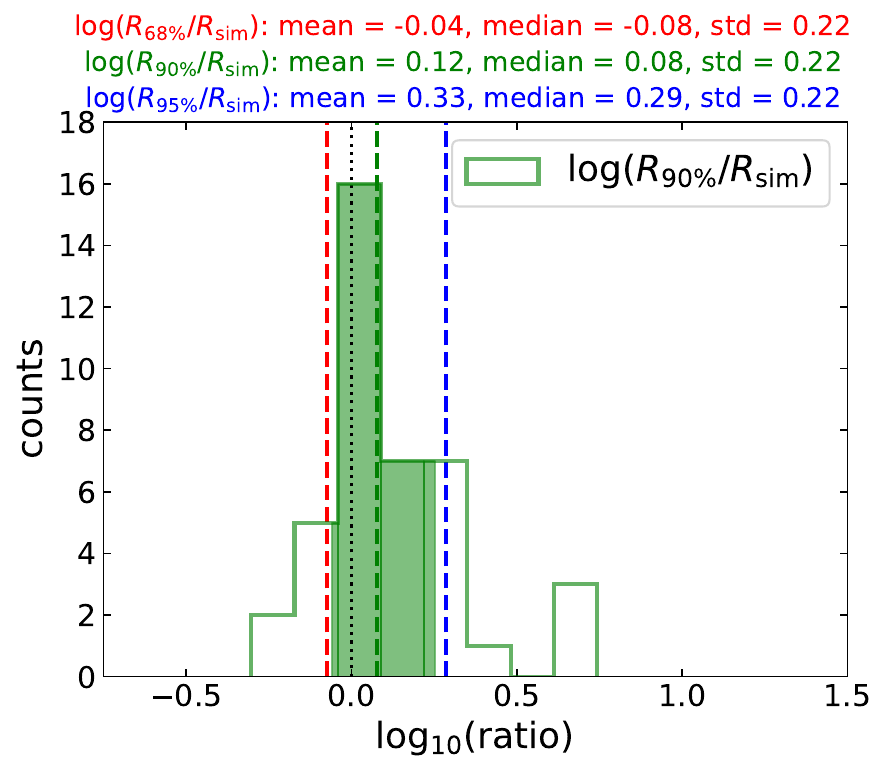}
    \includegraphics[width=0.3245\textwidth]{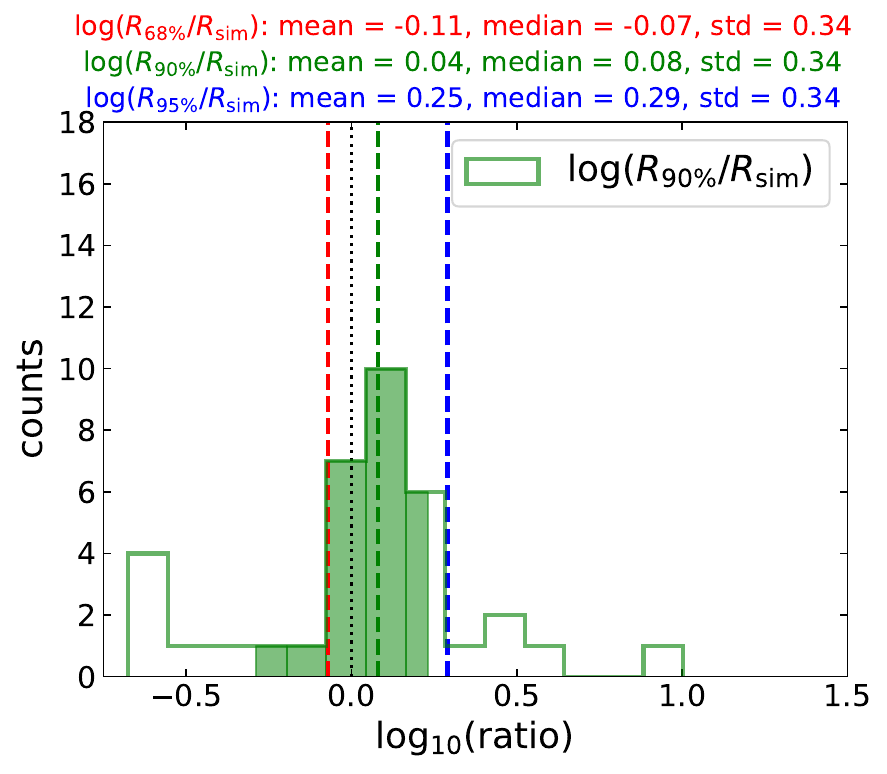}
    \end{center}
    \caption{Comparison of the disk sizes measured from the observations with three different angular resolutions ($0.41'', 0.266'', 0.887''$).{\it } $R_{90\%}$ radius of the observed disks versus their theoretical sizes for three configuration C43-3 (left), C43-4 (middle), and C43-6 (right), shown at the top. Histograms of the $R_{90\%}/R_{\rm sim}$ ratio for the reported configurations, shown at the bottom.}
    \label{fig:hist_radii}
\end{figure*}

Here, we focus on how different resolutions of the observations could affect the disk sizes obtained with our visibility modelling. For that purpose, we used the same model and fitting method on the data obtained with C43-3 and C43-6, with angular resolutions of $\theta_{\rm res} = 0.41'', 0.0887''$, respectively. Figure \ref{fig:hist_radii} shows the histogram of the ratio between the modelled radii over the simulations' radii for the data from C43-3 (left panel) and C43-6 (right panel), along with the results for C43-4 with intermediate resolution that we analyzed in the previous sub-section. We merged the projected disks in three projections into one single sample for each configuration to maximise the sample size. From C43-3 to C43-4, as the angular resolution of the observation increases, the mean and median of the $\log(R_{90\%}/R_{\rm sim})$ distribution get closer to 0, which represents unity in linear scale. The distribution for C43-4 also appears more centred and peaked, and the factor of uncertainty decreases from $\sim2.3$ to $\sim1.7$. From C43-4 to C43-6, the distribution keeps getting slightly better mean (factor of $1.1$ difference compared to $1.3$), but the uncertainty now slightly rises up to $\sim2.2$ as more outliers appear.

As suggested by the dotted red lines in the upper panels of Fig. \ref{fig:hist_radii} (displaying the angular resolution of the corresponding observations), the resolution achievable by C43-4, namely, $\theta_{\rm res}=0.266''$, allows us to sufficiently observe all the disks in this sample, whereas with C43-3, a significant portion of it falls under the resolution limit of the observation. This explains the right-skewness and the offsets of the mean and median of the $R_{90\%}/R_{\rm sim}$ ratio distribution for C43-3 and the smaller uncertainty in the case of C43-4. Furthermore, the fact that C43-4's angular resolution is very close to the smallest disk sizes in the sample means that a number of disks are only partially resolved. This is also the situation where the Gaussian profile describes best the disk emission due to the effect of the beam. C43-6, on the other hand, provides an angular resolution that is more than enough to resolve even the smallest disks in the sample and therefore suffers from larger uncertainty. This is because the well resolved disks would likely require a more complex description than a Gaussian profile. This is visible in the scatter plot for C43-6, where the few smaller disks that are not resolved (and thus overestimated) by C43-4 are now slightly better modelled with higher resolution observation; at the same time, a number of larger disks are severely underestimated as the Gaussian is trying to fit the more compact central peak rather than the more extended disk structure that is filtered out by the observation.

\subsubsection{Combined antenna arrays}

With these observations of different resolutions in hand, it is possible to alternatively combine the data during the fitting to have a better coverage of the $uv$ space and better separate the compact and extended emissions from the disk and the envelope, respectively, as done by \cite{2023A&A...676A...4C} to study the dust properties within the envelope. Here, we adopt the typical ALMA combination of the antenna arrays C43-3 and C43-6 for our modelling. The $\chi^2$ now is the sum of the values with the two data sets.

\begin{figure*}
\includegraphics[width=0.3225\textwidth]{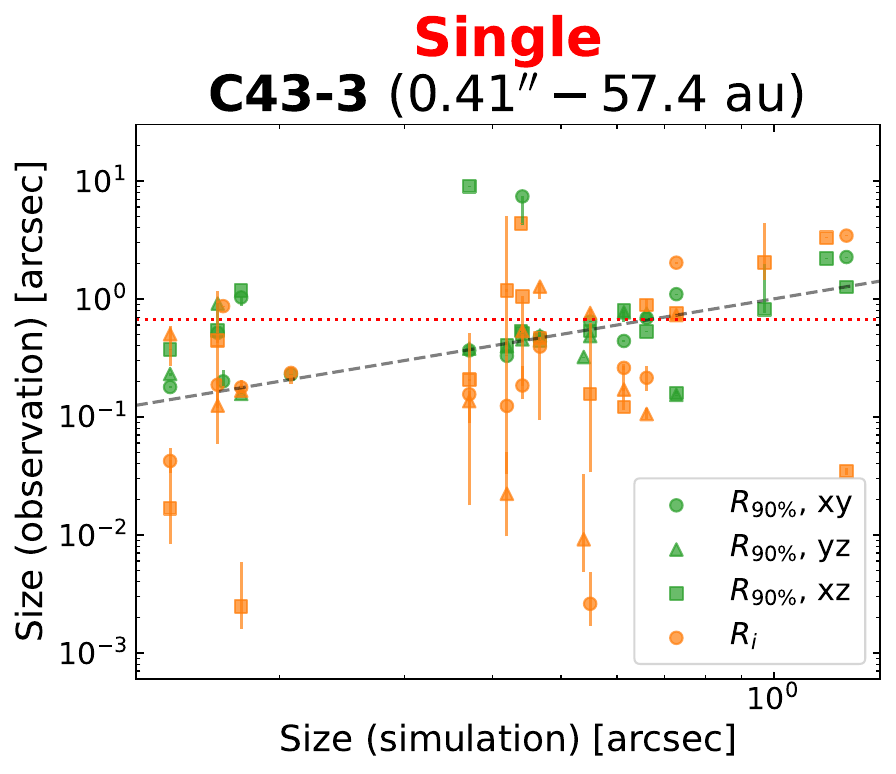}
\includegraphics[width=0.3225\textwidth]{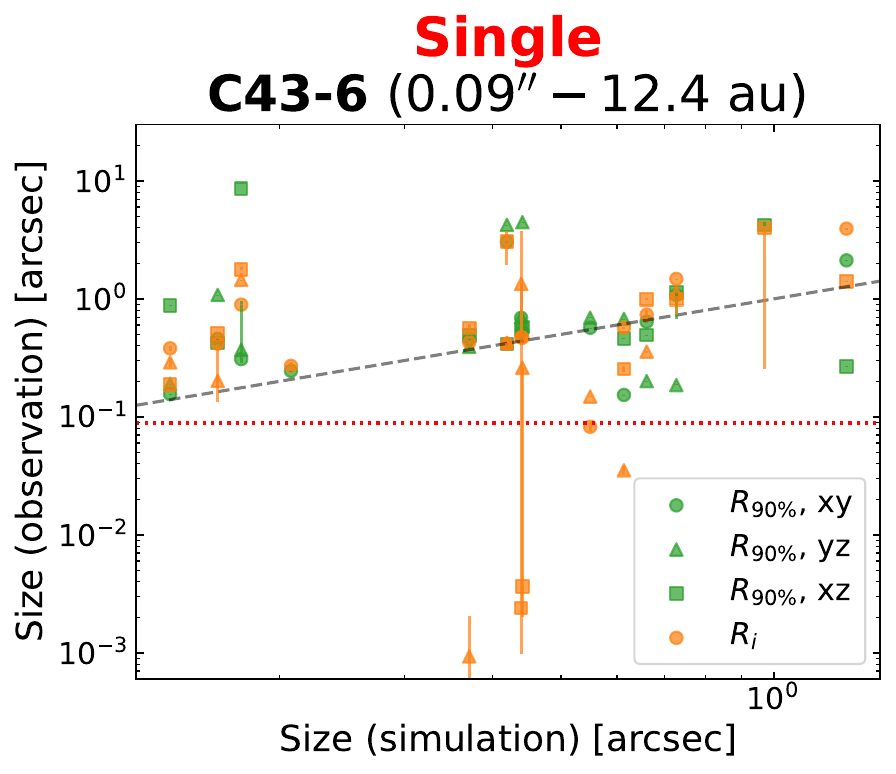}
\includegraphics[width=0.3225\textwidth]{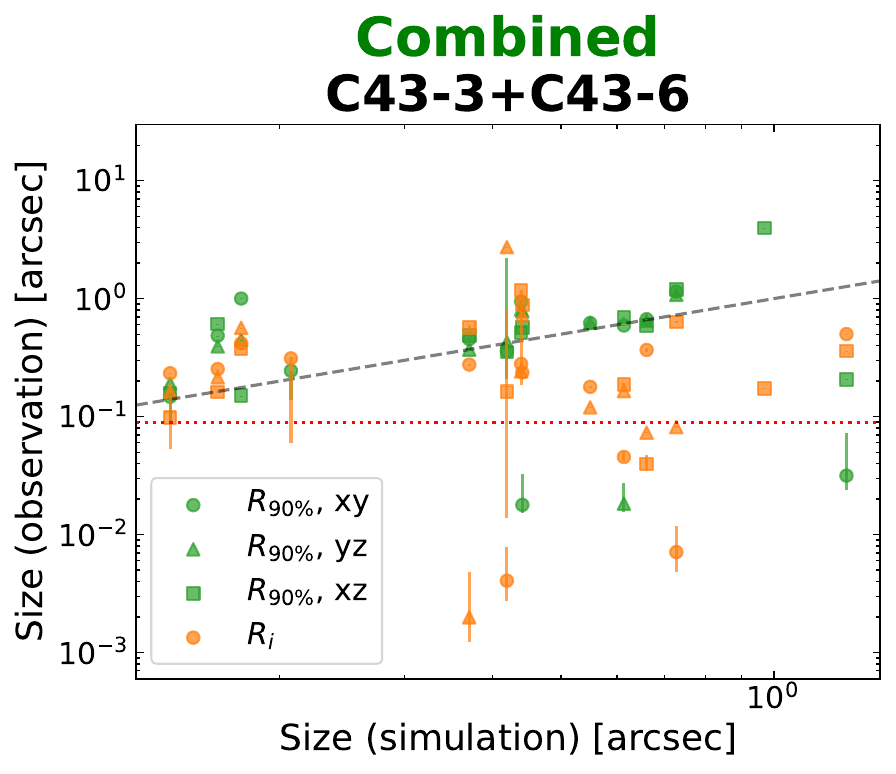}

\centering
\includegraphics[width=0.3245\textwidth]{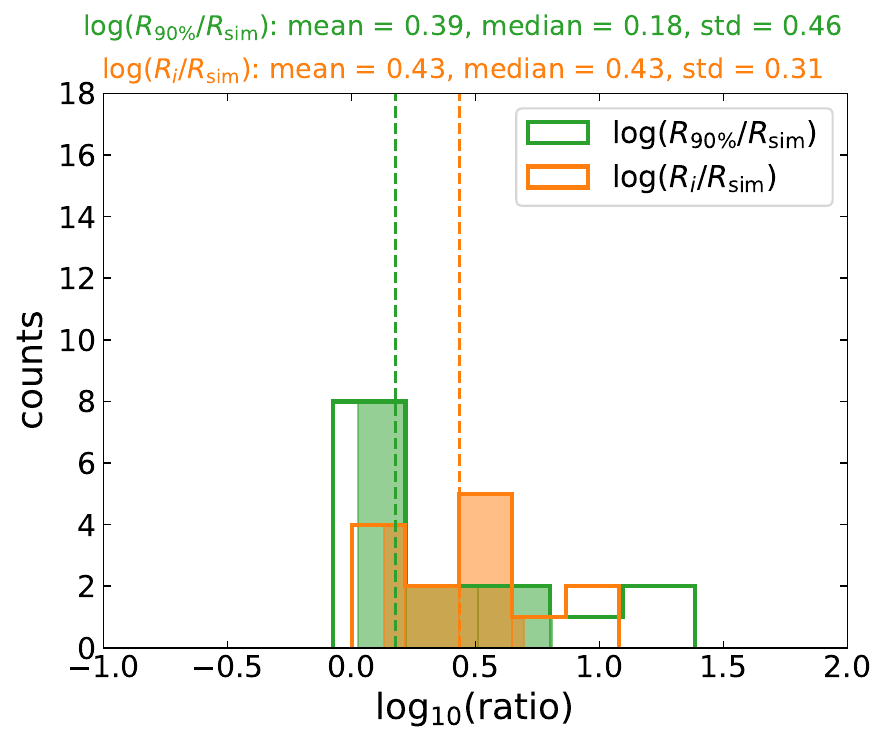}
\includegraphics[width=0.3245\textwidth]{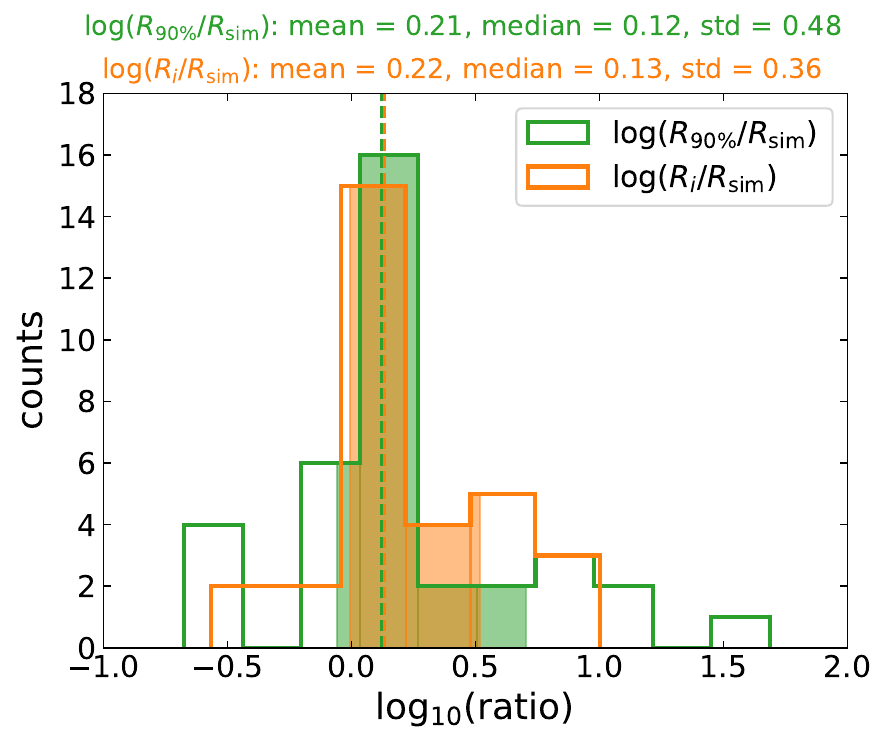}
\includegraphics[width=0.3245\textwidth]{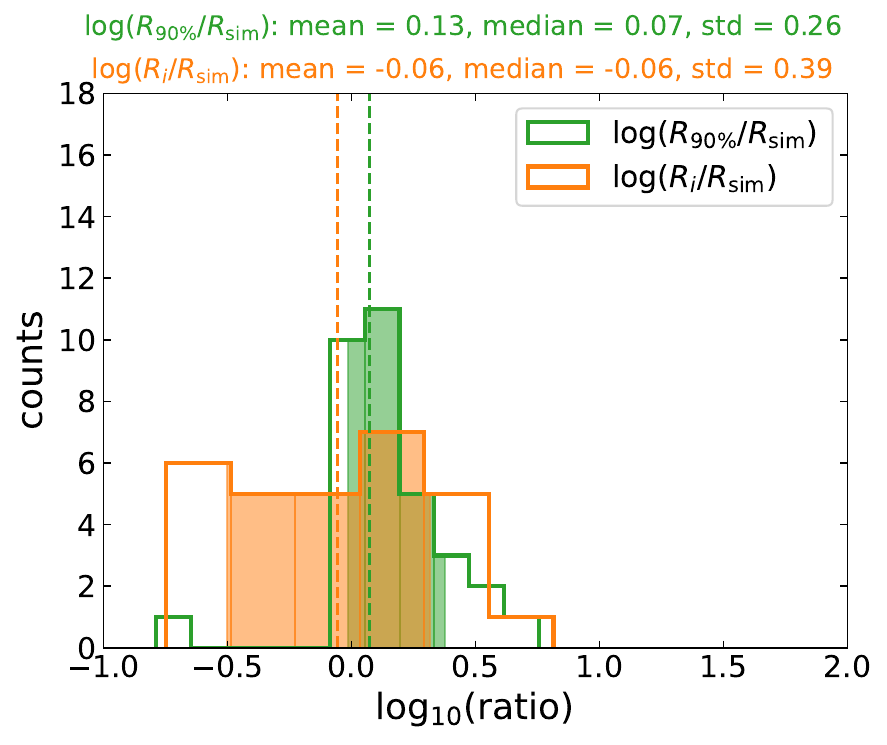}
\caption{Comparison of different fitting strategies: with data from individual configurations (left and middle panels for C43-3 and C43-6, respectively) and from combined antenna arrays (right panels for the combined C43-3 and C43-6).}
\label{fig:sigma_combined}
\end{figure*}

A comparison of the results obtained with this strategy and those with the separate visibilities from the two configurations in the previous sub-section can be found in Fig. \ref{fig:sigma_combined}. Since the disk radius, $R_{90\%}$, and the radius of the approximately uniform density region of the envelope, $R_i$, are two free parameters that are completely  unrelated to one another in the fittings (and both more or less trace the outer disk radius) we opted to compare them individually with the `true' disk size. This was done to evaluate the accuracy of the model in recovering and separating the two components, contrary to what is detailed in Sect. \ref{sec:radius_measure}. Therefore, we show (in the upper panels) the values of $R_{90\%}$ and $R_i$ radii versus the disk size from simulations and (in the lower panels) the histograms of the ratios of the two radii modelled from observations over the simulation sizes, without the fine-tuning criteria used in the two previous sub-sections. It can be seen from the scatter plots that there are cases where the values of $R_{90\%}$ and $R_i$ returned by the fits are much smaller than the angular resolution of the observations (marked by the dotted red lines). We consider those, namely, $\sigma_{\rm disk} < \theta_{\rm res}/(2\sqrt{2\ln2})$ or $R_i < \theta_{\rm res}/(2\sqrt{2\ln2})$, to be cases where the fitting failed to detect the disk (in the case of $\sigma_{\rm disk}$) or the transition between rotation and infall motions (in the case of $R_i$). Thus, we did not include them in the samples for the lower panels' statistics.

The histograms suggest that the combined data genuinely improve the measurements of the disk radii by eliminating both the underestimated points returned by the C43-6 fittings and the heavily overestimated ones by the C43-3 fittings due to the lack of resolution. It is worth keeping in mind that we obtain three non-detected Gaussian disks with the combined data; while that is not the case with C43-6, these can still be substituted with $R_i$ for the disk size (as described in the sub-sections above) since the values of $R_i$ for these disks are above the observation's resolution, which does not decrease nor increase the overall accuracy of the disk radius extraction. This indicates that the extension of the $uv$ coverage could play a significant role in subtracting the envelope's emission from that of the disk in the observations.  Consequently, this could help the Gaussian in detecting the correct disk structure, even when the statistics for the envelope's $R_i$ does not gain any enhancement. As a matter of fact, for C43-3, most of the results for $R_i$ fall below the resolution of the configuration. Thus, the distributions for C43-3 consist of very small samples which do not give meaningful statistical indications. C43-6 appears to have the best measurements for $R_i$, although they slightly overestimate the disk radius in general. The combined $R_i$ distributions have a broader spread within the 16th-84th percentile range, which displays quite a flat shape, without any real centralised peak. This may be an indication that $R_i$ might not be a meaningful parameter to study the envelope's structure with the Plummer density profile, but is nonetheless useful when employed in combination with the Gaussian radii, to provide a more accurate measurement of the disk sizes when the Gaussian encounters difficulties locating the right disk structure, as can be seen in the results for C43-6 (lower-right panel of Fig. \ref{fig:hist_radii} and lower-middle panel of Fig. \ref{fig:sigma_combined}).

\subsection{Disk masses}

Another important property of the disks is their masses, as these values offer insights into the mass budget for the formation of planets in their later evolutionary stages.

\begin{figure}
    \centering
    \includegraphics[width=0.45\textwidth]{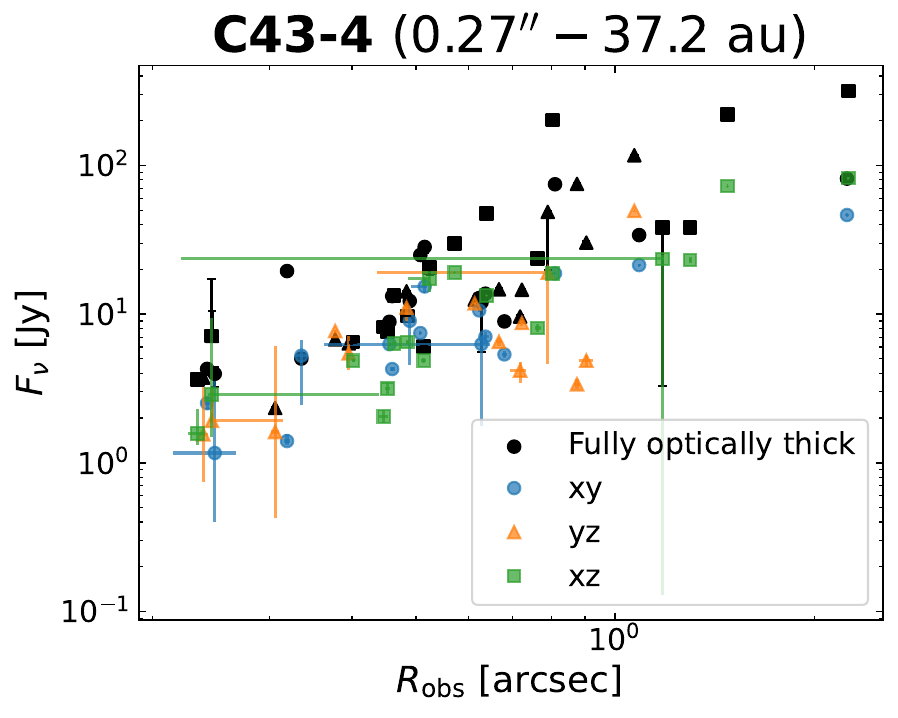}
    \caption{Integrated millimeter flux of the Gaussian disks as a function of the modelled disk radii from C43-4 data. Black points show the analytical fluxes in the optically thick regime.}
    \label{fig:flux}
\end{figure}

From the central peak intensity that we modelled with the Gaussian profile in Eq. (\ref{eq:model}), we performed an integration to get the millimeter flux, $F_{\nu}$, of the disks. These fluxes are shown in Fig. \ref{fig:flux} as a function of the modelled disk radii, $R_{\rm obs} = R_{\rm 90\%}$. As a reference, we plotted the fluxes derived analytically from $R_{\rm obs}$ while assuming that the disks are fully optically thick:
\begin{equation}
    F_{\nu, {\rm th}} = \frac{1}{d^2} \int^{R_{\rm obs}}_{0} B_{\nu}(T_{\rm dust}) 2\pi r dr.
    \label{eq:flux}
\end{equation}
Here, $d\sim140\pc$ is the distance to the source and $B_{\nu}$ is the Planck function. The dust temperature radial profile $T_{\rm dust} (r)$ for each disk is inferred from the 3D profile in the simulation by fitting a power-law function of $T(r) = T(R_0) (r / R_0)^{-v}$ to the radially averaged temperature shown in Fig. \ref{fig:temp_radial}. Using the 16th, 50th, and 84th percentiles of the fitted parameters, we found $\bar{T}(R_0) = 181_{-35}^{+74}\K, \bar{R_0} = 35_{-12}^{+36}\au$ and $\bar{v} = 0.52_{-0.05}^{+0.03}$. To derive the dependency of the temperature on the stellar luminosity, $T(R_0)$ for the disks in the sample is then fitted as the power-law function of their central stars' luminosities, $L_{\star}=L_{\rm int}+L_{\rm acc}$, such that $T(L_{\star},R_0) = T_0 (L_{\star}/L_{\star})^u$. This yields the following temperature description:

\begin{equation}
    T_{\rm dust}(L_{\star}, r) \simeq 71\K \left( {L_{\star} \over L_{\odot}} \right) ^{0.25} \left( {r \over 35 \, \au} \right) ^{-0.52}.
    \label{eq:Tscaling}
\end{equation}

\begin{figure}
    \centering
    \includegraphics[width=0.5\textwidth]{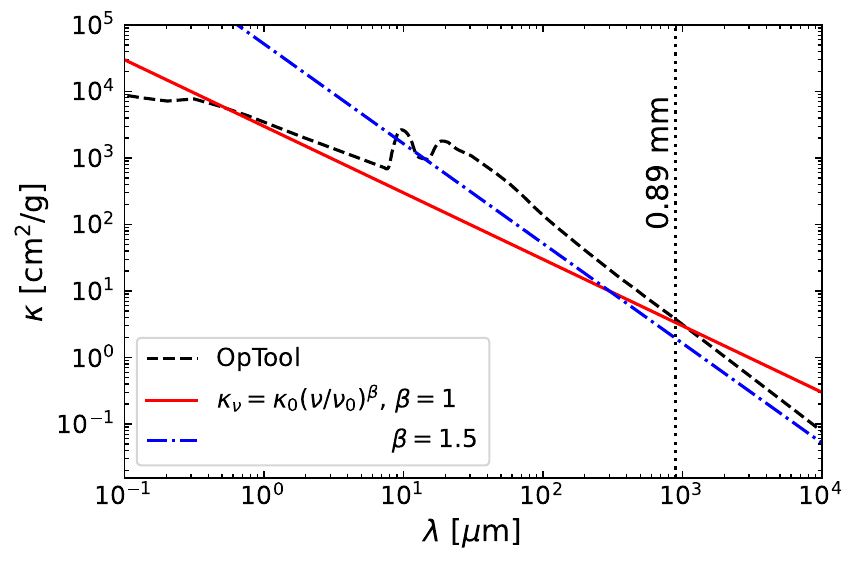}
    \caption{Absorption opacity from \textsc{OpTool} (dashed black curve) used in Sect. \ref{sec:radmc} and from the frequency-dependent power law approximation with the powers of 1 and 1.5 (red and blue lines, respectively).}
    \label{fig:opacity}
\end{figure}

Next, we compute the dust mass in each disk with the optically thin emission assumption:
\begin{eqnarray}
M_{\rm dust} = \frac{F_{\nu}d^2}{\kappa_{\nu}B_{\nu}(T_{\rm dust})},
\label{eq:mass}
\end{eqnarray}
where $\kappa_{\nu}$ is the dust opacity at the observations' frequency. This assumption likely does not hold for the young disks in consideration in our study. Indeed, Fig. \ref{fig:flux} indicates that the fluxes of our modelled disks are quite close to the optically thick fluxes, with a difference that is within one order of magnitude. However, we aim to pinpoint how far the results from this simple approximation could be from the masses of the identified disk material. For the dust opacity, we used the frequency-dependent power law approximation, $\kappa_{\nu}=\kappa_0(\nu/\nu_0)^{\beta}$, with $\beta$ being the dust emissivity index. Here, we adopted the commonly used value $\kappa_{0}=10~{\rm cm^2/g}$ for the normalisation coefficient and $\beta=1$ for the emissivity index in PPDs (\citealt{2006ApJ...636.1114D}; \citealt{2005ApJ...631.1134A}) in accordance with our chosen value of $q=3.5$ for the grain size distribution. Figure \ref{fig:opacity} compares the opacity from the power law approximation and the one from \textsc{OpTool}. We note that at the wavelength of interest in this study, $\lambda=0.89~{\rm mm}$, we get roughly the same value, namely, $\kappa_{\nu}\sim3.5~{\rm cm^2/g}$.

Another factor of uncertainty in the conversion comes from the average dust disk temperature ($T_{\rm dust}$). This temperature has been, for a long time, taken as a constant value for a specific evolutionary stage (see, e.g. \citealt{2020A&A...640A..19T}, \citealt{2005ApJ...631.1134A}, \citealt{2016ApJ...831..125P}, \citealt{2016ApJ...828...46A}, \citealt{2018ApJ...859...21A}, \citealt{2020A&A...633A.114S}). More recently, \cite{2020ApJ...890..130T} used a grid of radiative transfer models with \textsc{Radmc-3d} to sample the parameter space of a flared disk model embedded within a rotating envelope and found that for a protostar of $1~L_{\odot}$, a $50\au$ circumstellar disk has an average dust temperature of $43\K;  $  in addition, the general average temperature scales with the luminosity as $\propto L^{0.25}$, as well as with the disk radius, as $\propto R^{-0.5}$. We note that this dependency on the disk radius is close to what we found with the radial temperature profile in our simulation, to the power of $-0.52$; however, is generally lower by a factor of $\sim1.3$. In the following analyses, we chose to put some restriction on the uncertainty in the flux-to-mass conversion by assuming that we have a good knowledge of either the temperature scaling of our disk sample in relation to the disk radius and to the star's luminosity (e.g. from the literature), but not the radius itself; or, alternatively, of the median temperature of the sample. The latter is aimed at mimicking the use of a constant temperature for the mass determination of a disk population. With that in mind, we use the disk radius, $R_{\rm obs} = R_{90\%}$, from the synthetic modelling to estimate the disk temperature based on the power-law fits of the \textsc{Ramses} radial temperature profiles shown in Fig. \ref{fig:temp_radial}. In particular, we employed the temperature scaling in Eq. (\ref{eq:Tscaling}) in parallel with the median temperature of $\bar{T}_{\rm dust} = 122\K$ for the entire sample, which was determined by the median values of the fitted parameters.

\begin{figure*}
    \includegraphics[width=0.51\textwidth]{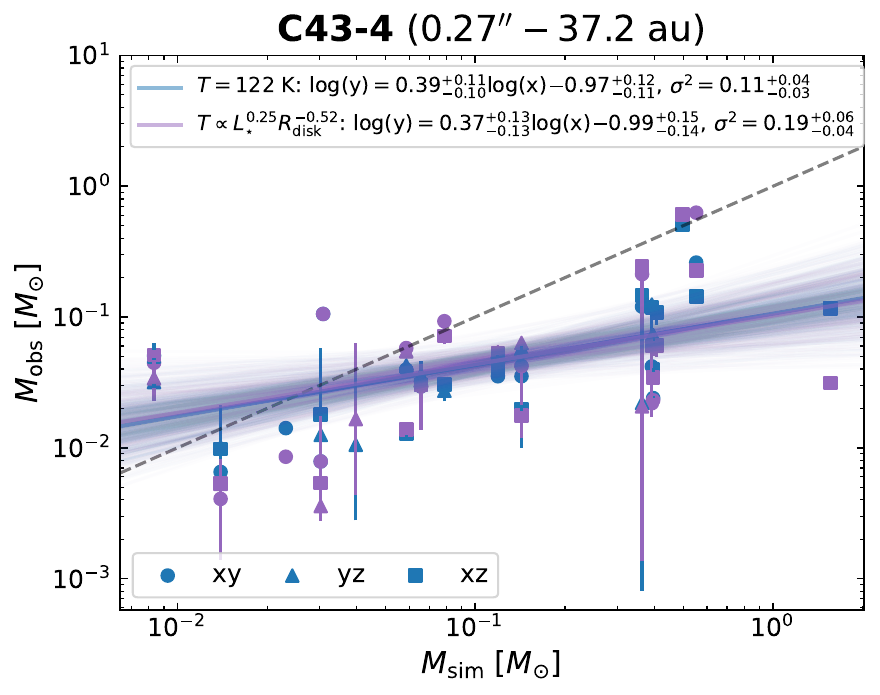}
    \includegraphics[width=0.49\textwidth]{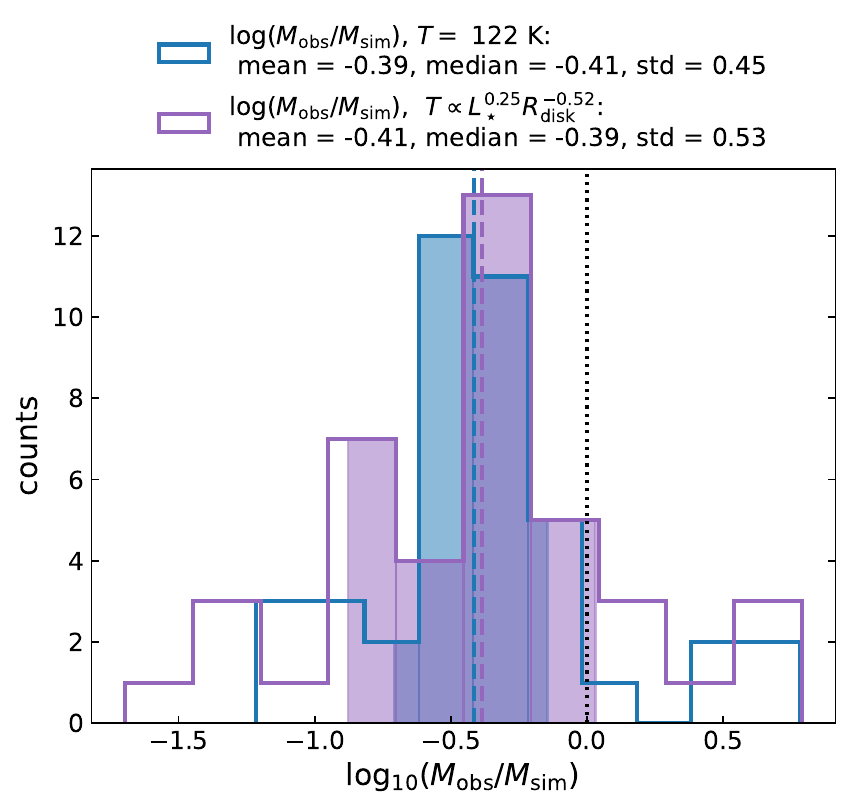}
    \caption{Scatter plots of the modelled disk masses using a constant dust temperature $T=123\K$ for all the disks (in blue) and a temperature scaling based on the disk radius $T_{\rm dust}(R_{\rm disk}) = 1122.76 (R_{\rm disk}/1~\au)^{-0.52} \K$ (in purple) versus the masses inferred from the simulation (shown on the left). The log-linear fits to the data points with corresponding $1\sigma$ confidence intervals based on the 16th and 84 percentiles are plotted with the solid lines. Histograms of the $M_{\rm obs}/M_{\rm sim}$ ratios in logarithmic scale. Vertical dashed blue and purple lines indicate the medians of the ratio distributions (shown on the right).}
    \label{fig:disk_mass}
\end{figure*}

Since the disk component is better recovered with the intermediate-resolution single-configuration setup as discussed in the previous sub-section, here we show only the results for C43-4, which gives the best values for the disk sizes (see Fig. \ref{fig:hist_radii}). We also include the plots for the other configurations in Appendix \ref{app:mass}.
Figure \ref{fig:disk_mass} presents the masses derived from this approximation, in comparison with the values inferred from the material that belongs to the disks based on the gas kinematics in the simulation. Similarly to the way we obtained the disk radii, we calculated the ratios of the masses in the modelling over the masses inferred from the simulations, $r_M = M_{\rm obs}/M_{\rm sim}$. The left panel shows the values obtained from our modelling, with constant and scaled temperatures versus those derived from the gas kinematics from the simulation; while the right panel the histograms of the logarithms of their ratios $r_M$. The error bars come from the uncertainties in the peak flux, $I_0$, and the flux fraction of the disk, $f$. Unlike the disk radius, the mass cannot be retrieved with a good level of accuracy.
We also provided  a simple log-linear fit to the data points in the scatter plots using the \texttt{linmix}\footnote{https://github.com/jmeyers314/linmix} library \citep{2007ApJ...665.1489K}. In general, we are underestimating the disk mass using the proposed temperature, as indicated by the shaded 16th-84th percentile uncertainty ranges in the histogram in the right panel. Notably, all of the log-linear fits return very shallow slopes ($\sim 0.37-0.39$), compared to the $x=y$ line which represents the perfect match between the values on the two axes. This $M_{\rm obs} \propto M_{\rm sim}^{\alpha}$ where $0 < \alpha < 1$ relation between the observed and the `real' masses would translate into larger negative offsets when we reach very high disk masses, suggesting a discrepancy of one order of magnitude toward the most massive disks ($\sim 1M_{\odot}$). We note that here we use $R_{90\%}$ for the disk radius to avoid the abuse of prior knowledge of real objects, which slightly overestimates their `actual' sizes and, hence, underestimates their temperatures. Otherwise speaking, the dust temperatures currently used to derive the masses are lower than the ones obtained with the power-law fits given by the black solid lines in Fig. \ref{fig:temp_radial}. The `correct' $T_{\rm dust}$, namely, the one that aptly describes the temperature profile of the disks in the simulation, could in fact reduce a little further the modelled disk masses. Interestingly, the results with the constant $T_{\rm dust}$ and $T_{\rm dust} \propto L_{\star}^{u}R_{\rm disk}^{-v}$ are not in too much of a disagreement, with the latter even giving a slightly larger spread.

Now, if we relax the assumption on the prior knowledge of the temperature, an alteration of $T_{\rm dust}$ can shift the masses of the sample upward or downward. For instance, a much lower temperature such as $T_{\rm dust} = 30\K$ used by \cite{2020A&A...640A..19T} or $T_{\rm dust} = 43\K$ for $50\au$ disks around $1~L_{\odot}$ protostars in the VLA/ALMA Nascent Disk and Multiplicity (VANDAM) survey by \cite{2020ApJ...890..130T} would increase  $M_{\rm obs}$ significantly; however, the change would be systematic throughout the population. 

In short, the disk masses cannot be easily recovered from the millimeter fluxes. This can be attributed to two possible sources of uncertainty, the conversion formula in Eq. (\ref{eq:mass}) and the optical thickness of the early Class 0/I disks in question. While the validity of the former is debatable, the latter can be confirmed by the small difference (within less than one order of magnitude) between the fully optically thick fluxes and the fluxes measured from our synthetic observations. The shallow slopes of the observed-versus-real-mass plots in Fig. \ref{fig:disk_mass} also point to the effects of high optical depth in hindering the estimates of the masses from the fluxes in the case of more massive disks, which is a confirmed obstacle in the modelling of young disks from millimeter observations (\citealt{2022ApJ...941L..23M}; \citealt{2023ApJ...954...69S}).

\subsection{Disk inclinations}

\begin{figure}
    \includegraphics[width=0.5\textwidth]{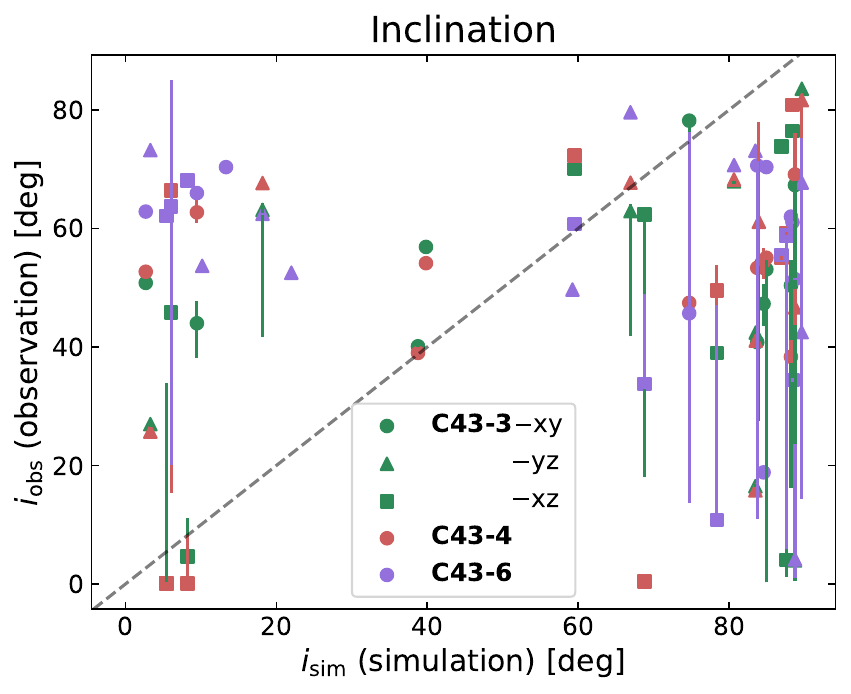}
    \caption{Modelled inclinations of the Gaussian disk versus the inclinations of the disks in simulation based on the angular momentum vector orientation.}
    \label{fig:disk_incl}
\end{figure}

Another property that we have access to in our modelling is the disk inclination, $i$, which can give insights into how the 2D projection of the disk on the plane of the sky affects the observed fluxes. In practical terms, this can be directly inferred from the semi-major axis, $a,$ and semi-minor axis, $b$, of the projected elliptic Gaussian disk as $\cos(i) = a/b$. In the simulation, we measured this quantity by determining the orientation of the angular momentum and calculated the angle between this vector and the observation plane.

These results are depicted in Fig. \ref{fig:disk_incl}. Though most of the disks in our simulation are close to either edge-on ($90^{\circ}$) or face-on ($0^{\circ}$) geometries when seen in the three native projections $xy$, $yz$, and $xz. $ This is due to the numerical tendency for the disks to align with the Cartesian axes of the grid, they are not correctly reconstructed in our modelling from observations. The face-on ones appears much more inclined, probably due to the fact that the original disks are mostly not circular and an elongated structure with some inclination is needed to account for the asymmetry and irregular shapes. In other words, the emission profile of an asymmetric, but face-on disk in the simulation, when it is modelled as a 2D elliptical Gaussian, may be erroneously interpreted in our analysis as a circularly symmetric, but inclined disk. This could potentially also be the case for an actual observation or a `real' disk modelled with the same method. The presence of strong non-symmetric structures, such as spirals or streamers, in some cases also interferes significantly with respect to how the disks’ elongation is identified (a typical example could be seen in the nearly face-on disk in Fig. \ref{fig:sink30}, which is modelled as a larger and much inclined disk from the C43-4's observation). Thus, such realistically well grounded scenario supported by observations (see, e.g., \citealt{2020NatAs...4.1158P}; \citealt{2016Natur.538..483T}) may have an important impact on the continuum modelling outcome of young disks using simple analytical prescriptions. Conversely, the edge-on disks are obtained with less inclination with respect to the face-on position. This is somewhat expected in our type of modelling due to the impossibility to obtain an infinitely flat Gaussian emission required for the $90^{\circ}$ edge-on inclination. 

Another possible explanation could be the increase in the apparent height when these disks are viewed from the side, especially for the points that are too far off in the plot where the $90^{\circ}$ inclination is reproduced with a model fit close to face-on $0-20^{\circ}$ structures. One example is the increase caused by the envelope back-warming that can heat the disk to higher temperature (\citealt{1991ApJ...376..636B}; \citealt{1994ApJ...420..326B}; \citealt{1997ApJ...474..397D}; see also \citealt{2019A&A...623A.147A}) and potentially enhance its flux in the perpendicular direction to blur its elongation. In that same vein, \cite{2023MNRAS.526.2566H} recently attempted to revisit the observational study of a subset of protostars in Orion from the VANDAM survey by \cite{2020ApJ...890..129K}, which excluded protostellar disks as a source of the observed irregular mm/sub-mm emission due to the lack of elongation in the supposed edge-on emission. In particular, the authors investigated the effect of the back-warming of the envelope on the gas pressure scale height of the disk. They reported that in the presence of an envelope around the disk, its irradiation onto the disk could significantly increase its millimeter flux, but only raise the scale height moderately. Here, we suggest an additional effect which acts on the disk scale height when looking at the geometry of the emission, also caused by the envelope back-warming, which could contribute to the arguments in favor of the disk geometry regarding the peculiar observed structures. It is worth pointing out that, contrary to the face-on cases, the edge-on results have greater error bars that indicate the possible range of values is very large, oftentimes covering the `original' inclination found in the simulation.

\section{Discussion}\label{sec:discussion}

\subsection{Implications for disk modelling from ALMA observations}

Our analyses with simple modelling of the synthetic ALMA observations with a Gaussian disk and a surrounding spherical envelope suggest that with angular resolution enough to resolve the disk component, we could reproduce their size properties with a good accuracy, within a factor of $1.6-2.2$ in uncertainty. This idealised model, despite being somewhat simplified, is still being extensively used in the flux modelling of protoplanetary disks and protostellar envelopes from continuum observations (\citealt{2019A&A...621A..76M}; \citealt{2023A&A...676A...4C}). It is now reassuring that with this method, we can still recover  the disk sizes within acceptable uncertainty relatively
well. Notably, the compact configuration C43-4 with which we are simulating for the close star-forming regions, whose physical resolution is $\approx37\au$, gives the best results in term of uncertainty for the disks of radii between $50-150\au,$ when those disks are modelled with a 2D Gaussian. This can be scaled for more distant regions to determine the most suitable antenna configuration and observing time for observations. 

Our further analyses with varied angular resolutions and combined antenna arrays showed that having not enough angular resolution, that is, a resolution lower than the `actual' sizes of the disks will result in over-estimating their radii. Very high resolutions, on the other hand, do not really improve the accuracy of the idealised Gaussian disk models and could potentially degrade the large-scale component (see the lower right panel of Fig. \ref{fig:hist_radii}). A more complex disk model, such as one with a power-law intensity profile and a rotating circumstellar envelope, potentially coupled with radiative transfer as in \cite{2022ApJ...929...76S}, would probably help in the case of more resolved disks. 

The disk masses, on the other hand, are poorly modelled, even when we have a reasonable estimation of the temperature, and overall underestimated with the optically thin conversion from the millimeter fluxes, which is a common method to this date to derive the mass in the modelling of young disks from continuum observations (\citealt{2018ApJS..238...19T}; \citealt{2020ApJ...890..130T}; \citealt{2020A&A...640A..19T}). As it stands, this would challenge our current understanding of the mass budget for planet formation. The poor correlation between the masses of the Gaussian disks and the masses estimated in the simulation suggests that either the conversion between the millimeter flux to the dust mass is to be revised, or the wavelength at which we performed our synthetic observations needs to be extended, so that the disks are less optically thick to satisfy the assumption of Eq. (\ref{eq:mass}). The fact that the fluxes that we modelled for the disks are relatively close to their fluxes in the optically thick regime (as suggested by Fig. \ref{fig:flux}) strongly supports the latter, while not excluding the former. We believe that observations at longer wavelengths, for example, those covered by ALMA Band 1 ($6-8.6~{\rm mm}$), could help us to probe better the disk mass, as \cite{2021MNRAS.506.2804T} showed that at $3.1~{\rm mm}$, the optically thick fraction of the fluxes of Class II disks in the Lupus star-forming region is already significantly reduced compared to at $0.9~{\rm mm}$.

\subsection{More simplistically: Considering whether a single Gaussian be adequate}

\begin{figure*}
    \includegraphics[width=0.495\textwidth]{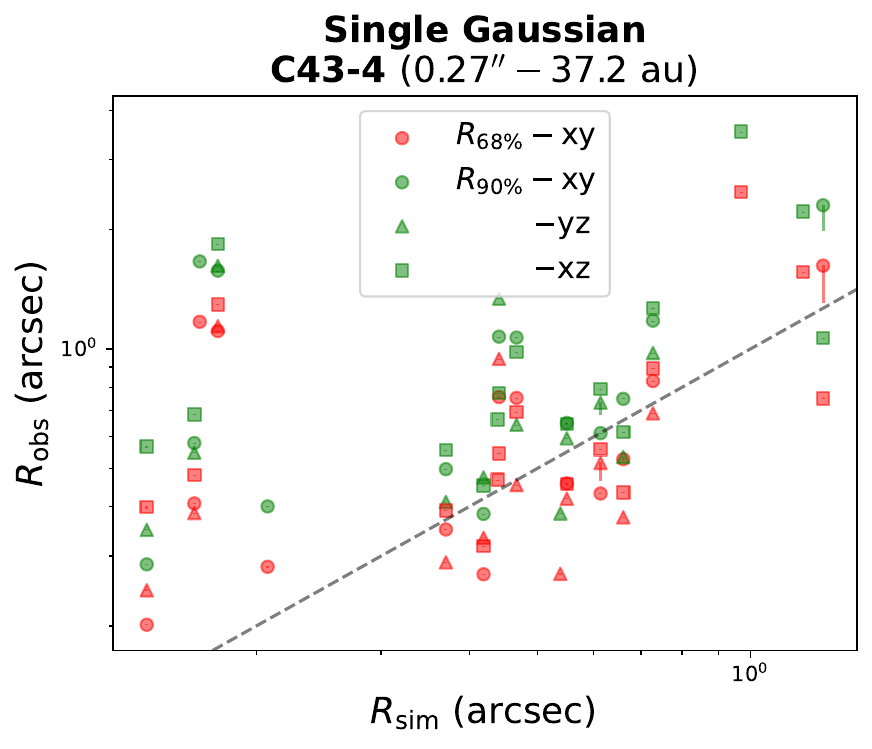}
    \includegraphics[width=0.505\textwidth]{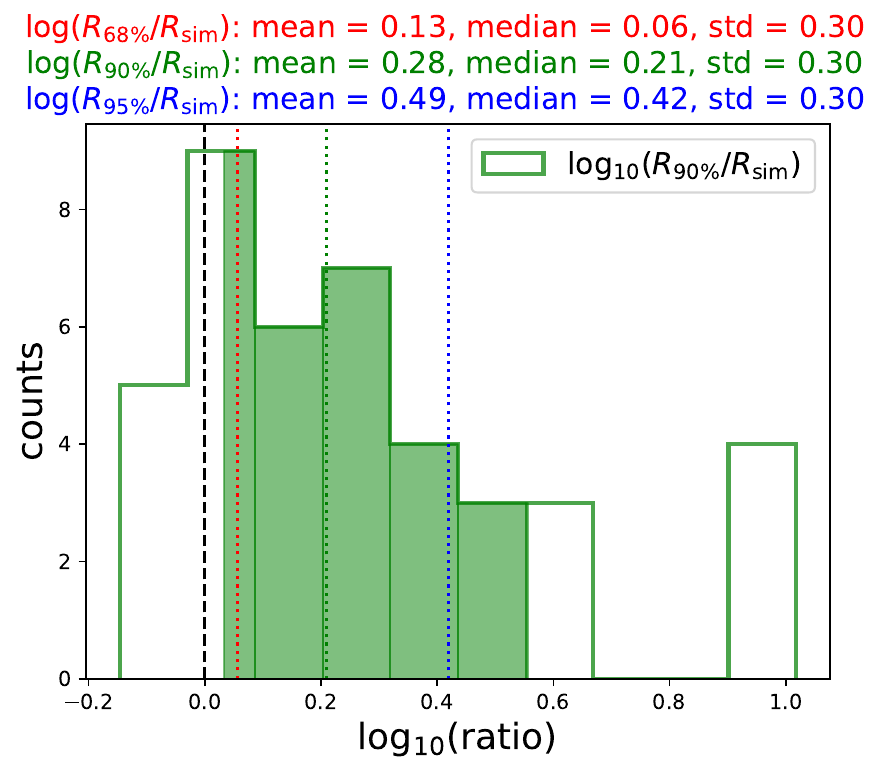}
    \caption{Disk sizes in arcsecs obtained from the modelling of C43-4 observations with a single Gaussian model vs. their radii as inferred from the simulation (left panel) and the distribution of the ratio between the observed and theoretical radii (right panel).}
    \label{fig:disk_size_Gaussian}
\end{figure*}

The simplest analytical prescription that often comes to mind when attempting to model the isolated disk emission is probably the single elliptic 2D Gaussian, thanks to its limited number of parameters. This limited set enables modellers to easily fit huge samples of sources in large surveys without massive computational trade-off, at the risk of missing more refined and complex structures. In the literature, the Gaussian has often been used for more evolved, isolated Class II disks (see, e.g. \citealt{2017A&A...606A..88T}; \citealt{2019A&A...628A..95M}; \citealt{2020A&A...633A.114S}), but sometimes also for embedded disks in younger class I or even class 0 stages (see, e.g. \citealt{2020ApJ...890..130T}; \citealt{2023ApJ...951....8O}).

With our simulation and synthetic observation data, we can substitute the two-component model presented in Sect. \ref{sec:method} by the single Gaussian and re-employ the fitting methodology to test the accuracy of this simplistic model. The results for the disk radii measured with C43-4's data, which provided us with the best estimations of the disk sizes with the disk-envelope model, are plotted in Fig. \ref{fig:disk_size_Gaussian}. Owing to the limitations of the single Gaussian in detecting the compact, embedded structures in the presence of the extended emission of the envelope, most of the disks (especially the partially resolved ones) are over-estimated in size; this drives the means and medians of the ratios for all three radii $R_{68\%}$, $R_{90\%}$, and $R_{95\%}$ above $1$, with an excess of very large $R_{90\%}/R_{\rm sim} \sim 10$ values on the far right. Combined with the fact that the $1\sigma$ confidence interval of $\log(R_{90\%})$ lying entirely on the positive side, this is an indication that the Gaussian tends to bias toward larger disk radii by an average factor of $\sim 2$. Thus, this overly simplistic model is inadequate for the type of embedded disks formed in our simulations  and studied in this paper.

In real applications of such methodology to modelling young sources where the contribution of both the disk and the envelope are equally important, we caution that the disk sizes may not be accurately measured. Hence, more complicated properties such as the masses (provided that the optical thickness allows for a better conversion from the millimeter fluxes) would consequently be wrongly determined. 

\subsection{Considering the comparison between simulations and observations of planet-forming disks}

The overarching goal of the post-processing and synthetic observation study presented in this paper is to provide a better connection between numerical simulations and interferometric observations of young disks in Class 0/I systems, with detailed physics of the transfer of radiation and various instrumental effects included. As a result, the outputs of our pipeline from the simulation results could then be used for the comparison of the disk properties with dust continuum observations. This could also help constraining the initial ingredients of the numerical simulations of disk formation.

Interestingly, there has been significant tension between simulations and observations regarding the theoretically and observationally constrained disk mass (\citealt{2018MNRAS.475.5618B}; \citealt{2020ApJ...890..130T}; \citealt{2023arXiv231019672L}), although the disk radii show relatively good agreements (\citealt{2021ApJ...917L..10L}; \citealt{2023arXiv231019672L}). This difference by one to two orders of magnitude  in the simulated and the observed disk mass adds another layer of complexity to our question, which specifically asks whether the mass budget of the disks is sufficient to form Solar-like planetary systems according to the Minimum Solar Mass Nebula (\citealt{1981PThPS..70...35H}; see also, e.g. \citealt{2007ApJ...671..878D}) and whether simulations and observations are correctly predicting these masses. Here, we have demonstrated that the masses modelled from the observed millimeter fluxes could be $2.5-10$ times lower than their actual masses. By extending this study to different simulation models and observations of various disk populations from various clouds, it is possible to constrain the choice of initial parameters that would be able to reconcile this current mismatch. 

Ultimately, our final goal would be to explore, with the standardised methodology presented in this paper, the larger parameter space to put more constraints on the physics in the simulations, and more broadly, theories of the evolution of PPDs. This includes the dust composition and distribution that determine the dust opacity, which in turn controls the results of the radiative transfer, as well as more complex processes such as dust growth and fragmentation (\citealt{2023MNRAS.518.3326L}). The changes in the opacity could consequently lead to a difference in the way the disk masses would be obtained, although it wold not be likely to optimally effect at the current $0.89~{\rm mm}$ wavelength, which urges a multi-wavelength extension to other ALMA bands of this study. The disk sizes would not potentially  be overly affected due to the constant dust-to-gas ratio assumption of $0.01$ throughout the simulation grid, which is another strong caveat that ought to be addressed. In addition, we must not neglect the criteria of the disk extraction from the gas kinematics, which can alter the theoretical disk properties that we are currently taking blindly as the `ground truth'. The same goes for the techniques used to model the observed fluxes of the disks in observations, as suggested by the difference between disk sizes obtained with the one-component and two-component models. Different tracers and observation techniques, such as molecular lines, channel maps, and PV diagrams, have also been shown to produce uncertainties in the disk physical quantities. This was demonstrated in
detail by \cite{2020ApJ...905..174A} with regards to the disk size using single core collapse simulation and synthetic observations.

Lastly, we would need also to address other shortcomings of our simulation models, such as the accretion luminosity, $L_{\rm acc}$, that is so far converted from the accretion gravitational energy with an efficiency factor, $f_{\rm acc}$, taken quite arbitrarily as $0.1$ in this study or additionally $0.5$ in \cite{2023arXiv231019672L}. The value of $f_{\rm acc}$, which eventually determines $L_{\rm acc}$, has been shown to impact not only the radiative transfer photon energy, but also the disk temperature and fragmentation from a numerical standpoint, even though it has little impact on the disk size and mass. Sink particles, which are a sub-grid modelling of stars due to the lack of resolution to resolve the disk-star interaction, with their own set of fine-tuning parameters including the accretion threshold, $n_{\rm thre}$, and sink radii, contribute further to the uncertainty of the disk properties modelled with our simulations. In particular, \cite{2020A&A...635A..67H}  showed that the choice of $n_{\rm thre}$ could significantly alter the disk mass, which has been studied extensively in this paper. As a side note, we refer to \cite{2023arXiv231019672L} for a complete discussion on the caveats of the simulation models. Such detailed investigation will be presented in our follow-up studies.

\section{Summary}\label{sec:summary}

In this paper, we study two critical disk parameters in Class 0/I systems, namely, the disk size and mass, by means of realistic radiative post-processing and synthetic ALMA observations of MHD simulations combined with observational modelling techniques. Our principal results can be summarised as follows:

\begin{enumerate}

\setlength\itemsep{1em}
  
\item We developed a procedure to compare disks' observational properties with the properties in the simulations, including 3D Monte Carlo radiative transfer with the \textsc{Radmc-3d} code, mock interferometric observations with the \textsc{Casa} software, and parameter fitting with the \texttt{galario} library.

\item Our analysis of the disk properties from synthetic observations suggests that observations with enough angular resolution to resolve the disks should allow us to recover their sizes with an uncertainties of a factor of $\sim 1.6-2.2$.

\item Their masses, on the other hand, are poorly reproduced, with increasingly larger negative offsets towards very high disk masses that could reach an order of magnitude difference at the higher end ($\sim 1 M_{\odot}$ disks). The potential problem might lie in the optically thin assumption for the conversion from the millimeter flux to the mass. This is due to the fact that at the wavelength we use to simulate the ALMA Band 7 observations $(\lambda=0.89~{\rm mm}$), which corresponds to a frequency of 345 GHz, the dust disks are still optically thick.

\item We varied the angular resolutions of our observations by using three antenna configurations C43-3 ($\theta_{\rm res} = 0.41''-57.4\au$), C43-4 ($\theta_{\rm res} = 0.266''-37.2\au$), and C43-6 ($\theta_{\rm res} = 0.0887''-12.4\au$) and found that different resolutions would return different disk properties. Low resolution (C43-3) would result in overestimating the disk sizes, a sufficient angular resolution (C43-4) to resolve the disks would be the most suitable for the Gaussian modelling of the disk fluxes, whereas overly high-resolution (C43-6) observations might need a more complex description for the disk intensity profile. 

\item Another experiment to use the visibility data from two antenna configurations C43-3 and C43-6 simultaneously for the fitting to extend the $uv$ coverage shows improvements in the results obtained for the disk, which suggests that by extending the $uv$ coverage of the observations, we can better separate  the compact (disk) and extended (envelope) emission.

\item Compared to the two-component model, a single Gaussian model tends to bias the disk size toward larger radii and overall over-estimate the disk radius by a factor of $\sim2$; therefore, it should be used with caution, especially for the modelling of young sources with significant contribution of the envelope to the total emission.

\end{enumerate}

In conclusion, we have demonstrated, via synthetic observations of self-consistent MHD simulations of disk population formation, how accurately the two important observable properties of Class 0/I disks, their sizes and masses, can or cannot be retrieved from ALMA observations, as well as the impact of certain observation designs and data analyses on the modelling results. Future studies making use of this methodology with potentially more advanced observational modelling techniques and greater parameter space would be essential to bridge the gap between the theoretically motivated numerical simulations and interferometric observations of young disks.

\begin{acknowledgements}
We are grateful to the referee for the very constructive and valuable report that improved immensely the overall quality of the paper, and Agnes Monod-Gayraud for her thorough proof-reading of the text. We thank Sylvie Cabrit for her suggestion on the inclusion of the single Gaussian fitting assessment, and Cornelis Dullemond for his tremendous help with the \textsc{Radmc-3d} code. This project has received funding from the European Union's Horizon 2020 research and innovation programme under the Marie Sklodowska-Curie grant agreement No 823823 (DUSTBUSTERS) and from the European Research Council (ERC) via the ERC Synergy Grant {\em ECOGAL} (grant 855130). This work was partly supported by the Italian Ministero dell Istruzione, Universit\`a e Ricerca through the grant Progetti Premiali 2012 – iALMA (CUP C$52$I$13000140001$).  We acknowledge PRACE for awarding us access to the JUWELS super-computer. MG acknowledges the support of the French Agence Nationale de la Recherche (ANR) through the project COSMHIC (ANR-20-CE31- 0009). RSK acknowledges financial support  from the German Excellence Strategy via the Heidelberg Cluster of Excellence (EXC 2181 - 390900948) ``STRUCTURES'', and from the German Ministry for Economic Affairs and Climate Action in project ``MAINN'' (funding ID 50OO2206). RSK also thanks for computing resources provided by the Ministry of Science, Research and the Arts (MWK) of the State of Baden-W\"{u}rttemberg through bwHPC and the German Science Foundation (DFG) through grants INST 35/1134-1 FUGG and 35/1597-1 FUGG, and for data storage at SDS@hd funded through grants INST 35/1314-1 FUGG and INST 35/1503-1 FUGG. GR acknowledges the support from the Fondazione Cariplo, grant no. 2022-1217, and the European Research Council (ERC) under the European Union’s Horizon Europe Research \& Innovation Programme under grant agreement no. 101039651 (DiscEvol). Views and opinions expressed are however those of the author(s) only, and do not necessarily reflect those of the European Union or the European Research Council Executive Agency. Neither the European Union nor the granting authority can be held responsible for them.

\end{acknowledgements}

\bibliographystyle{aa}
\bibliography{bibliography}

\begin{appendix}

\section{\textsc{Radmc-3d} and \textsc{Ramses} temperature profiles}\label{app:temperature}

To obtain the temperature profile with \textsc{Radmc-3d}, we set up the thermal Monte Carlo (MC) runs with $10^8$ photon packages for all the sources present in the specific \textsc{Ramses} output that we want to post-process. Due to the fact that the very few cells around the sink particles are a region of overdense gas and hence high optical depth ($\tau \sim 10$), the photon packages from the position of the star (of point-source type in \textsc{Radmc-3d}) would have to reprocess multiple times before being able to get out of this region, with the risk of being stuck there for a large amount of time. This translates into a long runtime for the MC runs. Therefore, the modified random walk (MRW; \citealt{2009A&A...497..155M}; \citealt{2010A&A...520A..70R}) mode is switched on to speed up the code. Additionally, we chose an embarrassingly parallel approach as an alternative to the parallelisation implemented in the code since its efficiency decreases when increasing the number of computer's threads in use. This is done by diving the total $10^8$ photon packages into $N$ non-parallelised runs, each with $10^8/N$ photon packages and different MC seeds, to calculate the dust temperature. Then, we averaged the temperature profiles from these $N$ runs according to their radiant heat energy:
\begin{equation}
    T_{\rm avg}^4 = \frac{1}{N}\sum_{i=1}^{N} T_i^4,
\end{equation}
where $T_{\rm avg}$ is the final average dust temperature for each cell in the grid and $T_i$ is the dust temperature in the same cell from the $i$-th non-parallelised run. We used $N=100$ runs to achieve the temperature profile with low computational cost.

\begin{figure*}
    \includegraphics[width=\textwidth]{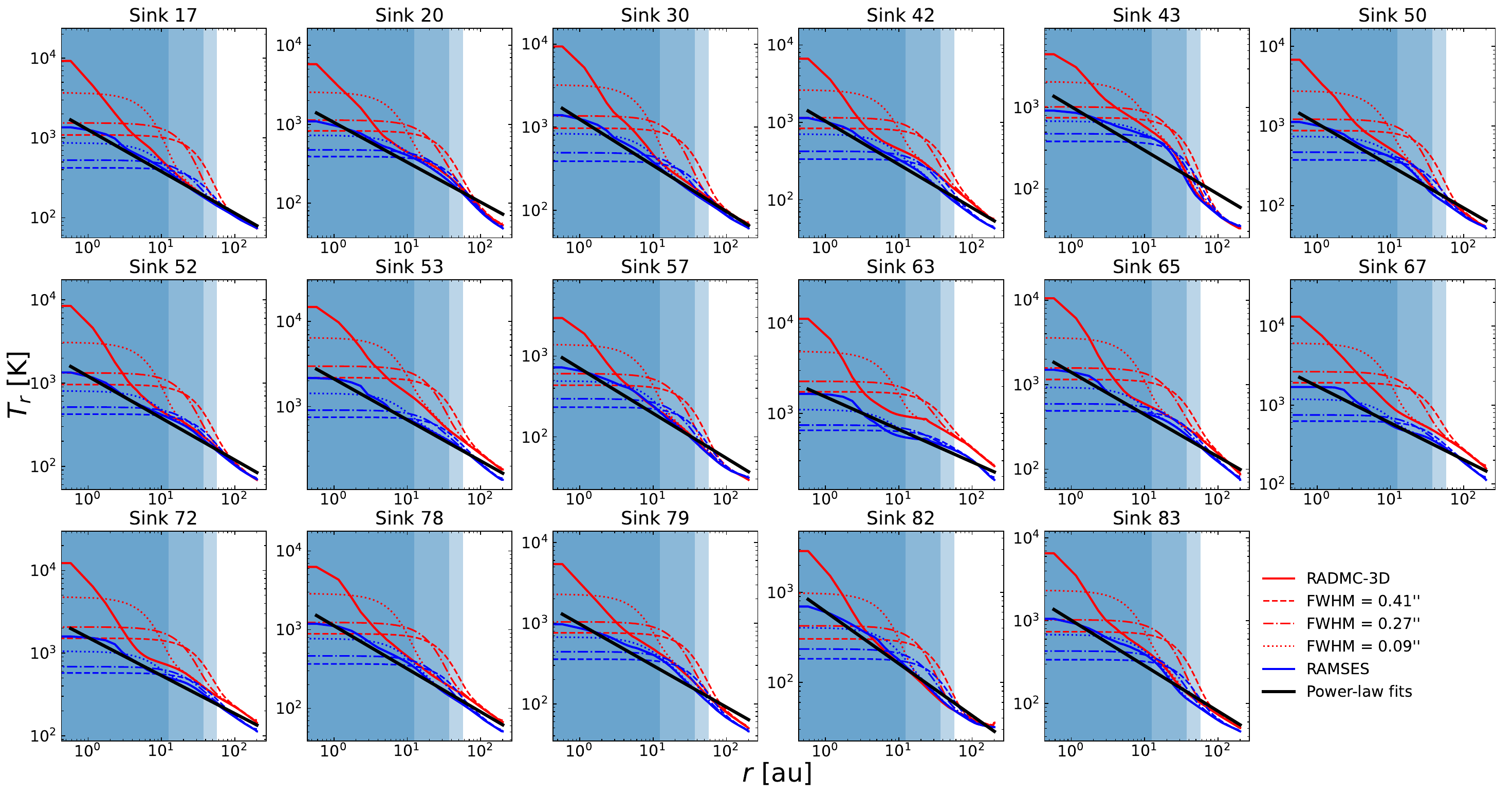}
    \label{fig:temp_radial}
    \caption{Radial temperature profiles from \textsc{Radmc-3d} (red curves) and \textsc{Ramses} (blue curves). The difference is  more pronounced within the sink particle radius ($r \lesssim 4\au$). Dashed, dotted, and dash-dotted lines shows the profiles smoothed by a Gaussian whose FWHM equal to that of the three antenna configurations in this study. The shaded areas marked the resolution limit corresponding to said configurations. The power-law fits to be used for the analytical calculations of the temperature are shown by the solid black line.} 
\end{figure*}

On the other hand, \textsc{Ramses}, with its radiative transfer solver using the gray FLD method, provides us with the temperature profile of the gas and dust in thermal equilibrium within the simulation grid. This radiative transfer approach includes approximation of the difussion of radiation, local production of heat, and gray approximation of the dust opacity. Unlike in \textsc{Radmc-3d}, the energy from the central source is smoothed by a Gaussian kernel centred around the star position and distributed within the radius of the sink particle ($r_{\rm sink} = 4\Delta_{\rm max} \sim 5\au$), and each cell emits like a black body. This is a possible explanation for the difference usually observed within the inner most $10\au$ of the disks, as suggested in the temperature profiles obtained with the two codes shown in Fig. \ref{fig:temp_radial}, which causes difficulties in the parameter fitting and modelling process as the Gaussian is mistakenly fitted to the over-fluxed region instead of the disk. The lack of dust sublimation in \textsc{Radmc-3d}, which is implemented in \textsc{Ramses} could also account for the discrepancy. The same difference was reported in \cite{2020A&A...635A..42M} for their model with high opacity $\tau = 100$ and high stellar temperature $T_{\star}=15000\K$ (see Fig. 4 in \citealt{2020A&A...635A..42M}).

\begin{figure*}
    \includegraphics[width=0.33\textwidth]{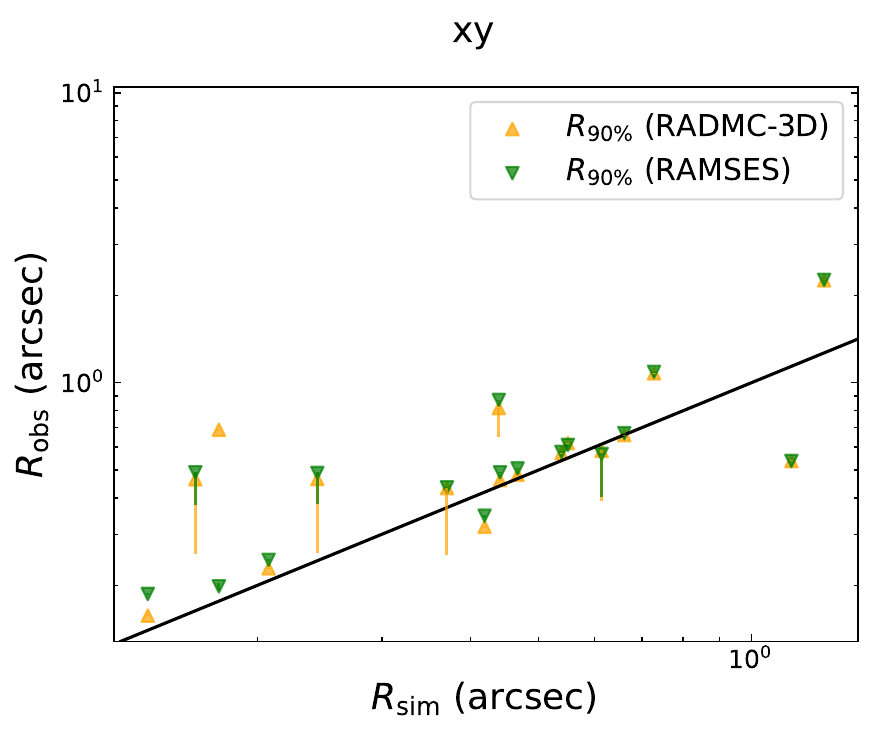}
    \includegraphics[width=0.33\textwidth]{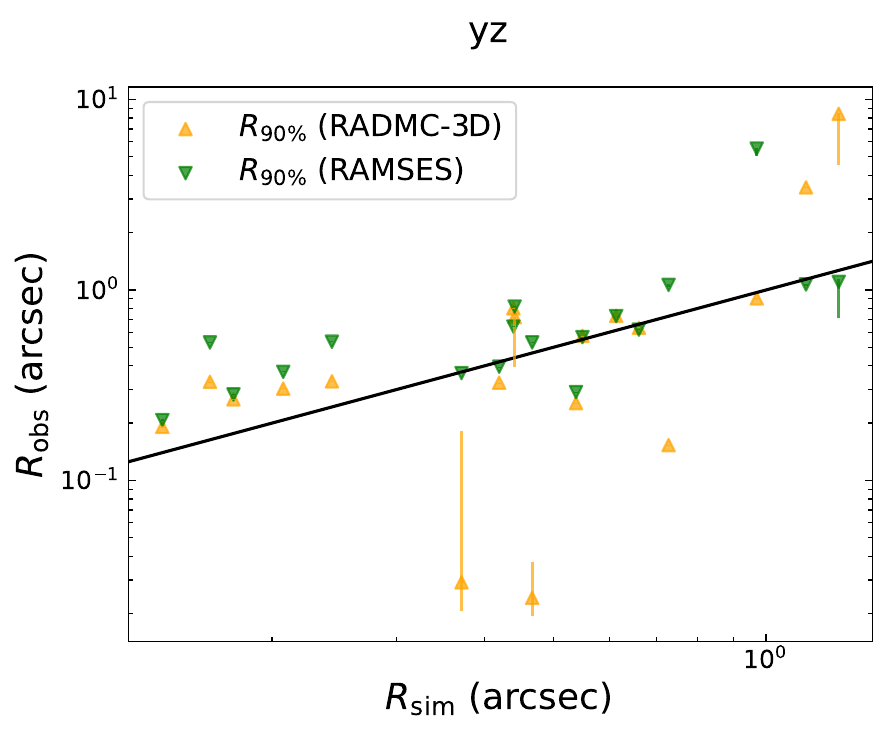}
    \includegraphics[width=0.33\textwidth]{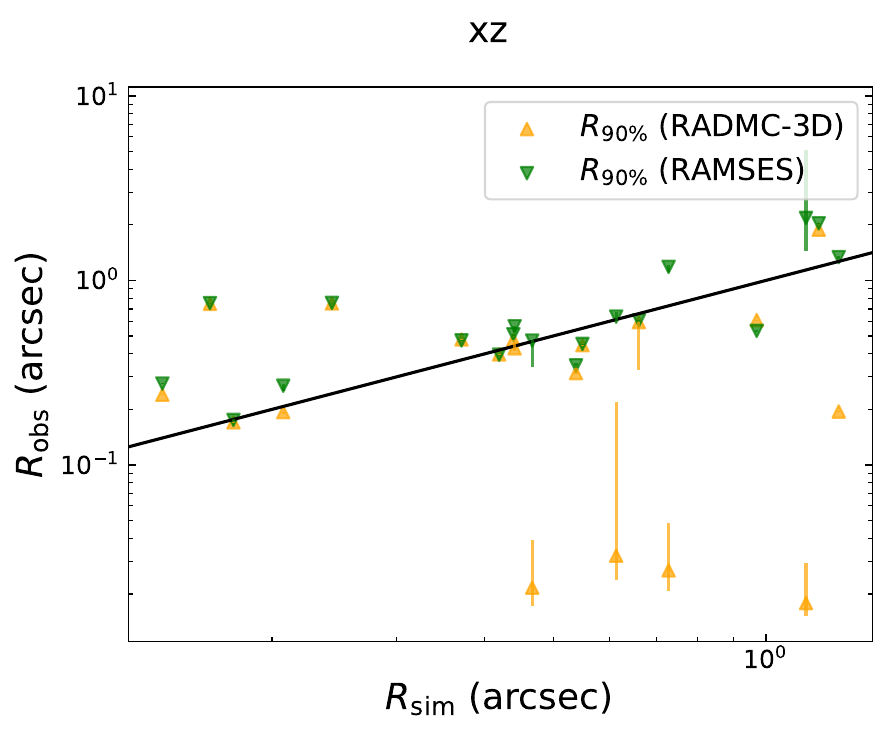}
    \caption{Scatter plots for $R_{90\%}$ extracted with the temperature profiles from \textsc{Radmc-3d} (green dots) and \textsc{Ramses} (orange triangles) for all the single disks in one population versus the disk sizes extracted from the simulation. The results are similar in most cases, except notably for the disks where the central flux produced by \textsc{Radmc-3d} are modelled as the very narrow Gaussian.}
    \label{fig:size_compare}
\end{figure*}

We continue and use both temperature profiles, which we found to be relatively similar in the outer disk (where $r \gtrsim 10\au$) and we performed the ray-tracing, mock interferometric observations, and synthetic modelling to extract the disks in the fiducial case with C43-4. Our goal is to test whether the final results change significantly when changing the profiles as the input for the imaging processes. Figure \ref{fig:size_compare} compares the disk sizes obtained with the temperature from \textsc{Radmc-3d} and \textsc{Ramses}. In most cases, the results are identical, except for the disks in which the central flux resulted from \textsc{Radmc-3d} temperature profile is mistaken by a narrow Gaussian, and very few other disks which are not properly modelled either with \textsc{Radmc-3d} or \textsc{Ramses} profiles.

We note, however, that with C43-6, using \textsc{Ramses} would only eliminate a good portion of the problematic cases, but not all. In other words, some disks still have the central flux mistaken by the Gaussian component. In those very few cases, we could alternatively use the $R_i$ parameter of the envelope, which describes  the disk-envelope structure well, compared to the value obtained from the simulation (see, e.g. Fig. \ref{fig:sigma_combined}), as the radius of the disk.

\section{Synthetic observations and modelling results for the disk population}\label{app:images}

Here, we present the results of our modelling from synthetic observations with \textsc{Casa} with three antenna configuration C43-3, C43-4, and C43-6 for all the disks around single stars seen in one projection $xy$, in the same fashion as Fig. \ref{fig:model_images}. It is worth noting that the selection of single systems is based on the images of the observations obtained with \texttt{tclean}, not the column density plot from \textsc{Ramses} or the sky model from \textsc{Radmc-3d}; that is, without any knowledge of the real objects.

\begin{figure*}
    \centering
    \includegraphics[width=0.9\textwidth]{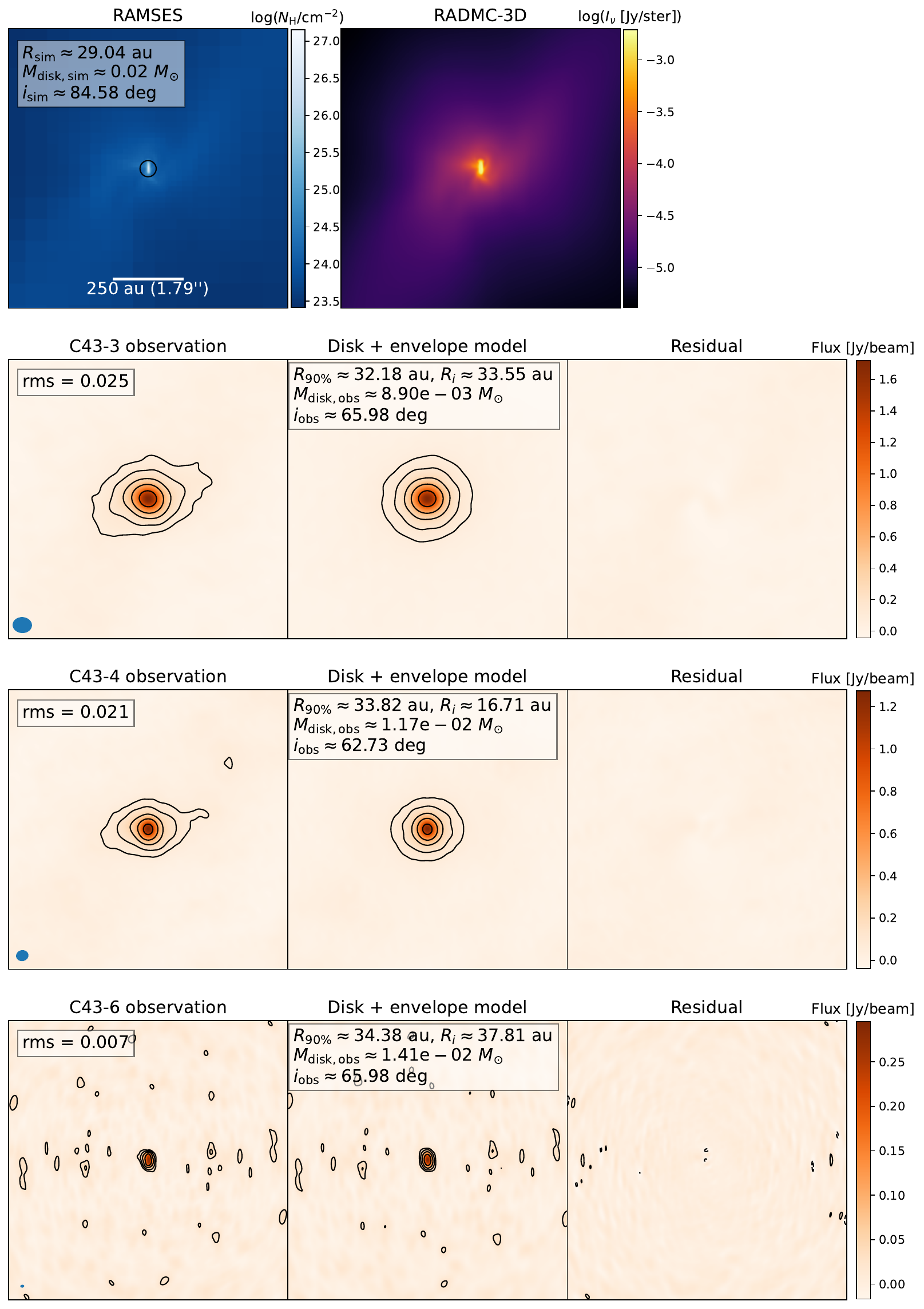}
    \caption{Same as Fig. \ref{fig:model_images}, but for sink 17.}
    \label{fig:sink17}
\end{figure*}

\begin{figure*}
    \centering
    \includegraphics[width=0.9\textwidth]{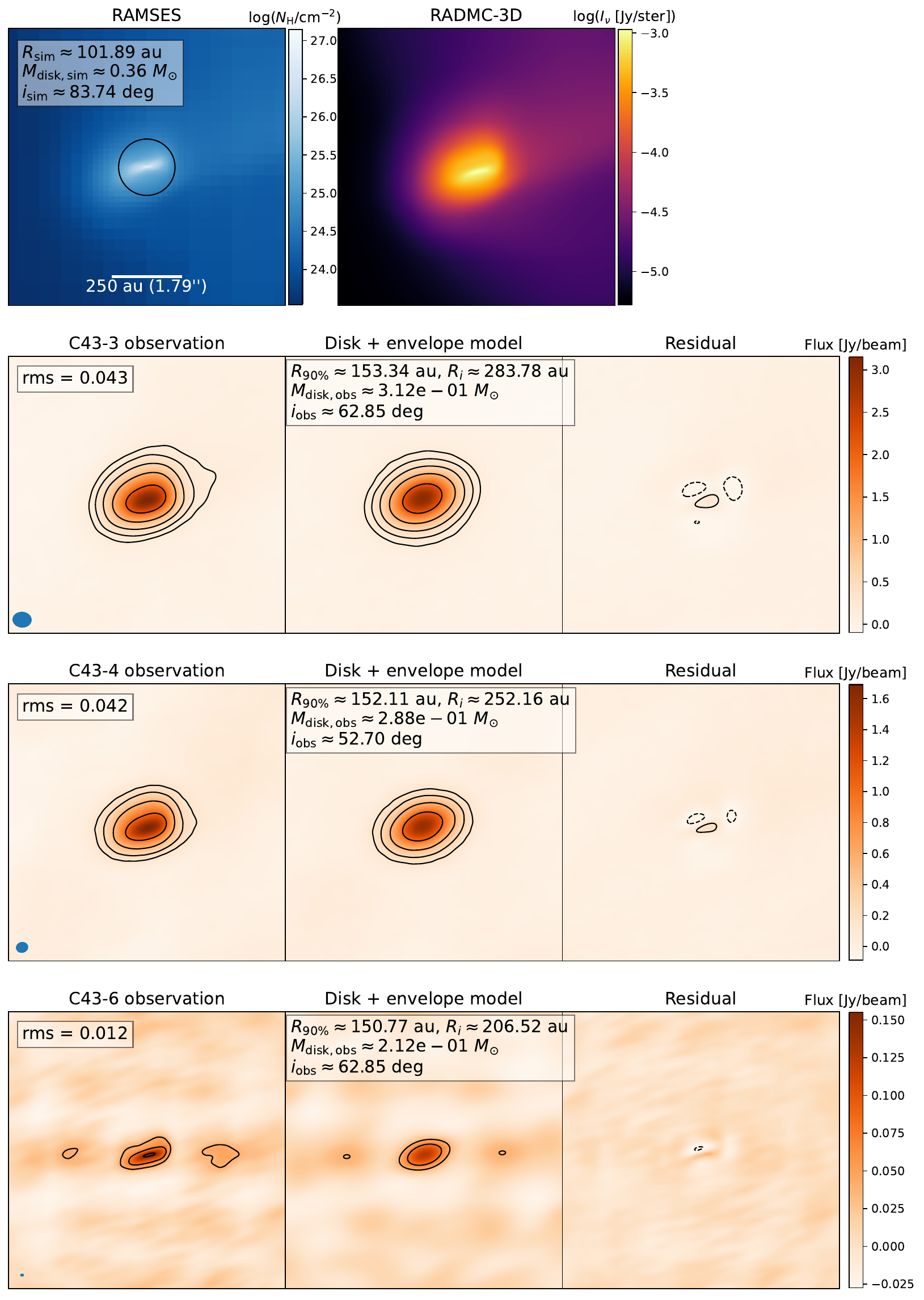}
    \caption{Same as Fig. \ref{fig:model_images}, but for sink 20.}
    \label{fig:sink20}
\end{figure*}

\begin{figure*}
    \centering
    \includegraphics[width=0.9\textwidth]{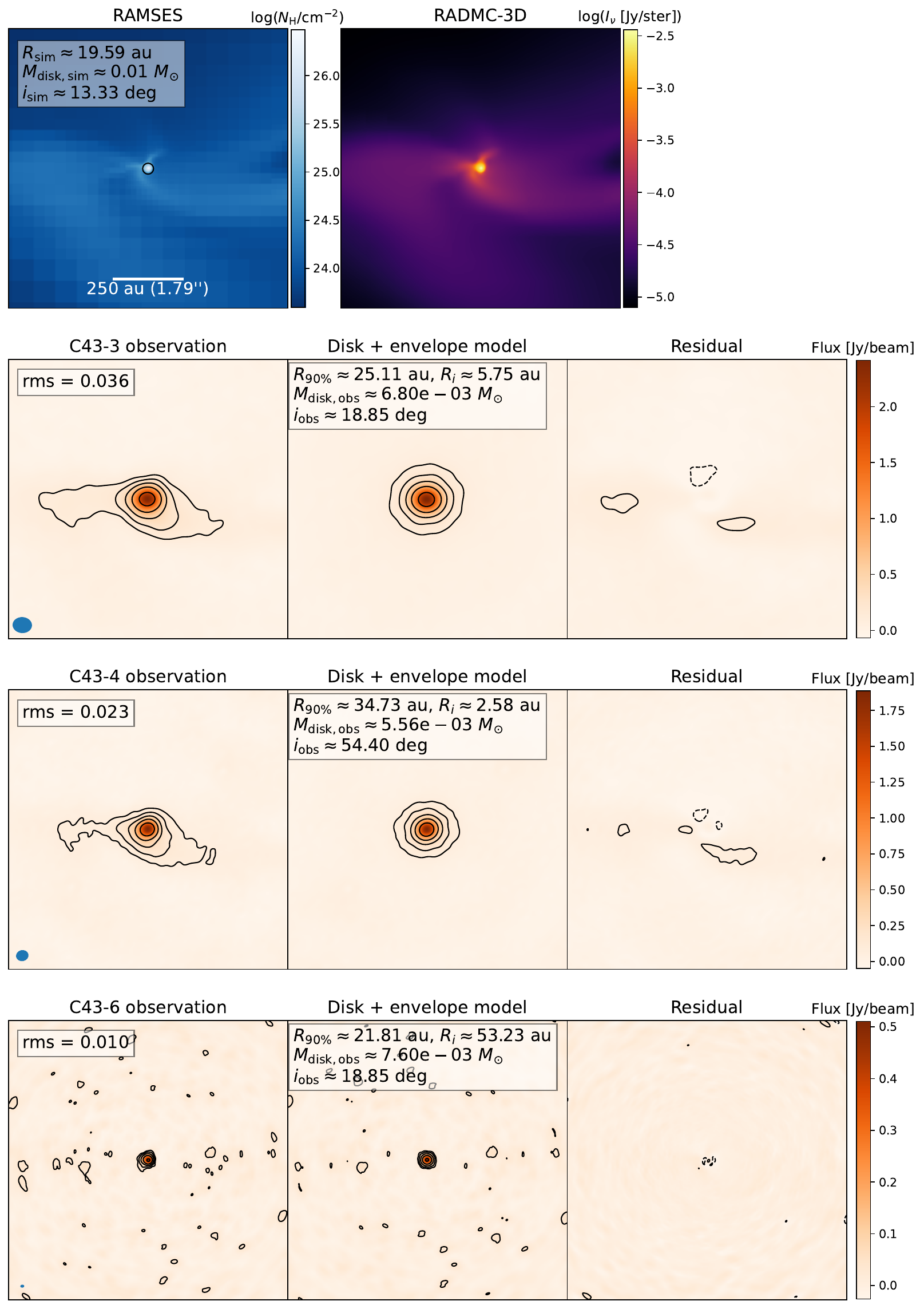}
    \caption{Same as Fig. \ref{fig:model_images}, but for sink 30.}
    \label{fig:sink30}
\end{figure*}

\begin{figure*}
    \centering
    \includegraphics[width=0.9\textwidth]{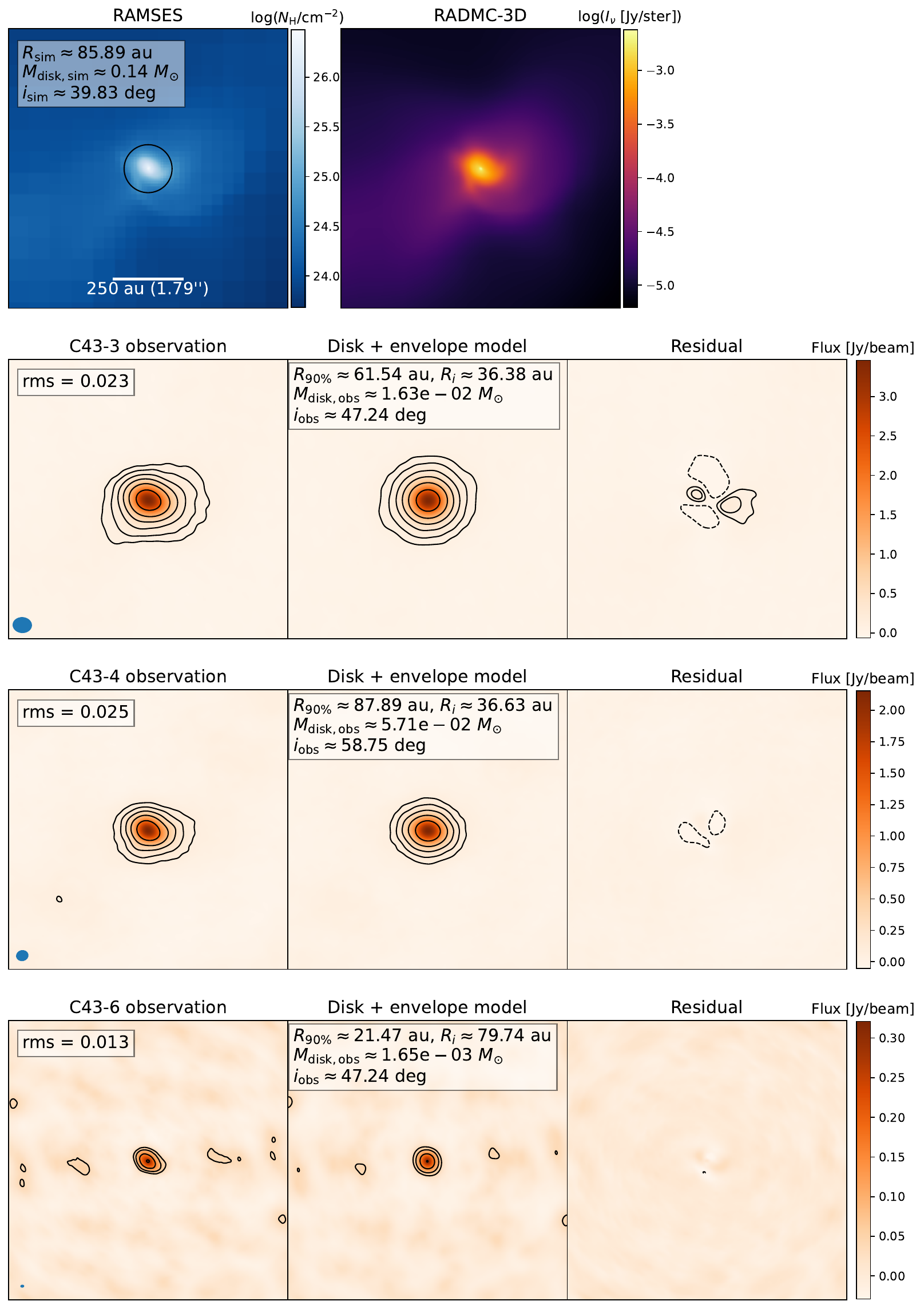}
    \caption{Same as Fig. \ref{fig:model_images}, but for sink 42.}
    \label{fig:sink42}
\end{figure*}

\begin{figure*}
    \centering
    \includegraphics[width=0.9\textwidth]{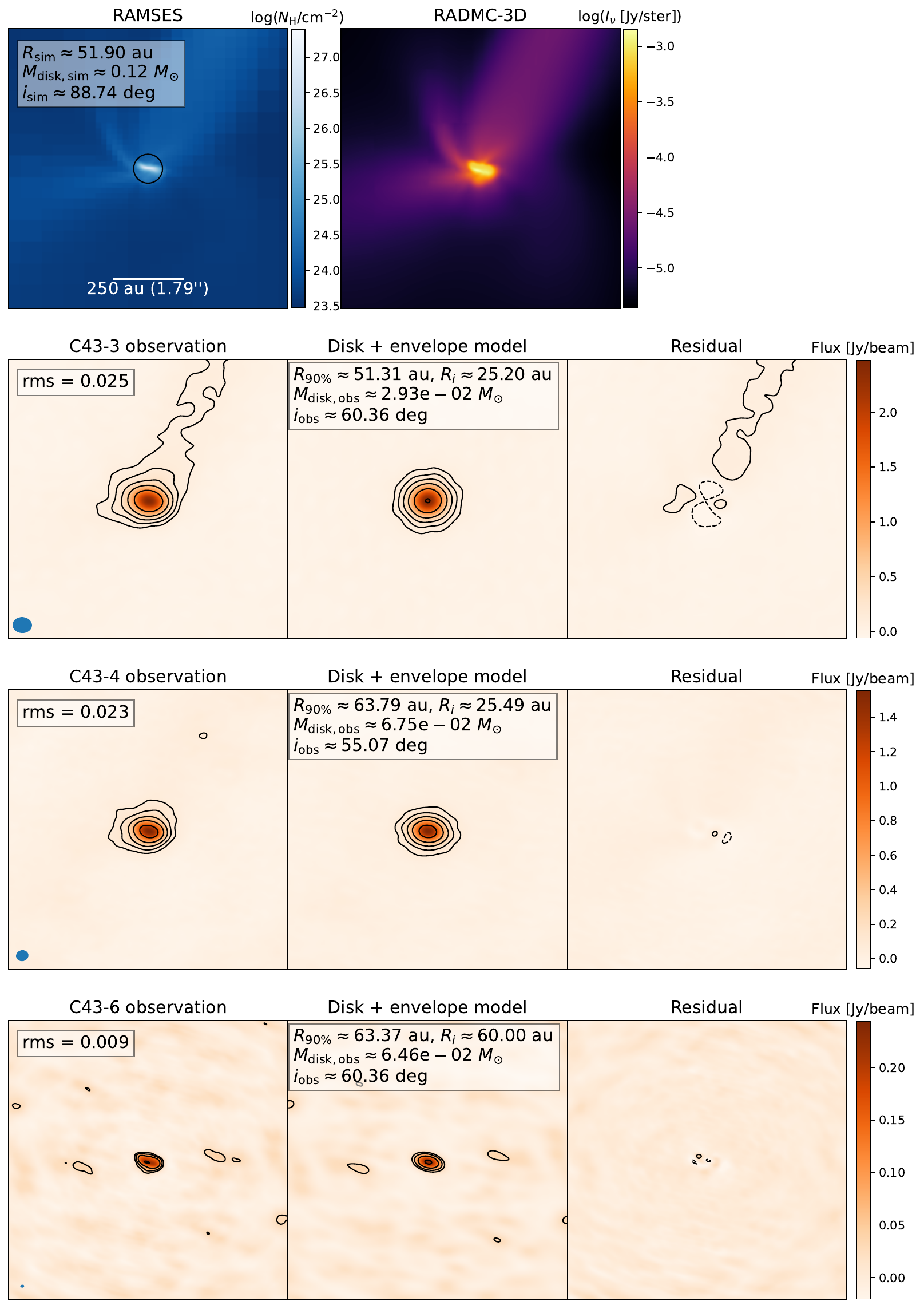}
    \caption{Same as Fig. \ref{fig:model_images}, but for sink 50.}
    \label{fig:sink50}
\end{figure*}

\begin{figure*}
    \centering
    \includegraphics[width=0.9\textwidth]{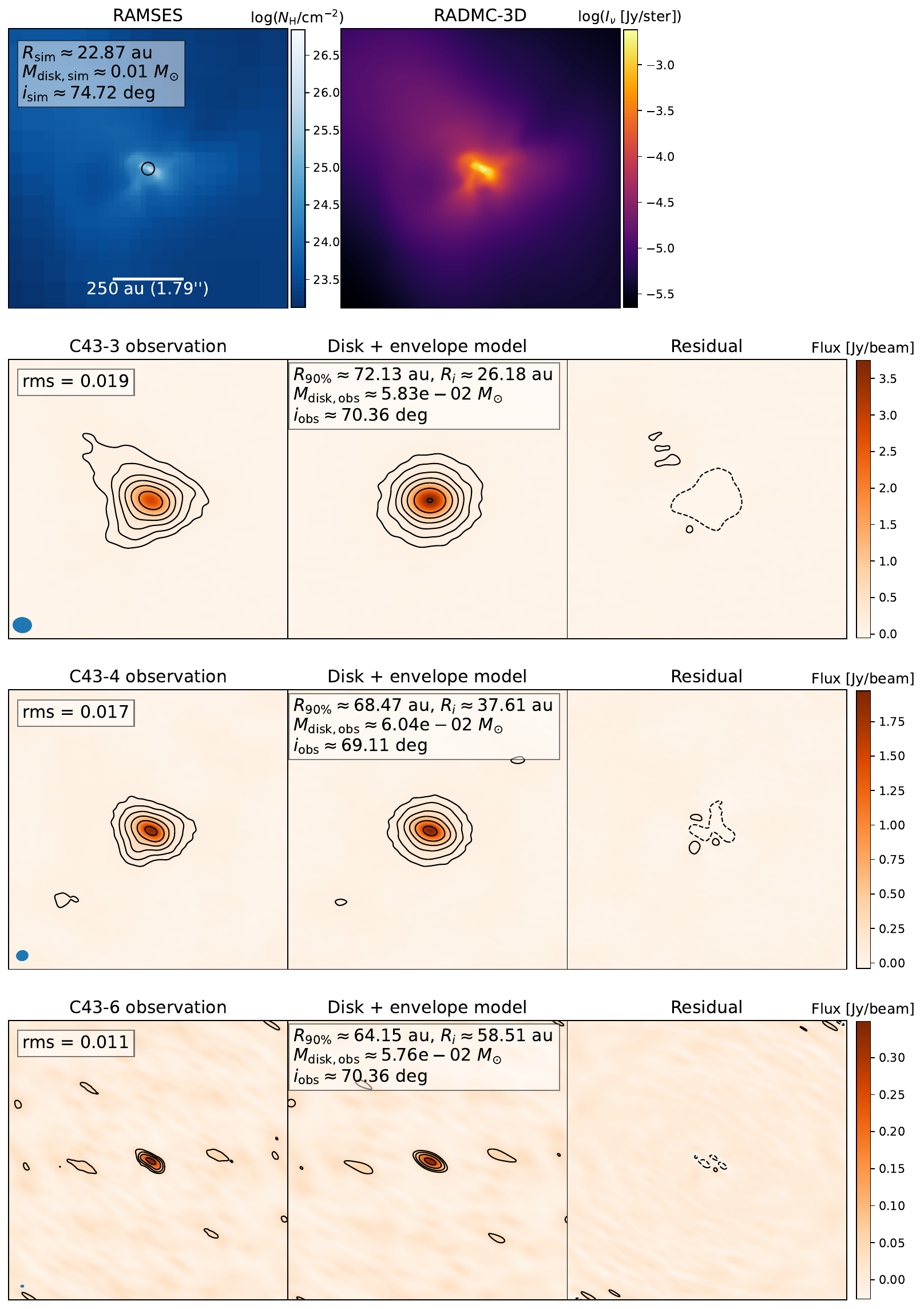}
    \caption{Same as Fig. \ref{fig:model_images}, but for sink 52.}
    \label{fig:sink52}
\end{figure*}

\begin{figure*}
    \centering
    \includegraphics[width=0.9\textwidth]{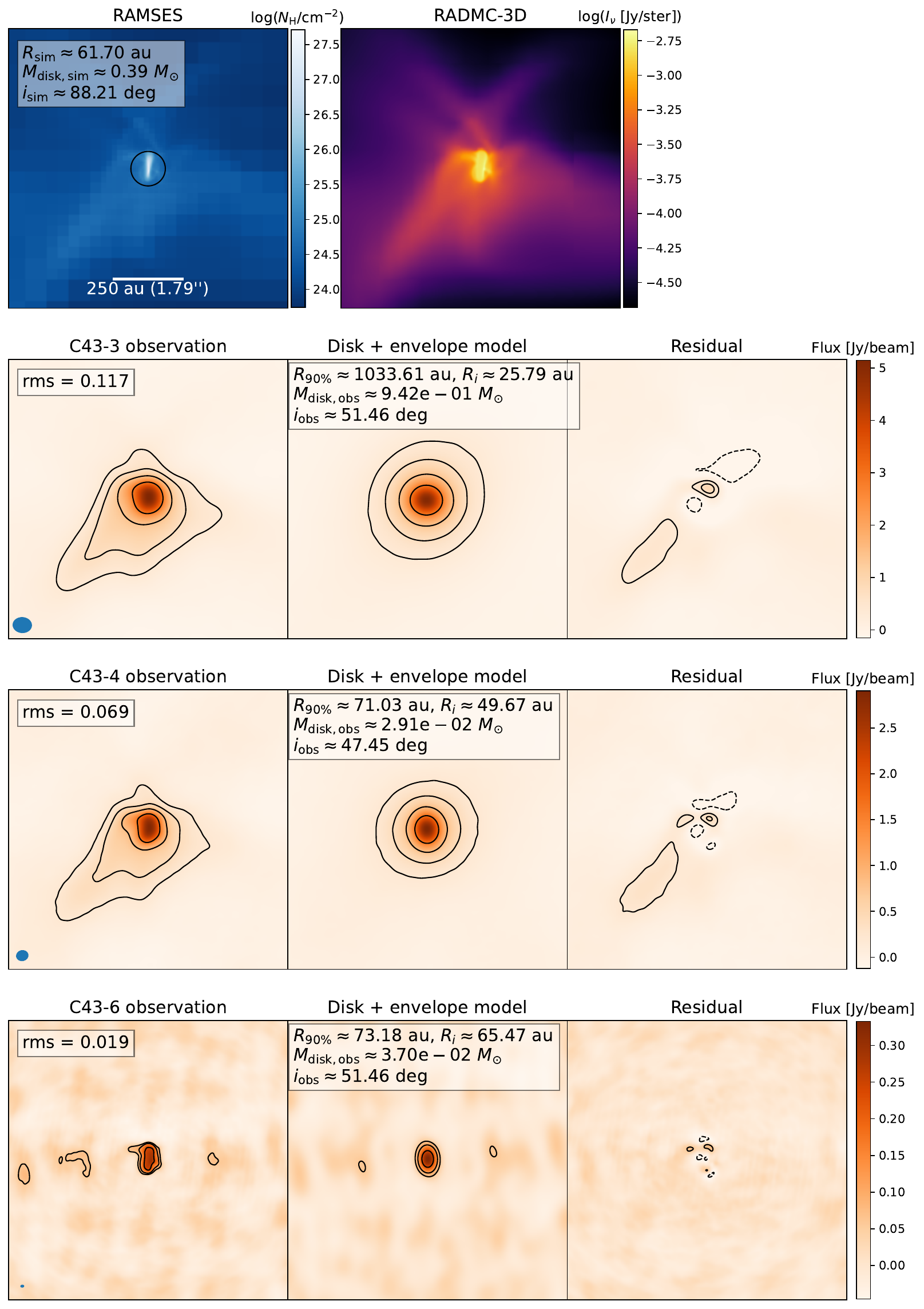}
    \caption{Same as Fig. \ref{fig:model_images}, but for sink 53.}
    \label{fig:sink53}
\end{figure*}

\begin{figure*}
    \centering
    \includegraphics[width=0.9\textwidth]{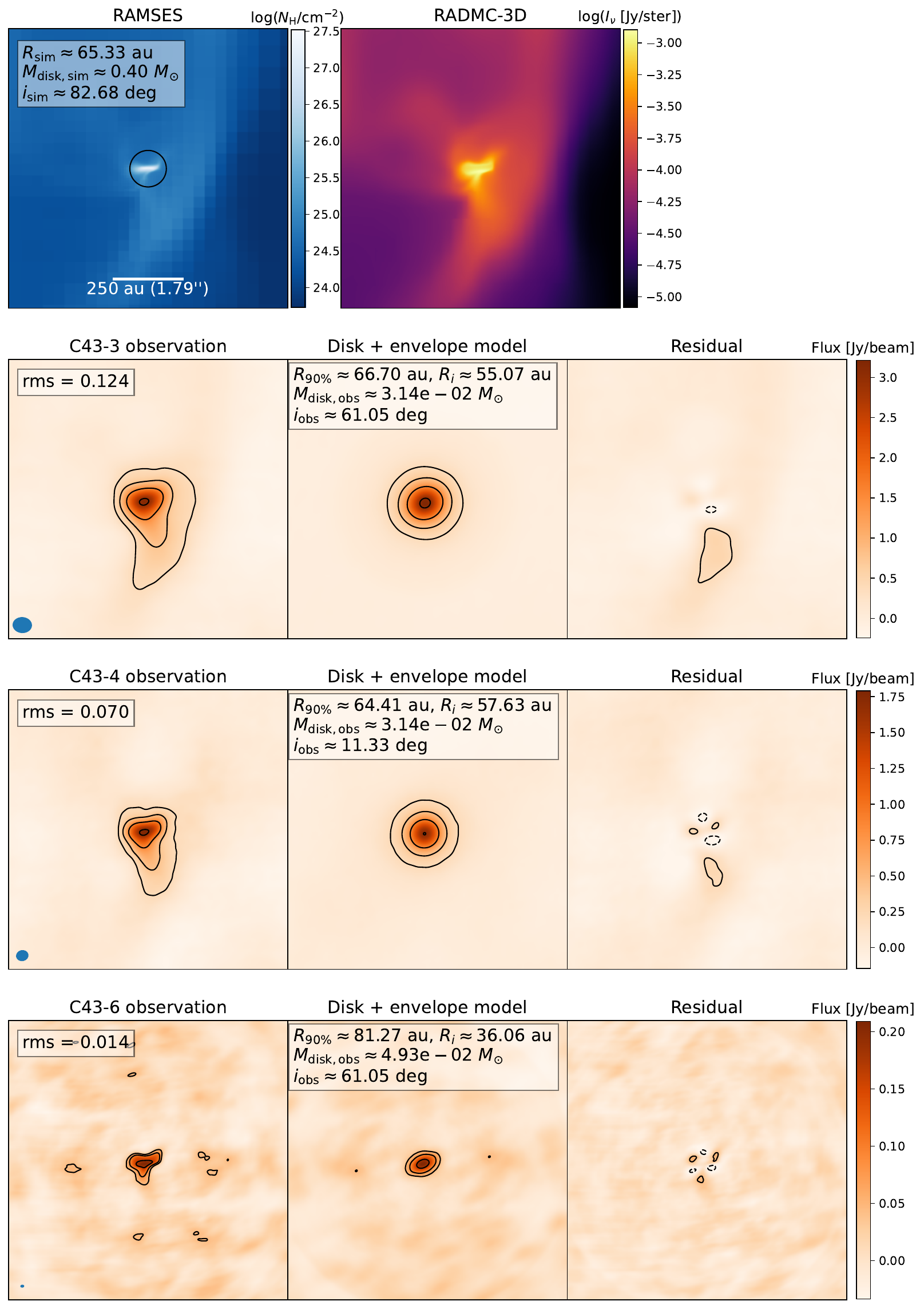}
    \caption{Same as Fig. \ref{fig:model_images}, but for sink 65.}
    \label{fig:sink65}
\end{figure*}

\begin{figure*}
    \centering
    \includegraphics[width=0.9\textwidth]{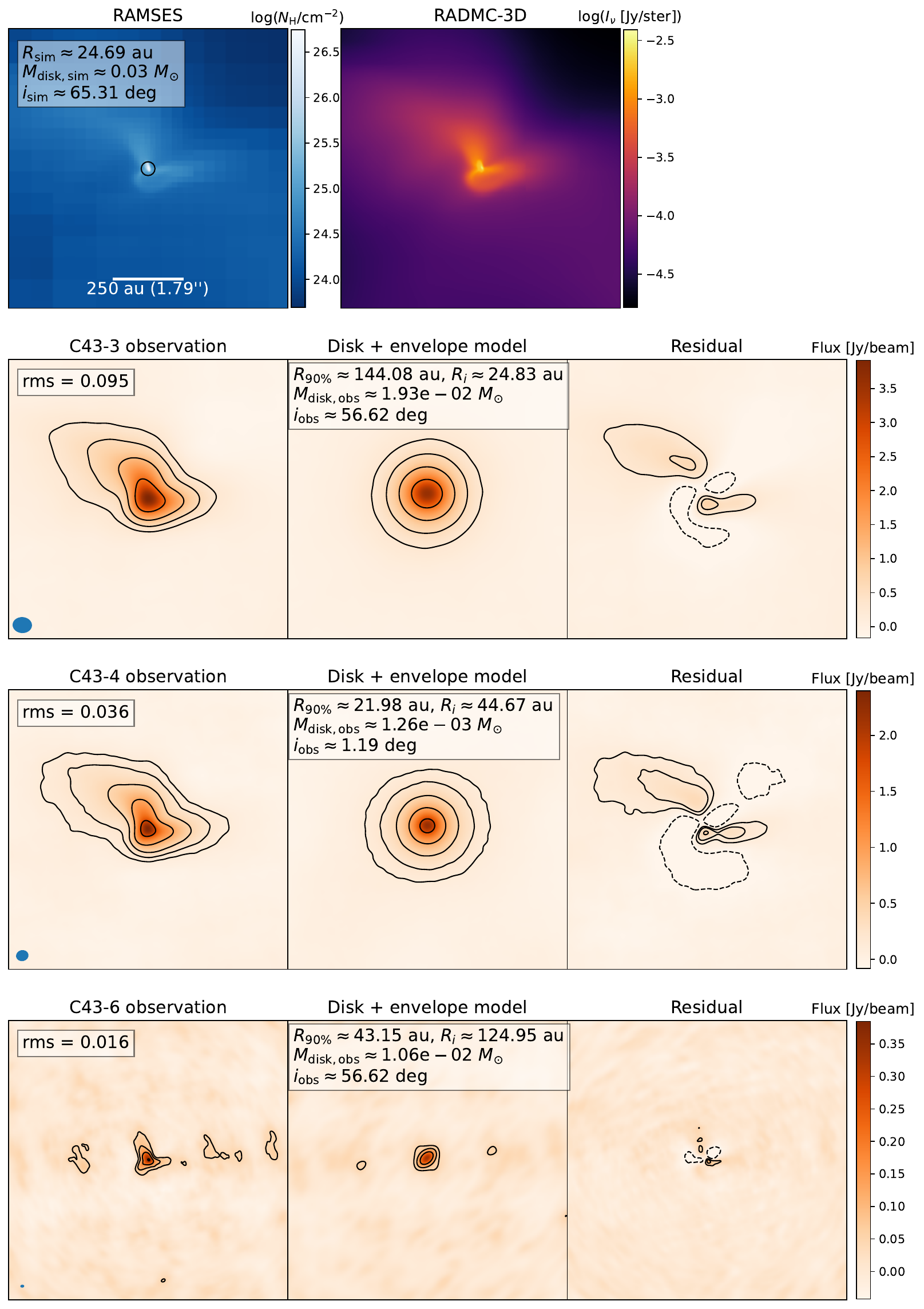}
    \caption{Same as Fig. \ref{fig:model_images}, but for sink 67.}
    \label{fig:sink67}
\end{figure*}

\begin{figure*}
    \centering
    \includegraphics[width=0.9\textwidth]{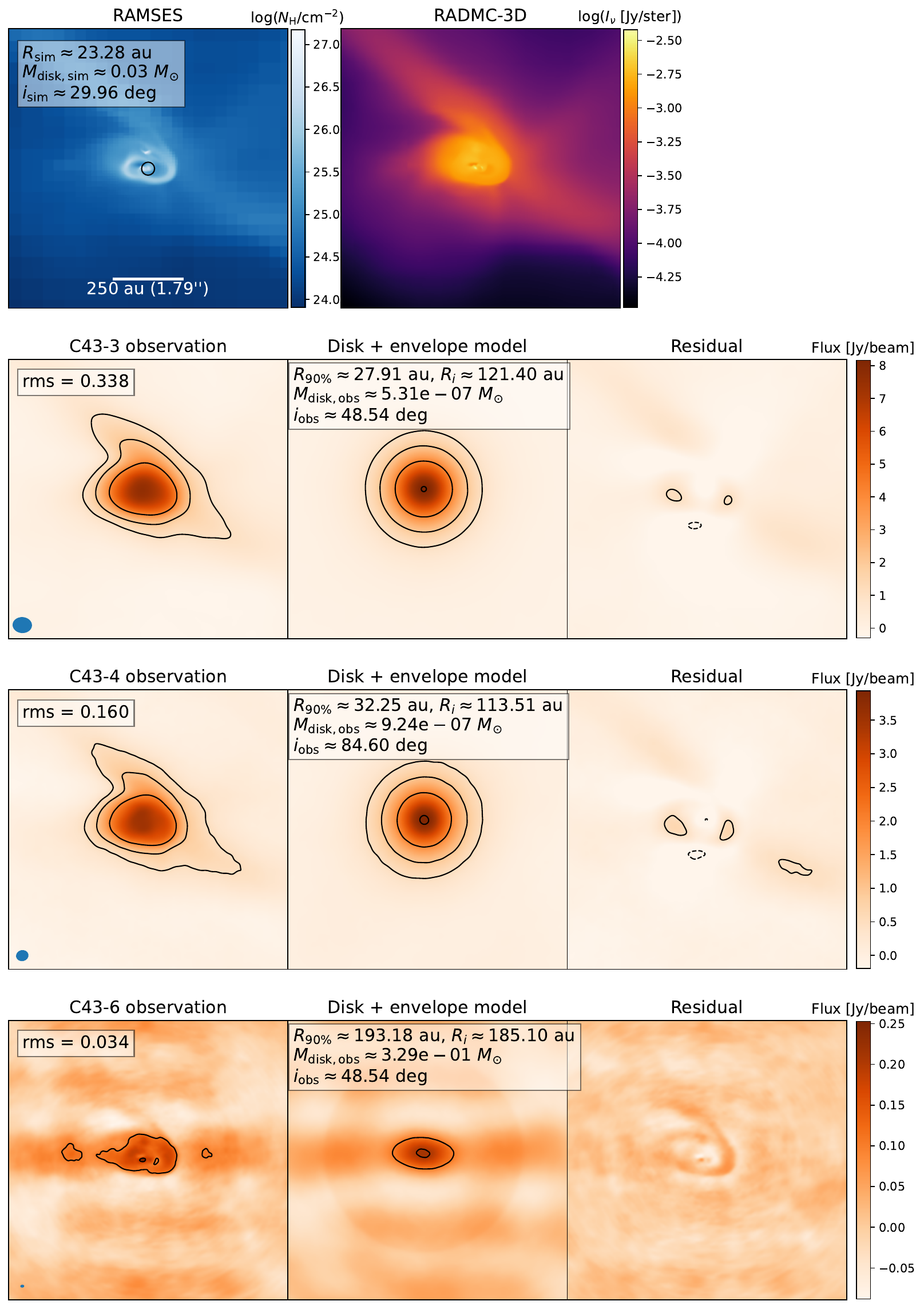}
    \caption{Same as Fig. \ref{fig:model_images}, but for sink 70. For configurations C43-3 and C43-4, the imaged disk appear as a single system; therefore, it is included in the statistic, whereas for C43-6 it is not the case.}
    \label{fig:sink70}
\end{figure*}

\begin{figure*}
    \centering
    \includegraphics[width=0.9\textwidth]{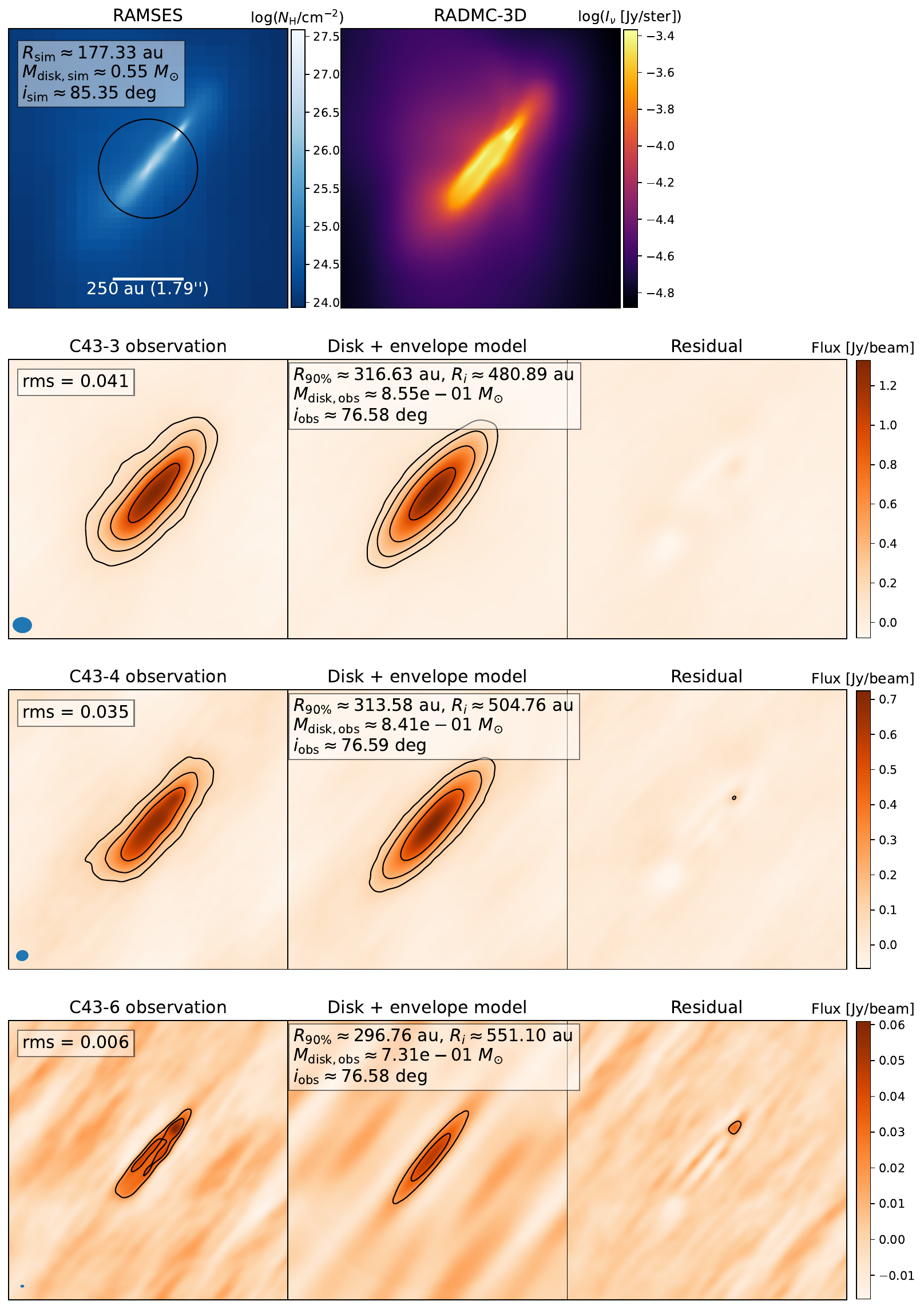}
    \caption{Same as Fig. \ref{fig:model_images}, but for sink 78.}
    \label{fig:sink78}
\end{figure*}

\begin{figure*}
    \centering
    \includegraphics[width=0.9\textwidth]{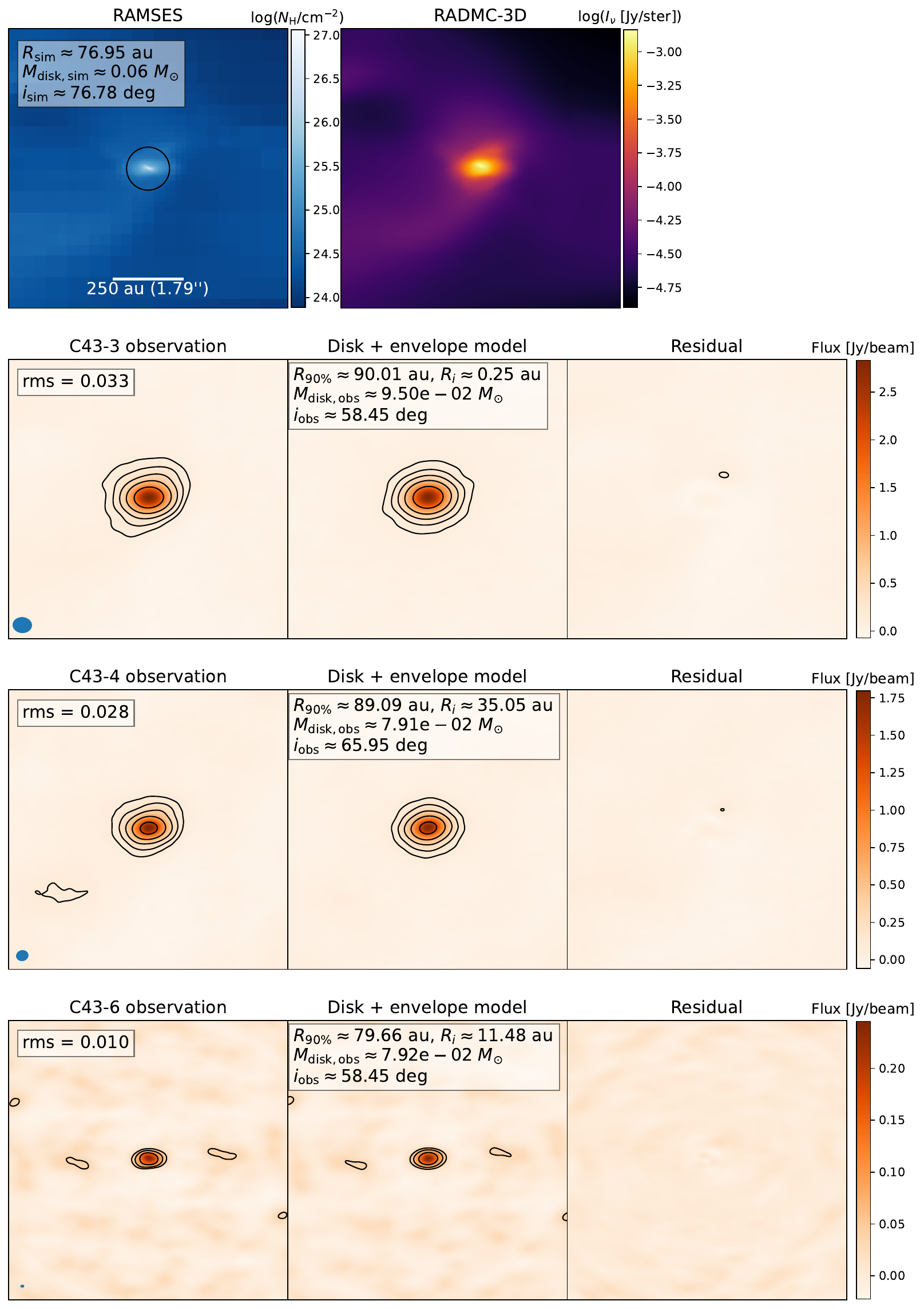}
    \caption{Same as Fig. \ref{fig:model_images}, but for sink 79.}
    \label{fig:sink79}
\end{figure*}

\section{Disk masses modelled from C43-3's and C43-6's data}\label{app:mass}

Below, we show the disk masses we modelled from the observations with C43-3 and C43-6. Similarly to the C43-4 results, the masses are not correctly measured, with shallow slope and large spread in the log-linear fits. With a disk temperature loosely derived from the simulation temperature profiles, we tend to underestimate the mass by a factor of $2-3$ (and in extreme cases: $10)$.

\begin{figure*}
    \includegraphics[width=0.5\textwidth]{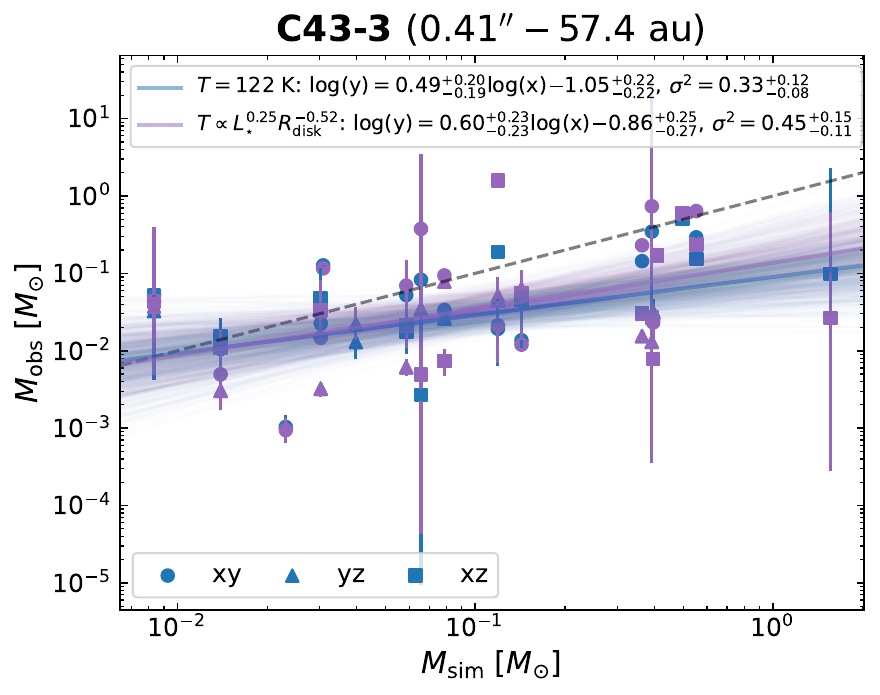}
    \includegraphics[width=0.5\textwidth]{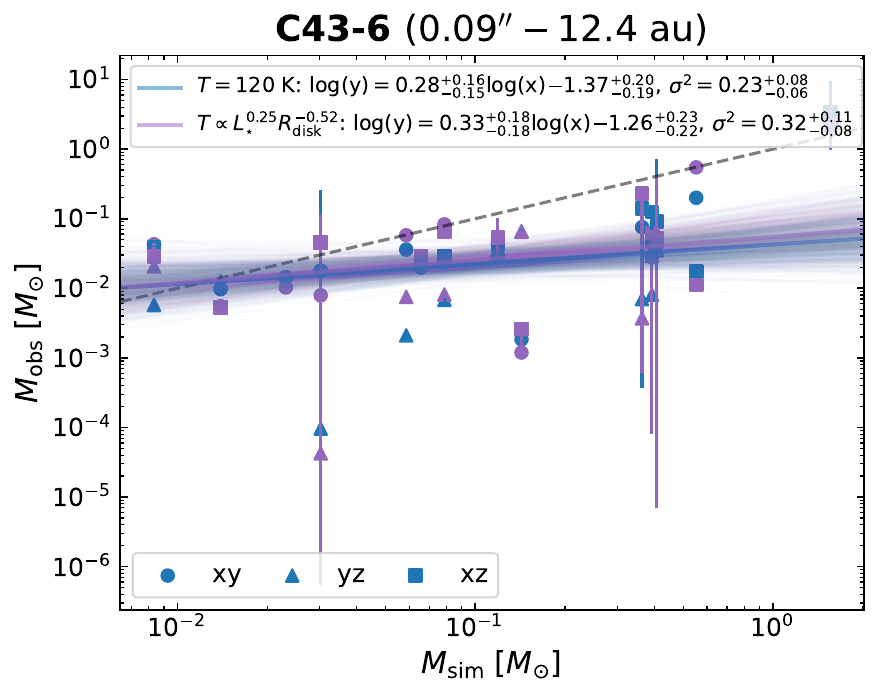}
    \includegraphics[width=0.5\textwidth]{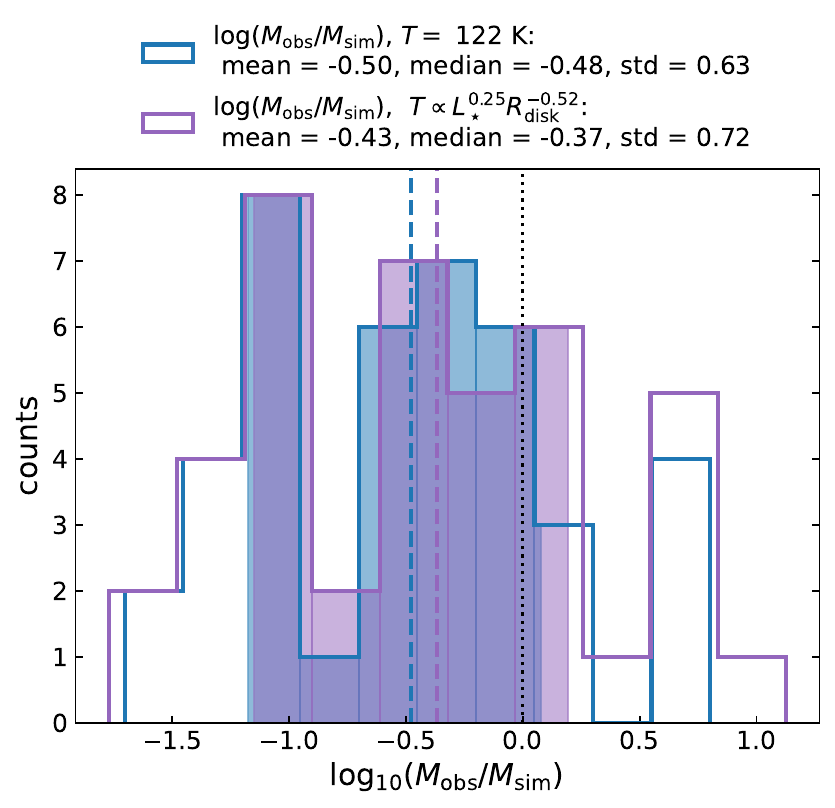}
    \includegraphics[width=0.5\textwidth]{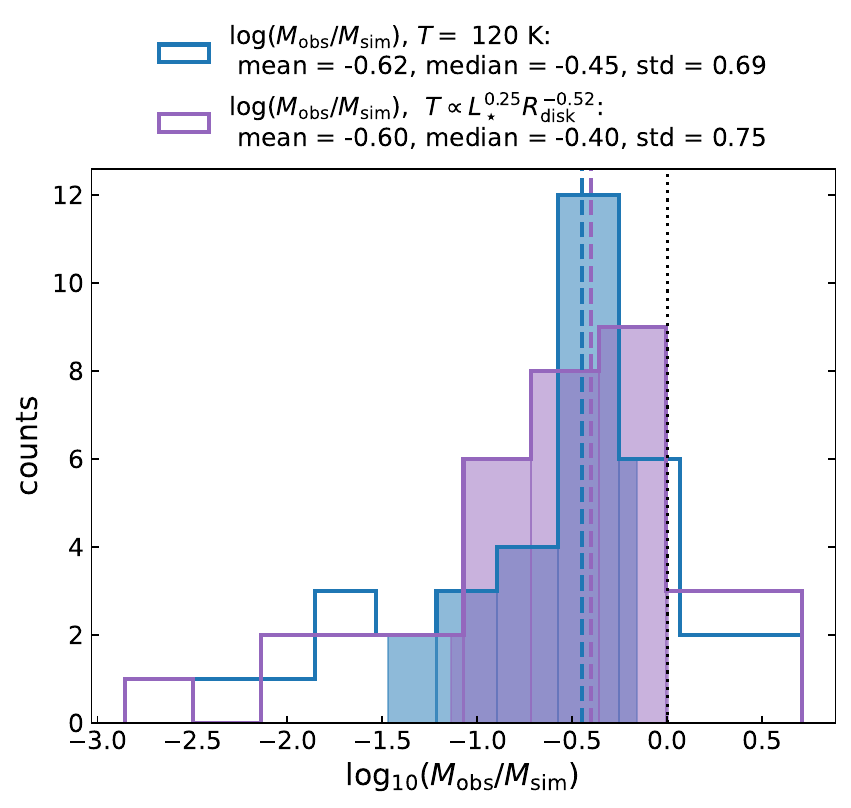}
    \caption{Scatter plots of the modelled disk masses versus the ones inferred from the simulation for C43-3 (left) and C43-6 (right) using the constant median (in blue) and variable disk temperature (in purple), shown at the top. Solid colored show the log-linear fits for the data points and the shaded areas indicate the corresponding confidence intervals. Histograms of the $M_{\rm obs}/M_{\rm sim}$ ratios in logarithmic scale. Vertical dashed lines indicate the medians of the ratio distributions, shown at the bottom.}
    \label{fig:disk_mass_apendix}
\end{figure*}

\end{appendix}

\end{document}